\documentclass{article}

\usepackage{arxiv}

\usepackage[utf8]{inputenc} % allow utf-8 input
\usepackage[T1]{fontenc}    % use 8-bit T1 fonts
\usepackage{hyperref}       % hyperlinks
\usepackage{url}            % simple URL typesetting
\usepackage{booktabs}       % professional-quality tables
\usepackage{amsfonts}       % blackboard math symbols
\usepackage{nicefrac}       % compact symbols for 1/2, etc.
\usepackage{microtype}      % microtypography
\usepackage{graphicx}       % graphics package
\usepackage{natbib}         % package for customising citations
\usepackage{doi}            % create correct hyperlinks for DOI numbers
\usepackage{xcolor}         % color package
\usepackage{CJK}            % Japanese font
\usepackage{enumerate}      % enumerate package
\usepackage{indentfirst} 

\hypersetup{
  colorlinks=true,
  linkcolor=red,
  citecolor=orange,
  urlcolor=blue,
}

\graphicspath{{../figure}}

\title{Numerical simulation for axis switching of pulsating jet issued from rectangular nozzle at low Reynolds number
}

\author{\href{https://orcid.org/0000-0002-4875-8174}{\includegraphics[scale=0.06]{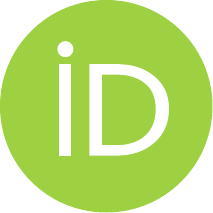}\hspace{1mm}Hideki Yanaoka (柳岡英樹)}
\thanks{Email address for correspondence: yanaoka@iwate-u.ac.jp} \\
	Department of Systems Innovation Engineering, \\
    Faculty of Science and Engineering, Iwate University, \\
    4-3-5 Ueda, Morioka, Iwate 020-8551, Japan \\
	\And
	Yoshitomo Hatakeyama (畠山嘉智) \\
    Mechanical Engineering, \\
    Graduate School of Engineering, Iwate University, \\
    4-3-5 Ueda, Morioka, Iwate 020-8551, Japan
	\texttt{} \\
}

\renewcommand{\shorttitle}{Axis switching of pulsating jet issued from rectangular nozzle at low Reynolds number}

\makeatletter
\hypersetup{
pdfusetitle,
bookmarksnumbered,
pdftitle={\@title},
pdfsubject={},
pdfauthor={Hideki Yanaoka, Yoshitomo Hatakeyama},
pdfkeywords={Axis switching, Mixing, Vortex, Low Reynolds number, Rectangular nozzle, Pulsating jet, Unsteady flow, Numerical simulation},
}
\makeatother

\begin{document}

\begin{CJK*}{UTF8}{ipxg} % use Japanese font
\maketitle
\end{CJK*}

%%% Abstract
\begin{abstract}
Axis switching of a jet ejected from a rectangular nozzle affects flow mixing characteristics. 
To elucidate such a mixing mechanism, 
the axis switching and vortex structure deformation should be investigated in detail. 
This study performed a numerical analysis of the axis switching of a pulsating jet 
ejected from a rectangular nozzle at a low Reynolds number. 
At all aspect ratios, 
a rectangular vortex ring similar to the shape of the nozzle cross-section is periodically shed downstream, 
and the side of the vortex ring deforms into a hairpin shape downstream. 
A vortex pair is generated inside the vortex ring downstream of the nozzle corner. 
When the aspect ratio is $AR = 1.0$, the vortex pair consists of symmetrical vortices, 
while as $AR$ increases, the asymmetry of the vortex pair enlarges. 
At $AR = 1.0$, regeneration of a vortex ring occurs downstream. 
For $AR = 2.0$, alternately on the long and short sides of the nozzle, 
an upstream vortex ring overtakes a downstream vortex ring. 
Regardless of $AR$, downstream near the nozzle, 
as the vortex pair existing inside the vortex ring distorts the vortex ring, 
the positions of the side and corner of the vortex ring exchange, 
resulting in a 45-degree axis switching. 
For $AR > 1.0$, further downstream, 
the hairpin part of the vortex ring on the long side develops 
away from the jet center compared to the short side, 
causing a 90-degree axis switching. 
As a result, high turbulence occurs over a wide area, 
strengthening the mixing action. 
As $AR$ increases, intensive interference between the vortex rings 
on the upstream and downstream sides diffuses the vortices downstream. 
Then, as turbulence by the diffused vortices widely occurs, 
the mixing effect is further strengthened.
\end{abstract}

% Keywords

%##############################################################################
\section{Introduction}
%##############################################################################

Mixing and diffusion by jets ejected from circular nozzles are widely used 
in various engineering devices such as combustors and chemical reactors. 
Therefore, many studies have been conducted on the mixing characteristics 
in jets ejected from circular nozzles 
\citep{Hussain&Zaman_1981,Suto_et_al_2002,Gohil_et_al_2012}. 
In recent years, miniaturization of equipment has been demanded 
from the viewpoint of material cost reduction, weight saving, and efficiency improvement. 
The Reynolds number of the flow in a device decreases due to the miniaturization of the device. 
As the flow at a low Reynolds number becomes laminar, 
the mixing due to turbulence cannot be expected. 
Therefore, efficient mixing promotion technology in low Reynolds number flow fields is required.

Regarding a jet ejected from a circular nozzle, 
research has been conducted to control the flow by pulsation of the jet. 
As the active control of a jet, a method of exciting vortex structure 
by the jet outlet velocity is utilized. 
A blooming jet is generated by applying a streamwise velocity 
combined with axial and helical excitation. 
Existing experiments \citep{Reynolds_et_al_2003} and calculations \citep{Tyliszczak_2018} 
have confirmed the phenomenon of bifurcation into multi-armed jets. 
In addition, a study on controlling flames formed on circular nozzles 
has also been carried out. 
\citet{Tyliszczak&Wawrzak_2022} performed a large eddy simulation 
of a lifted H$_2$/N$_2$ flame excited by an axial and flapping forcing. 
The lift-off height of the flame and its global shape changed 
depending on the method of forcing and its frequency. 
The excitation increased the level of temperature fluctuations 
caused by an intensified mixing process.

One of the mixing promotion techniques is to use a non-circular nozzle. 
Jets ejected from non-circular nozzles are known to exhibit flow axis switching phenomena. 
For jets ejected from rectangular nozzles 
\citep{Sforza_et_al_1966,Sfeir_1976,Krothapalli_et_al_1981,Tsuchiya&Horikoshi_1986,Quinn_1992,Toyoda_et_al_1992,Miller_et_al_1995,Zaman_1996,Grinstein_2001,Rembold_et_al_2002,Chen&Yu_2014} 
and elliptical nozzles \citep{Ho&Gutmark_1987,Hussain&Husain_1989,Miller_et_al_1995,Quinn_2007,Zhang&Chua_2012}, 
we can observe a phenomenon in which the major and minor axes of the cross-section 
of the jet are interchanged downstream. 
In addition, for jets ejected from square nozzles \citep{Gutmark_et_al_1989,Toyoda_et_al_1992,Miller_et_al_1995,Grinstein_et_al_1995,Grinstein_2001,Chen&Yu_2014,Gohil_et_al_2015} 
and triangular nozzles \citep{Koshigoe_et_al_1989,Toyoda_et_al_1992,Miller_et_al_1995,Azad_et_al_2012}, 
the axis and diagonal positions of the cross-section of the jet are interchanged downstream. 
\citet{Zaman_1996} defined the switching of the major and minor axes 
in the transverse cross-section of a jet as the 90-degree axis switching 
for elliptical or rectangular nozzles 
and the switching of the axis and diagonal position in the cross-section 
as the 45-degree axis switching for the square nozzle. 
It has been confirmed that when such a phenomenon of axial switching occurs, 
the diffusion and mixing of the jets ejected from non-circular nozzles are better 
than those of circular nozzles \citep{Miller_et_al_1995}. 
Experiments on jets ejected from elliptical nozzles have shown 
that the elliptical vortex ring formed in the jet has a non-uniform curvature, 
and such three-dimensional deformation of the vortex promotes fluid mixing, 
causing the axis switching due to the induced flow by the vortex 
\citep{Hussain&Husain_1989,Ho&Gutmark_1987}. 
\citet{Toyoda_et_al_1992} conducted experiments on jets emitted from square, 
equilateral triangular, and rectangular nozzles. 
The deformation of the non-circular vortex structure and the elongation and splitting of vortex structures 
in the interaction process between that non-circular vortex and other vortices 
produced the mixing-promoting effect, 
which was most pronounced when using a rectangular nozzle.

As the characteristics of jets are governed by the behavior of vortices in the flow field, 
many studies have focused on vortex structures in jets at high Reynolds numbers 
\citep{Ho&Gutmark_1987,Hussain&Husain_1989,Toyoda_et_al_1992,Grinstein_et_al_1995,Zaman_1996,Grinstein_2001}. 
On the other hand, the relationship between vortex structures 
and the phenomena of axis switching at low Reynolds numbers has not been investigated in detail. 
At low Reynolds numbers, free jets ejected from non-circular nozzles exhibited no flow axis switching 
\citep{Gohil_et_al_2015}. 
However, when the jet pulsated, even at low Reynolds numbers, 
a square nozzle caused axis switching \citep{Gohil_et_al_2015}. 
Therefore, it is considered that pulsating jets ejected from rectangular nozzles 
with various aspect ratios also show axial switching. 
\citet{Straccia&Farnsworth_2021} experimentally investigated 
axis switching in rectangular orifice synthetic jets for aspect ratios $AR = 2-19$. 
For $AR = 2-6$, the jet axis switches two to three times. 
However, at $AR = 13$ and 19, the jet axis only switches once. 
The absence of additional axis switching is due to the collision of the vortex ring with itself 
and the bifurcation of the vortex ring. 
The axial profiles of centerline velocity for the $AR = 4-19$ jets exhibited two local peaks. 
These peaks are due to the dynamics of the primary vortex ring and not the secondary vortex. 
A pulsating jet ejected from a rectangular nozzle is a complicated flow, 
even at a low Reynolds number. 
Further investigations on the axis switching and vortex structure deformation are required 
to elucidate the flow mixing characteristics with a non-circular nozzle.

In this study, to elucidate the relationship between vortex structures and axis switching, 
we perform a numerical analysis of the pulsating jet ejected from a rectangular nozzle 
at a low Reynolds number 
and investigate the formation of vortex structures in the jet, 
deformation process, interaction between vortices, 
and generated local turbulence.

%##############################################################################
\section{Numerical procedures}
%##############################################################################

\begin{figure}[!t]
\centering
\includegraphics[trim=0mm 0mm 0mm 0mm, clip, width=70mm]{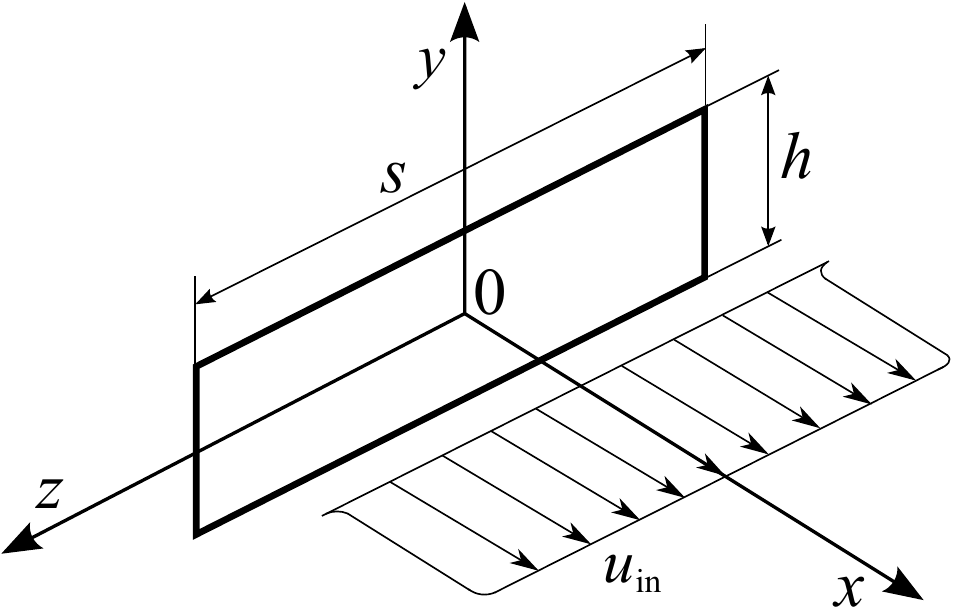} \\
%\vspace*{-0.5\baselineskip}
\caption{Flow configuration and coordinate system.}
\label{flow_model}
\end{figure}

Figure \ref{flow_model} shows the flow configuration and coordinate system. 
The origin is on the center axis at the nozzle exit, 
and the $x$-, $y$-, and $z$-axes are the streamwise, cross-streamwise, 
and spanwise directions, respectively. 
The velocities in each direction are denoted as $u$, $v$, and $w$, respectively. 
The velocity at the nozzle exit is $u_\mathrm{in}$. 
The height and width of the nozzle are $h$ and $s$, respectively, 
and the aspect ratio is defined as $AR = s/h$.

%------------------------------------------------------------------------------
% Figure 1
%------------------------------------------------------------------------------
%------------------------------------------------------------------------------

This study deals with a three-dimensional flow of an incompressible viscous fluid 
with constant physical properties. 
The fundamental equations are the continuity and Navier-Stoke equations 
in Cartesian coordinates. 
These equations are non-dimensionalized by the height $h$ of the nozzle exit 
and time-averaged maximum velocity $\overline{U}_\mathrm{max}$ at the nozzle exit, 
and given as
%------------------------------------------------------------------------------
\begin{equation}
   \nabla \cdot {\bf u} = 0,
   \label{continuity}
\end{equation}
\begin{equation}
   \frac{\partial {\bf u}}{\partial t} + \nabla \cdot ({\bf u} \otimes {\bf u}) 
   = - \nabla p + \frac{1}{Re} \nabla^{2} {\bf u},
   \label{navier-stokes}
\end{equation}
%------------------------------------------------------------------------------
where $t$ is the time, ${\bf u} = (u, v, w)$ is the velocity vector 
at the coordinates ${\bf x} = (x, y, z)$, $p$ is the pressure, 
$Re = \overline{U}_\mathrm{max} h/\nu$ is the Reynolds number, 
and $\nu$ is the kinematic viscosity.

The governing equations are solved using the simplified marker and cell (SMAC) method \cite{Amsden&Harlow_1970}. 
The Crank--Nicolson method is used to discretize the time derivatives, 
and time marching is performed. 
The second-order central difference scheme is used for the discretization 
of the space derivatives. 
Similar to existing studies \citep{Yanaoka&Inafune_2023,Yanaoka_2023}, 
we used a simultaneous relaxation method for velocity and pressure. 
This study analyzed the flow field by direct numerical simulation.

%##############################################################################
\section{Calculation conditions}
%##############################################################################

In this study, the aspect ratio of the nozzle is changed to $AR = 1.0$, 1.5, and 2.0, 
and we investigate the influence of the axis-switching phenomenon 
on turbulence characteristics. 
The calculation region is 0 to $20h$ in the $x$-direction, 
$-15h$ to $15h$ in the $y$-direction and $-15h$ to $15h$ in the $z$-direction.

At the nozzle exit, the velocity is given as shown in the following equation \cite{Gohil_et_al_2015}, 
and the velocity at the nozzle exit is pulsated to generate the axis switching of the flow:
%------------------------------------------------------------------------------
\begin{equation}
  u_\mathrm{in}(y,z,t) = \overline{u}_\mathrm{in}(y,z) 
  \left[
    1 + A \sin \left( 2 \pi St \frac{\overline{U}_{\rm max}}{h} t \right) 
  \right],
\end{equation}
\begin{equation}
\renewcommand{\arraystretch}{1.8}
  \overline{u}_\mathrm{in}(y,z) = \left\{ \begin{array}{ll}
    \displaystyle 
    \frac{\overline{U}_\mathrm{max}}{2} 
    \left[ 1 + \tanh \left( \frac{h-2|y|}{4\theta_\mathrm{in}} \right) \right] 
      \quad \mbox{if} \quad |y| \geq \left| \frac{z}{AR} \right|, \\
    \displaystyle 
    \frac{\overline{U}_\mathrm{max}}{2} 
    \left[ 1 + \tanh \left( \frac{s-2|z|}{4\theta_\mathrm{in}} \right) \right] 
      \quad \mbox{if} \quad |y| < \left| \frac{z}{AR} \right|, \\
   \end{array} \right.
   \renewcommand{\arraystretch}{1}
\end{equation} 
%------------------------------------------------------------------------------
where $\theta_\mathrm{in}$ and $A$ are the momentum thickness of the velocity profile 
and the velocity amplitude at the nozzle exit. 
$St$ is the pulsating Strouhal number defined as $St = f h/\overline{U}_\mathrm{max}$, 
and $f$ is the pulsating frequency. 
As the inlet boundary condition, we set the time-averaged streamwise velocity 
given by Michalke and Hermann\cite{Michalke&Hermann_1982}. 
We also gave the excitation set by Danaila and Boersma\cite{Danaila&Boersma_2000} 
to the streamwise velocity. 
A no-slip condition is given at the upstream boundary outside the nozzle exit. 
A convective boundary condition is applied to the downstream boundary, 
and for the convective velocity, the average value obtained from velocities higher 
than half the maximum velocity at the downstream boundary is used. 
The convection velocity was determined so that vortices did not stagnate at the exit boundary. 
As for transverse boundary conditions, 
slip boundary conditions and traction-free boundary conditions are considered. 
Similar to previous studies \cite{da_Silva&Metais_2002,Gohil_et_al_2015}, 
periodic boundary conditions were applied in the vertical and spanwise directions. 
The transverse boundary conditions of the jet and the computational domain 
greatly affect the entrainment of the fluid. 
Existing studies \cite{da_Silva&Metais_2002,Gohil_et_al_2015} set 
the transverse computational region length to $7h$ \cite{da_Silva&Metais_2002} 
and $10.65h$ \cite{Gohil_et_al_2015} to reduce the effects of these conditions. 
This study further expanded the calculation area to calculate 
using the nozzle with a large aspect ratio.

This study performs calculations under the Reynolds number $Re = 1000$ 
and the Strouhal number $St = 0.4$. 
In addition, the momentum thickness at the nozzle exit is set to $\theta_{\rm in} = 0.025h$, 
and the velocity amplitude is set to $A = 0.15$, 
which is 15\% of $\overline{u}_\mathrm{in}(y,z)$. 
These conditions are the same as an existing study \cite{Gohil_et_al_2015} 
in which axis switching was confirmed. 
Gohil et al. \cite{Gohil_et_al_2015} reported 
that the critical Reynolds number for the free jet in a square nozzle is $Re = 875-900$ without perturbations. 
Therefore, the flow at $Re = 1000$ with $AR = 1$ is not steady, 
and $Re = 1000$ is close to the critical Reynolds number. 
The Strouhal number of the free jet at $Re = 1000$ was $St = 0.27$. 
The pulsating jet frequency $St = 0.4$ is higher than this value. 
Referring to the results of free jets in the previous study \cite{Gohil_et_al_2015}, 
this study only dealt with pulsating jets without analyzing free jets. 
However, as shown later, 
the present results for the pulsating jet were also compared with 
the existing results for the free jet.

We use three grids to confirm the grid dependency on the calculation results. 
The numbers of grid points are $192\times125\times125$ (grid1), 
$243\times157\times157$ (grid2), and $309\times197\times197$ (grid3). 
The grid is set to be dense near the nozzle exit at $x/h = 0$ to capture the development of the jet. 
Furthermore, to capture the shear layer of the jet accurately, 
the grid resolution was increased near the cross-sections of $y/h = \pm 0.5$ and $z/h = \pm AR/2$. 
The minimum grid widths in grid1, grid2, and grid3 are $0.08h$, $0.04h$, and $0.02h$, respectively. 
For $AR = 1$ and 2, we investigated the difference in calculation results 
depending on the number of grid points. 
We will discuss grid dependence later. 
In this study, we mainly show the calculation results using grid3 
to clarify turbulences due to vortex structures in more detail. 
The time intervals used for calculation are $\Delta t/(h/\overline{U}_\mathrm{max}) = 0.005$, 
0.0025, and 0.00125 for grid1, grid2, and grid3, respectively.

The previous studies \cite{Tsuchiya&Horikoshi_1986,Jiang_et_al_2007,Straccia&Farnsworth_2021} 
used the nozzle height as a reference length 
and compared the results at each aspect ratio. 
It is also possible to use the equivalent diameter as a reference length. 
This study used the same nozzle height for each aspect ratio 
to facilitate a comparison of the behavior and size of the vortex structure 
for each aspect ratio. 
Therefore, as in the existing research, 
the reference length was set to the height of the nozzle outlet, 
and the dimensionless number was defined using the height. 
When comparing entrainment at each aspect ratio with existing results, 
the circular-equivalent diameter $D_e$ is used to make the coordinates dimensionless. 
The Reynolds and Strouhal numbers based on the equivalent diameter are denoted as $Re_D$ and $St_D$, respectively.

%------------------------------------------------------------------------------
% Figure 2
%------------------------------------------------------------------------------
\begin{figure}[!t]
\begin{minipage}{0.48\linewidth}
\begin{center}
\includegraphics[trim=0mm 0mm 0mm 0mm, clip, width=70mm]{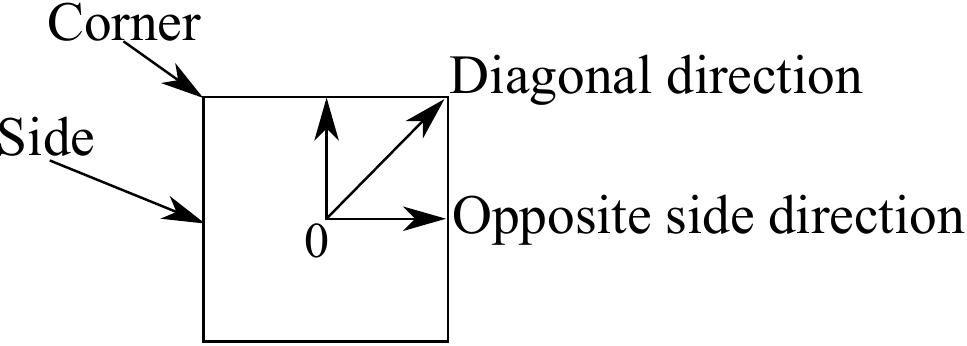} \\
(a)
\end{center}
\end{minipage}
\hspace{0.02\linewidth}
\begin{minipage}{0.48\linewidth}
\begin{center}
\includegraphics[trim=0mm 0mm 0mm 0mm, clip, width=80mm]{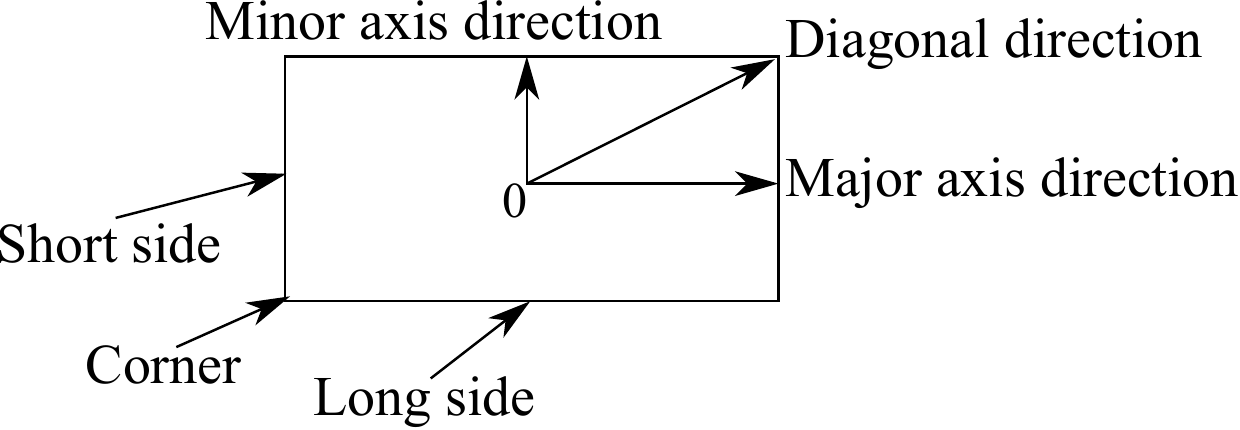} \\
(b)
\end{center}
\end{minipage}
\caption{Schematic of nozzle exit: (a) $AR = 1.0$ and (b) $AR > 1.0$.}
\label{schema}
\end{figure}
%------------------------------------------------------------------------------

Figure \ref{schema} shows a schematic diagram of the nozzle exit. 
For $AR = 1.0$ in Fig. \ref{schema} (a), 
the nozzle outlet shape is a square with four corners and four sides. 
When discussing the obtained results, 
the horizontal and vertical directions of the nozzle exit are defined as 
the opposite side direction, 
and the direction of the line connecting the diagonal points is defined as 
the diagonal direction. 
For $AR > 1.0$ in Fig. \ref{schema} (b), 
the nozzle outlet shape is a rectangle with four corners, two long sides, and two short sides. 
The horizontal direction passing through the origin is the major axis, 
and the vertical direction is the minor axis. 
The horizontal direction of the nozzle exit is defined as the major axis direction, 
the vertical direction is defined as the minor axis direction, 
and the direction connecting the diagonal points is defined as the diagonal direction.

%##############################################################################
\section{Results and discussion}
%##############################################################################

%++++++++++++++++++++++++++++++++++++++++++++++++++++++++++++++++++++++++++++++
\subsection{Flow near nozzle exit}
%++++++++++++++++++++++++++++++++++++++++++++++++++++++++++++++++++++++++++++++

\begin{figure}[!t]
%
%%% AR=1.0
\begin{minipage}{0.48\linewidth}
\begin{center}
\includegraphics[trim=0mm 0mm 0mm 0mm, clip, width=65mm]{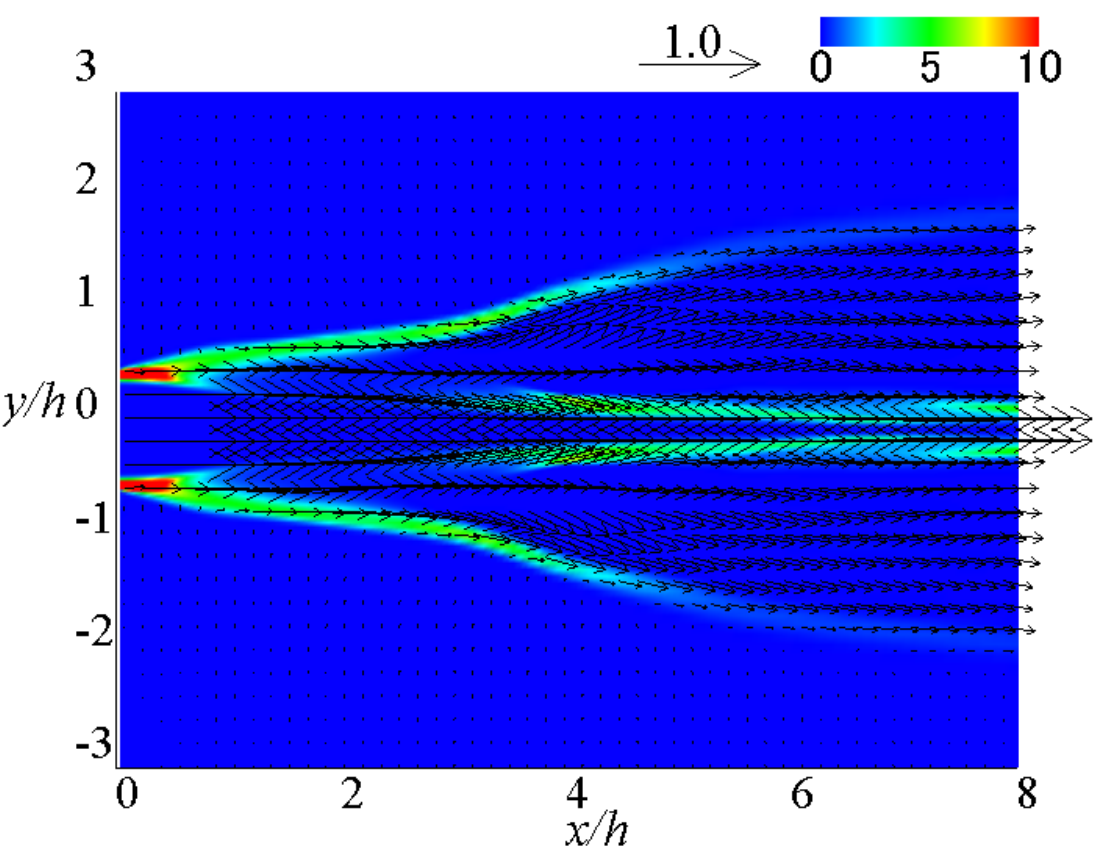} \\
\end{center}
\end{minipage}
\hspace{0.02\linewidth}
\begin{minipage}{0.48\linewidth}
\begin{center}
\includegraphics[trim=0mm 0mm 0mm 0mm, clip, width=55mm]{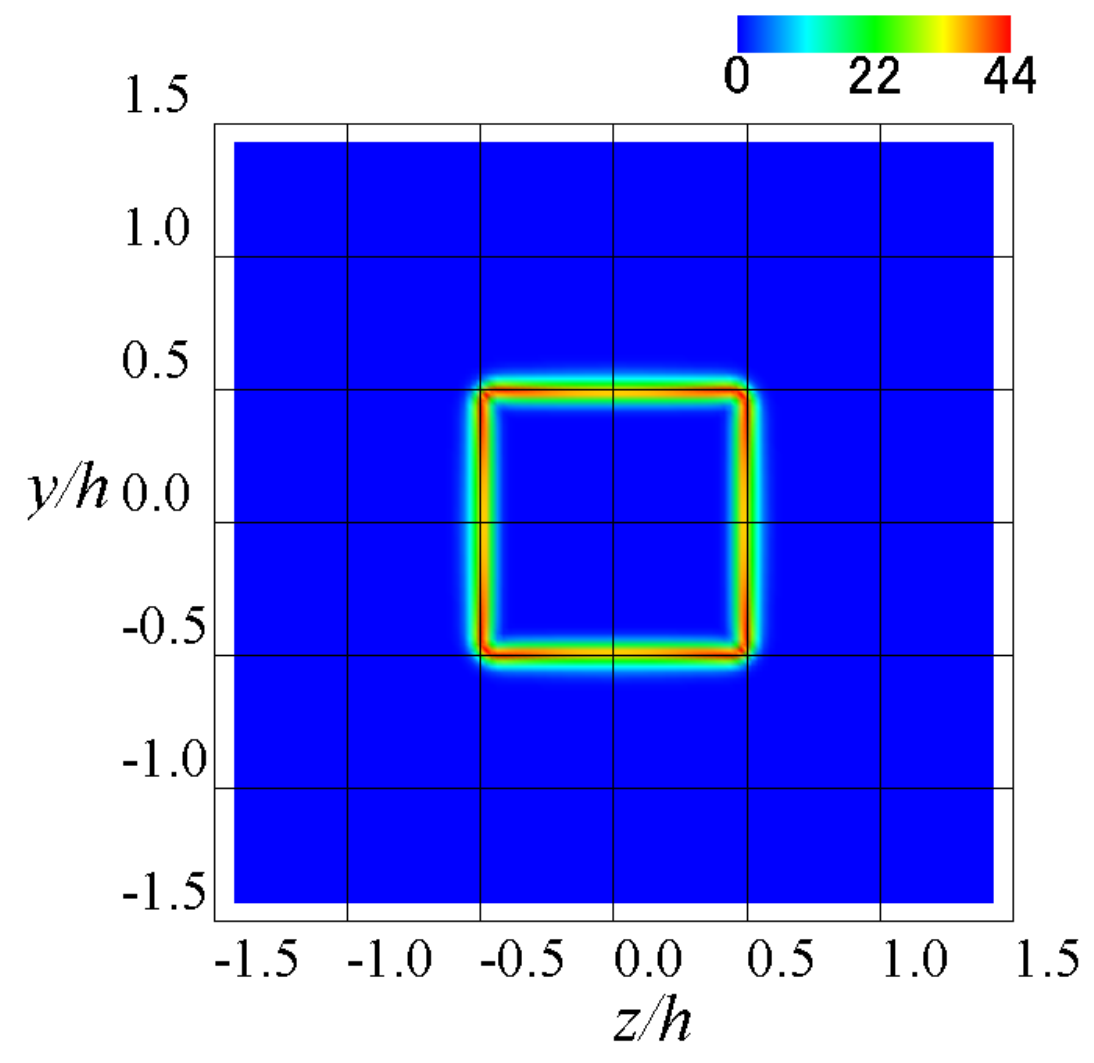} \\
\end{center}
\end{minipage}
\centering
(a) \\

%%% AR=1.5
\begin{minipage}{0.48\linewidth}
\begin{center}
\includegraphics[trim=0mm 0mm 0mm 0mm, clip, width=65mm]{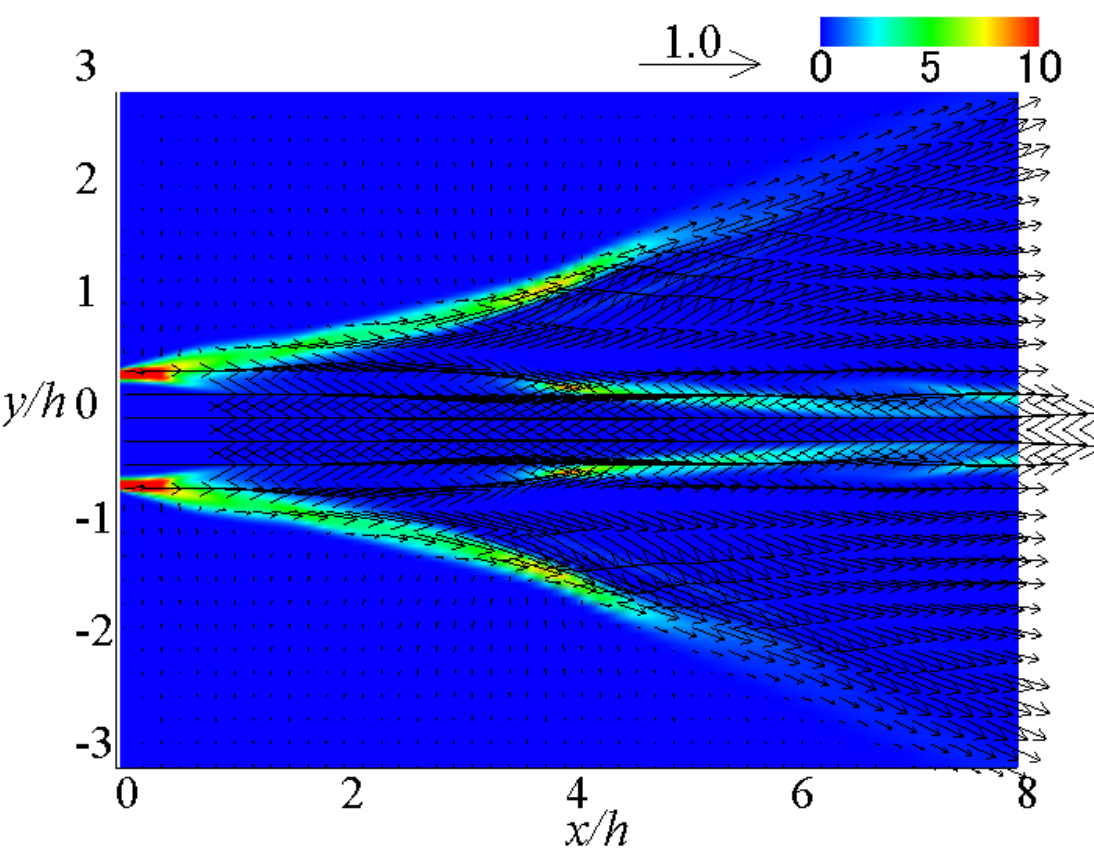} \\
\end{center}
\end{minipage}
\hspace{0.02\linewidth}
\begin{minipage}{0.48\linewidth}
\begin{center}
\includegraphics[trim=0mm 0mm 0mm 0mm, clip, width=55mm]{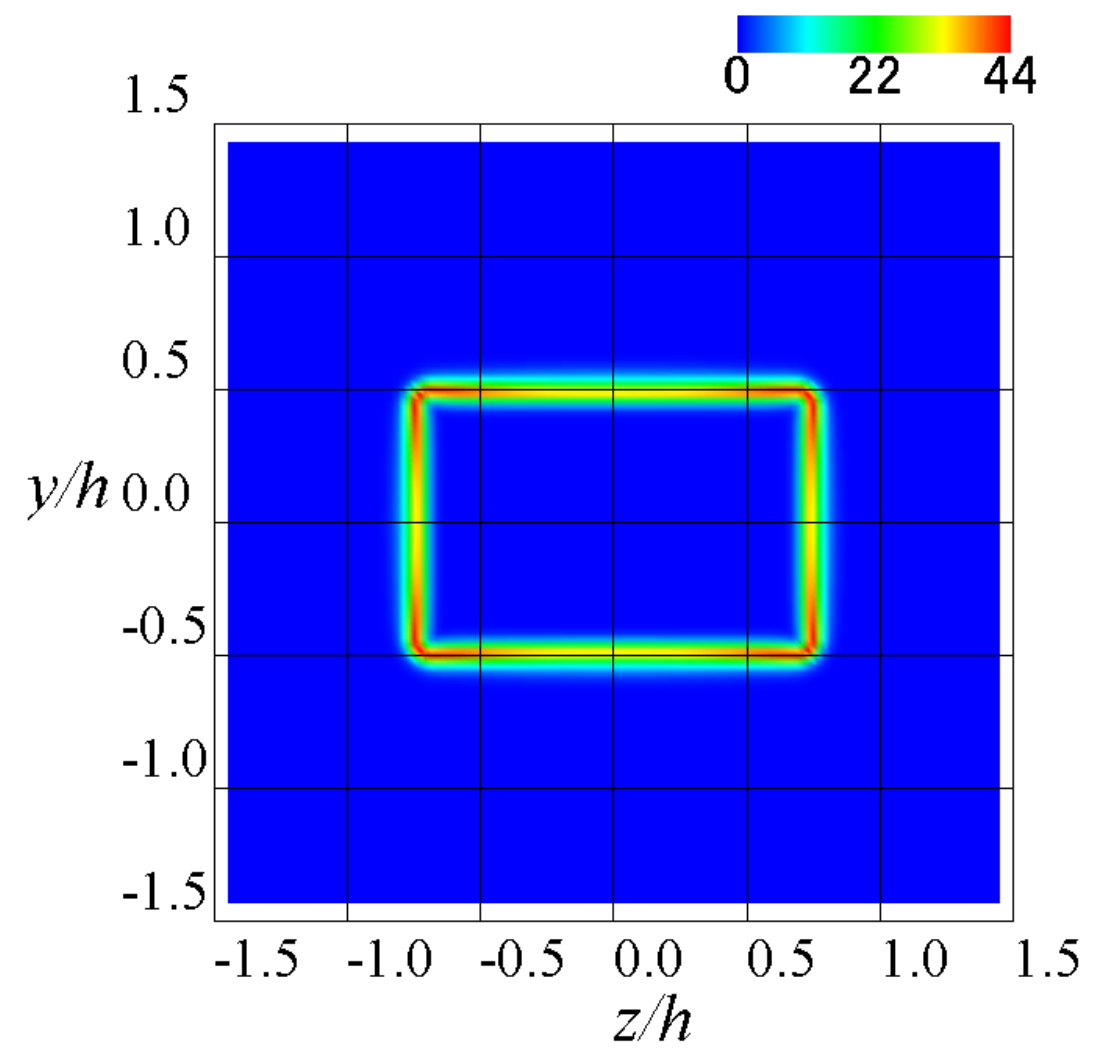} \\
\end{center}
\end{minipage}
\centering
(b) \\

%%% AR=2.0
\begin{minipage}{0.48\linewidth}
\begin{center}
\includegraphics[trim=0mm 0mm 0mm 0mm, clip, width=65mm]{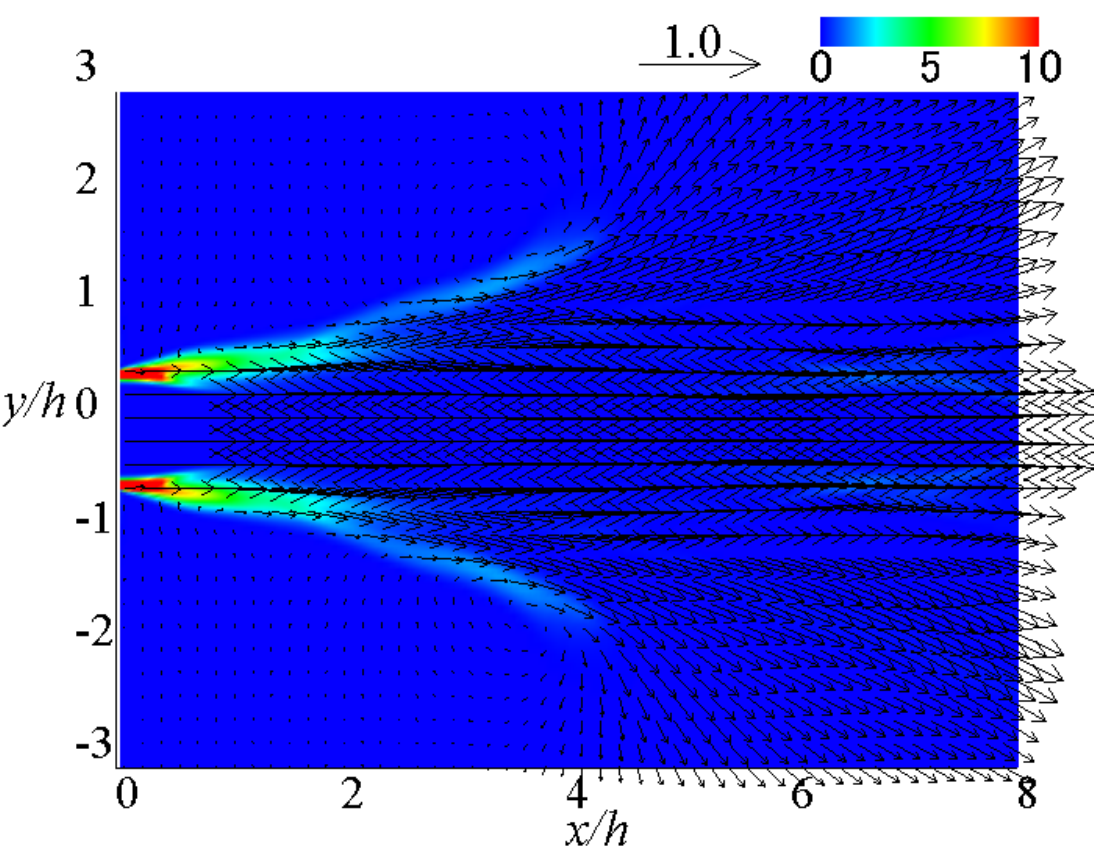} \\
\end{center}
\end{minipage}
\hspace{0.02\linewidth}
\begin{minipage}{0.48\linewidth}
\begin{center}
\includegraphics[trim=0mm 0mm 0mm 0mm, clip, width=55mm]{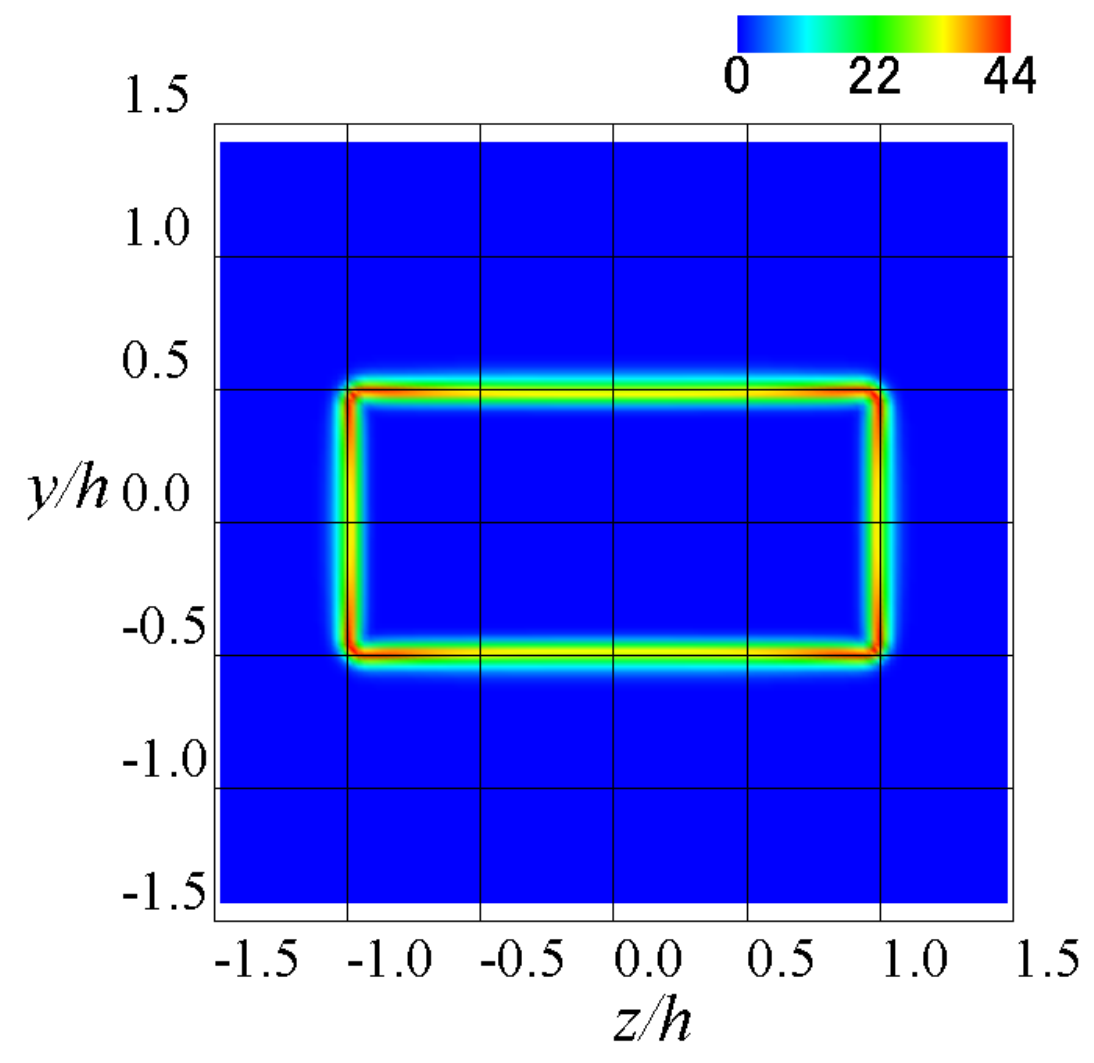} \\
\end{center}
\end{minipage}
\centering
(c) \\
\vspace*{-0.5\baselineskip}
\caption{Time-averaged enstrophy contour and velocity vectors 
in $x$-$y$ plane at $z/h = 0$ (left), 
and time-averaged enstrophy contours in $y$-$z$ plane at $x/h = 0.25$ (right): 
(a) $AR = 1.0$, (b) $AR = 1.5$, and (c) $AR = 2.0$.}
\label{ens_vec_xy&ens_yz}
\end{figure}

Figure \ref{ens_vec_xy&ens_yz} shows the time-averaged enstrophy distribution 
and velocity vector in the $x$-$y$ plane at $z/h = 0$ 
and the time-averaged enstrophy distribution in the $x$-$z$ plane at $x/h = 0.25$. 
As can be seen from the $x$-$y$ plane, 
a shear layer is generated between the ejected fluid and the surrounding fluid 
for all the aspect ratios $AR$ and extends downstream. 
The jet spreads with increasing $AR$. 
For $AR = 1$ and 1.5, a high enstrophy distribution appears near the jet center axis. 
As explained later, this is caused by a vortex pair around the jet center 
or by a regenerated vortex ring. 
When looking at the $y$-$z$ plane, regardless of $AR$, 
the shear layer with the same shape as the nozzle cross-section exists. 
The enstrophy increases near the corners of the rectangular shear layer 
because the flow of the surrounding fluid entrained by the jet concentrates there.

%------------------------------------------------------------------------------
% Figure 3
%------------------------------------------------------------------------------
%------------------------------------------------------------------------------

%------------------------------------------------------------------------------
% Figure 4
%------------------------------------------------------------------------------
\begin{figure}[!t]
%
%%% AR=1.0
\begin{minipage}{.325\linewidth}
\centering
\includegraphics[trim=0mm 0mm 0mm 0mm, clip, width=55mm]{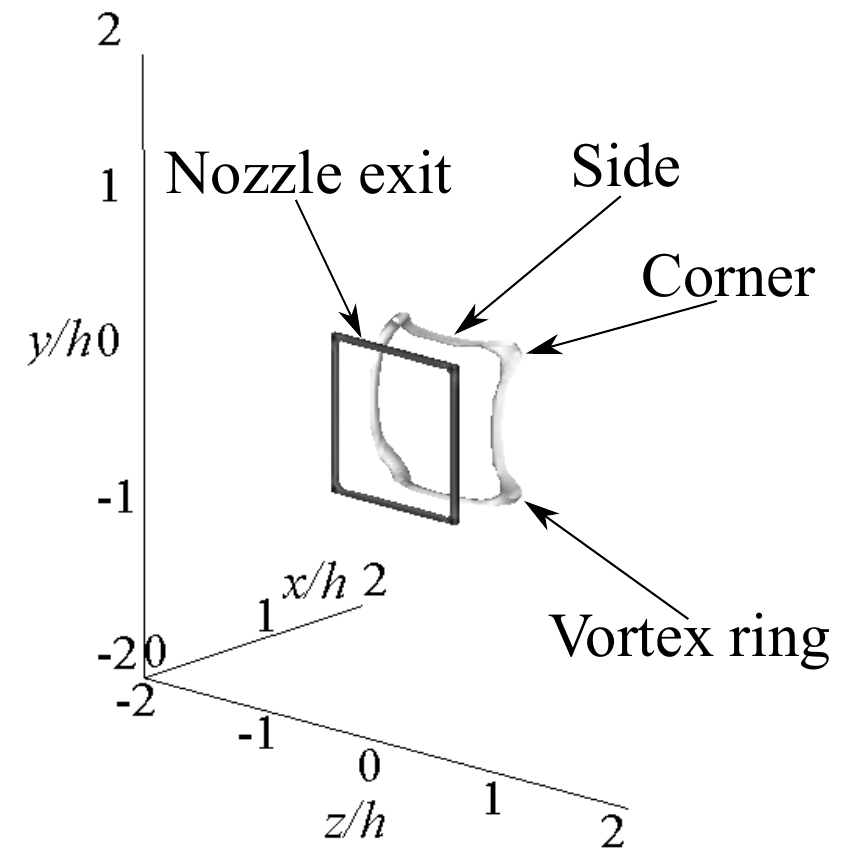} \\
(a)
\end{minipage}
%
%%% AR=1.5
\begin{minipage}{.325\linewidth}
\centering
\includegraphics[trim=0mm 0mm 0mm 0mm, clip, width=55mm]{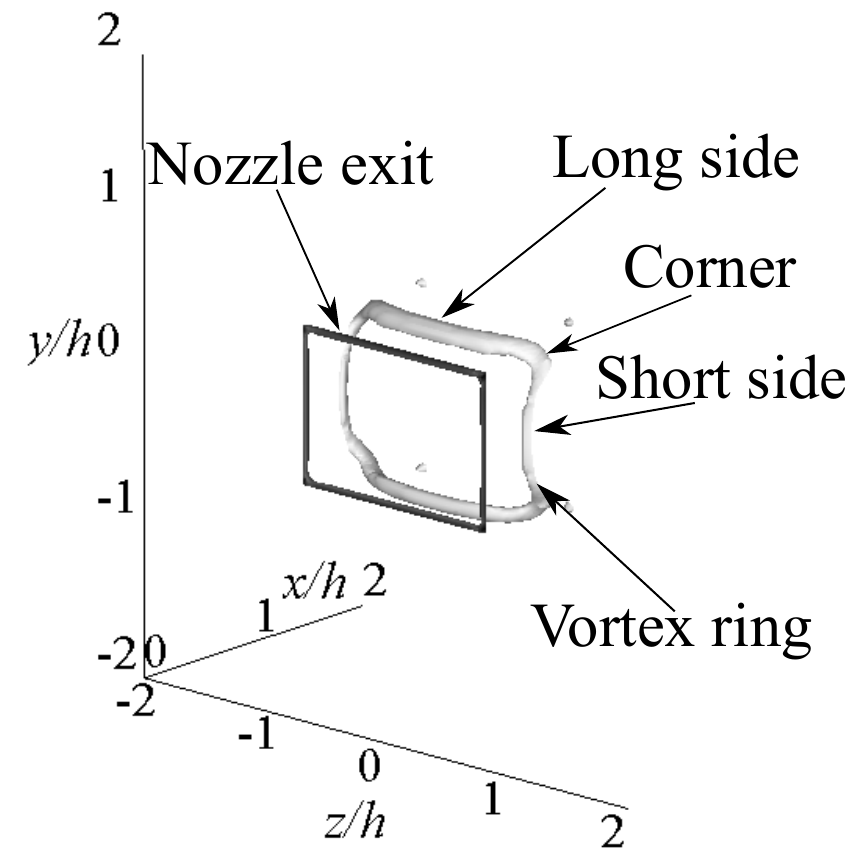} \\
(b)
\end{minipage}
%
%%% AR=2.0
\begin{minipage}{.325\linewidth}
\centering
\includegraphics[trim=0mm 0mm 0mm 0mm, clip, width=55mm]{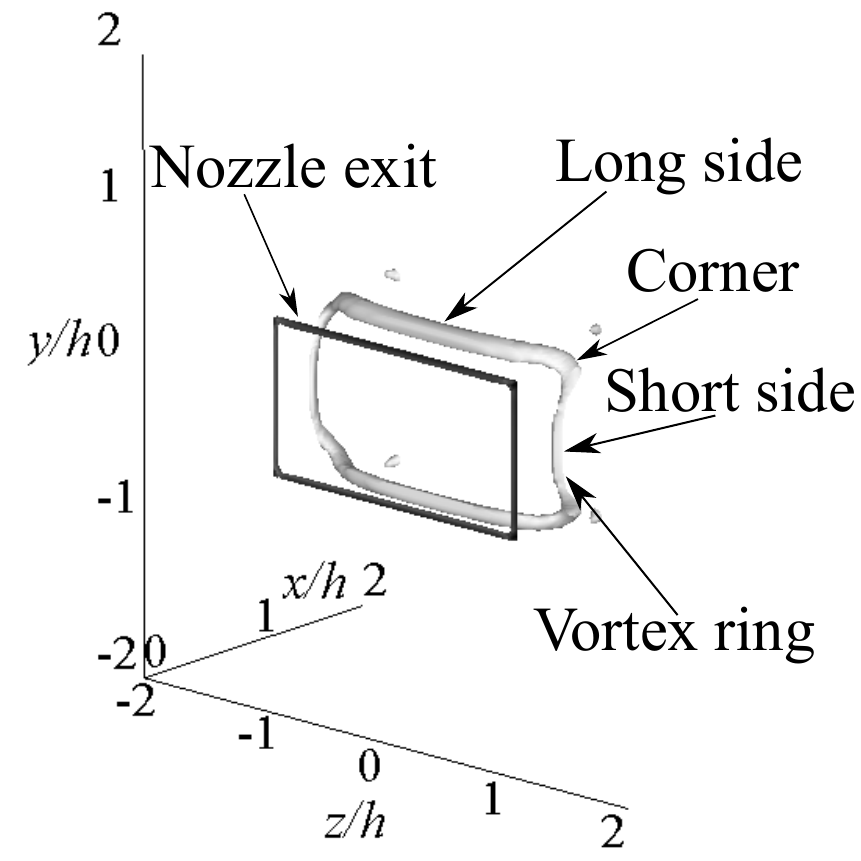} \\
(c)
\end{minipage}
\caption{Perspective view of isosurface of curvature of equipressure surface at $T = 1$: 
Isosurface value is $-20/h$. 
(a) $AR = 1.0$, (b) $AR = 1.5$, and (c) $AR = 2.0$.}
\label{ring}
\end{figure}
%------------------------------------------------------------------------------

%++++++++++++++++++++++++++++++++++++++++++++++++++++++++++++++++++++++++++++++
\subsection{Vortex structure}
%++++++++++++++++++++++++++++++++++++++++++++++++++++++++++++++++++++++++++++++

To clarify vortex structures existing in an instantaneous flow field, 
Figs. \ref{ring} to \ref{curv2} show the isosurface of the curvature 
calculated from an equipressure surface. 
In the results shown below, 
the dimensionless start time of the data sampling is $T = 0$. 
At $T = 0$, the jet velocity at the inlet starts to accelerate. 
This method of visualizing vortex structures was used in our previous research 
\citep{Yanaoka_et_al_2007b, Yanaoka_et_al_2007c, Yanaoka_et_al_2008b}. 
The curvature is defined as $\kappa = -\nabla \cdot \hat{\bf n}$. 
Here, $\hat{\bf n}$ is a unit normal vector, 
and it is determined using pressure as $\hat{\bf n} = \nabla p/|\nabla p|$. 
In Fig. \ref{ring}, the nozzle exit, vortex ring, and the side and corner of the vortex ring are labeled. 
In Figs. \ref{curv1} to \ref{curv2}, the labels are shown for a vortex ring, 
the hairpin part of the vortex ring, a longitudinal vortex, and a vortex pair. 
In addition, vortices forming vortex pairs are labeled as A to F. 
For the results with $AR = 1.0$, 1.5, and 2.0, the labels have subscripts 10, 15, and 20, respectively. 
As can be seen from Fig. \ref{ring}, 
a shear layer near the nozzle exit becomes unstable in all $AR$ 
and rolls up into a vortex. 
This vortex is a rectangular vortex ring with the side and corner of almost the same shape as the nozzle shape 
and is periodically shed downstream. 
This vortex ring is similar to the results of pulsating jets ejected from a square nozzle 
for $Re = 1000$ \citep{Gohil_et_al_2015} and high Reynolds numbers \citep{Grinstein_et_al_1995,Grinstein_2001}. 
The vortex shedding frequency of the vortex ring is $f = 0.40\overline{U}_\mathrm{max}/h$ 
regardless of $AR$, 
which agrees with the pulsation frequency of the velocity at the nozzle exit.

In each $AR$ in Figs. \ref{curv1} to \ref{curv2}, 
as the vortex ring moves downstream, 
the corners of that vortex ring move downstream faster than the sides. 
At this time, the corner approaches the central axis of the jet, 
and the side deforms into a hairpin shape. 
Further downstream, the hairpin part of the vortex ring moves toward the surrounding, 
away from the jet center axis. 
At $x/h = 1$ downstream near the nozzle corner, 
a vortex pair is generated inside the vortex ring. 
For example, when $AR=1.0$, two vortices A$_{10}$ and B$_{10}$ 
and two vortices C$_{10}$ and D$_{10}$ form vortex pairs, respectively. 
This vortex pair exists so as to connect the vortex rings upstream and downstream, 
and around $x/h = 1-2$, it is found that the vortex pair elongated 
in the streamwise direction forms a rib structure. 
This rib structure was also observed in the previous results on pulsating jets ejected 
from square or rectangular nozzles \citep{Grinstein_et_al_1995,Grinstein_2001,Gohil_et_al_2015}. 
As the vortex ring expands downstream, 
the vortex pair also expands in the $y$- and $z$- directions. 
In this manner, the interaction between the vortex ring and vortex pair occurs, 
causing axis switching. 
It is considered that the hairpin part of the vortex ring and the vortex pair promote 
the mixing of ejected and surrounding fluids. 
In Fig. \ref{curv1} (b), 
the distribution of the second invariant $Q$ of the velocity gradient tensor is shown. 
Using the velocity ${\bf u}$, 
$Q$ is defined as $Q = -(\partial {\bf u})(\partial {\bf u})^T/2$. 
For incompressible flow, we can rewrite $Q$ as $Q = \nabla^2 p/2$. 
The definition of the curvature $\kappa$ of the equipressure surface is similar to that of $Q$. 
By increasing the $Q$ value, we can extract narrow vortex tubes, 
but downstream vortex structures disappear. 
If the $Q$ value is small, we can visualize vortex structures downstream, 
but vortex rings and vortex pairs do not show clear structures because of thick vortex tubes. 
In Fig. \ref{curv1} (a), we can see that the thin vortex tubes are captured over a wide area. 
The reciprocal of this curvature magnitude $|\kappa|$ corresponds to the diameter of the vortex tube.

%------------------------------------------------------------------------------
% Figure 5
%------------------------------------------------------------------------------
\begin{figure}[!t]
\centering
\begin{minipage}{0.48\linewidth}
\centering
\includegraphics[trim=0mm 2mm 0mm 0mm, clip, width=80mm]{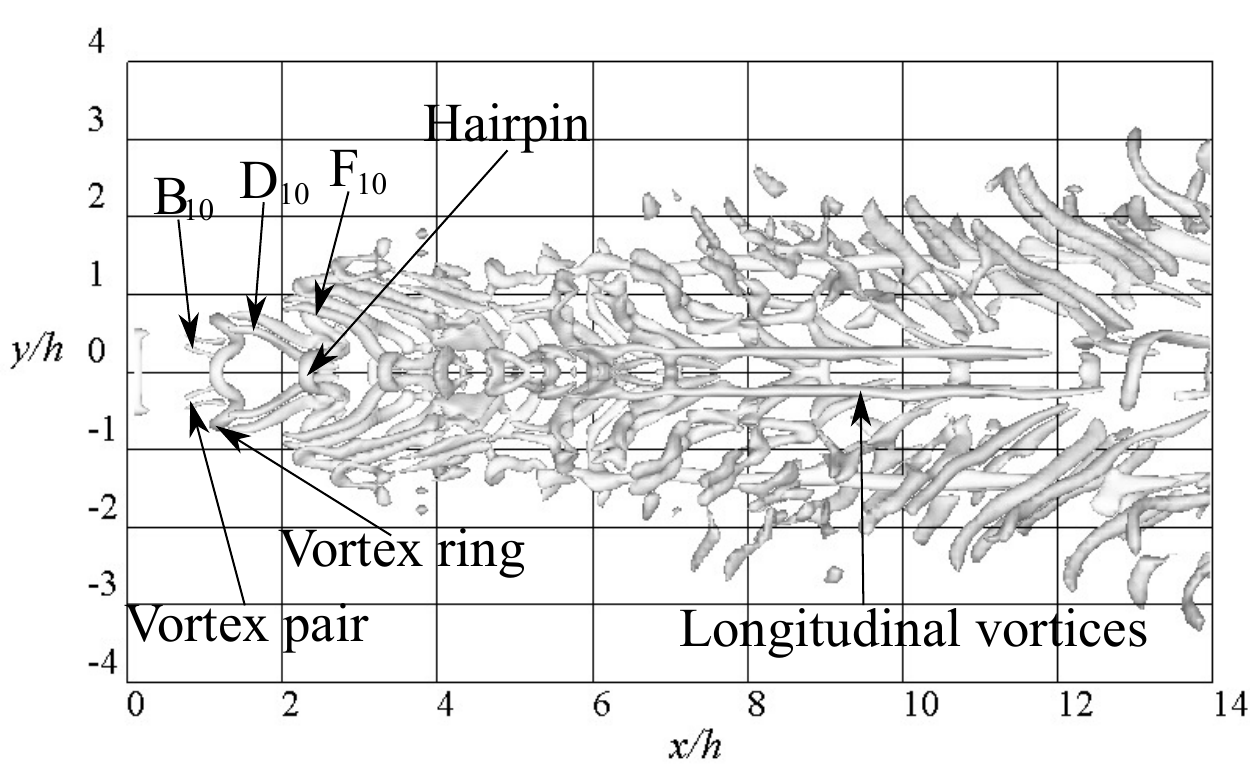} \\
(a) \\
\end{minipage}
\hspace{0.02\linewidth}
\begin{minipage}{0.48\linewidth}
\centering
\includegraphics[trim=0mm 2mm 0mm 0mm, clip, width=80mm]{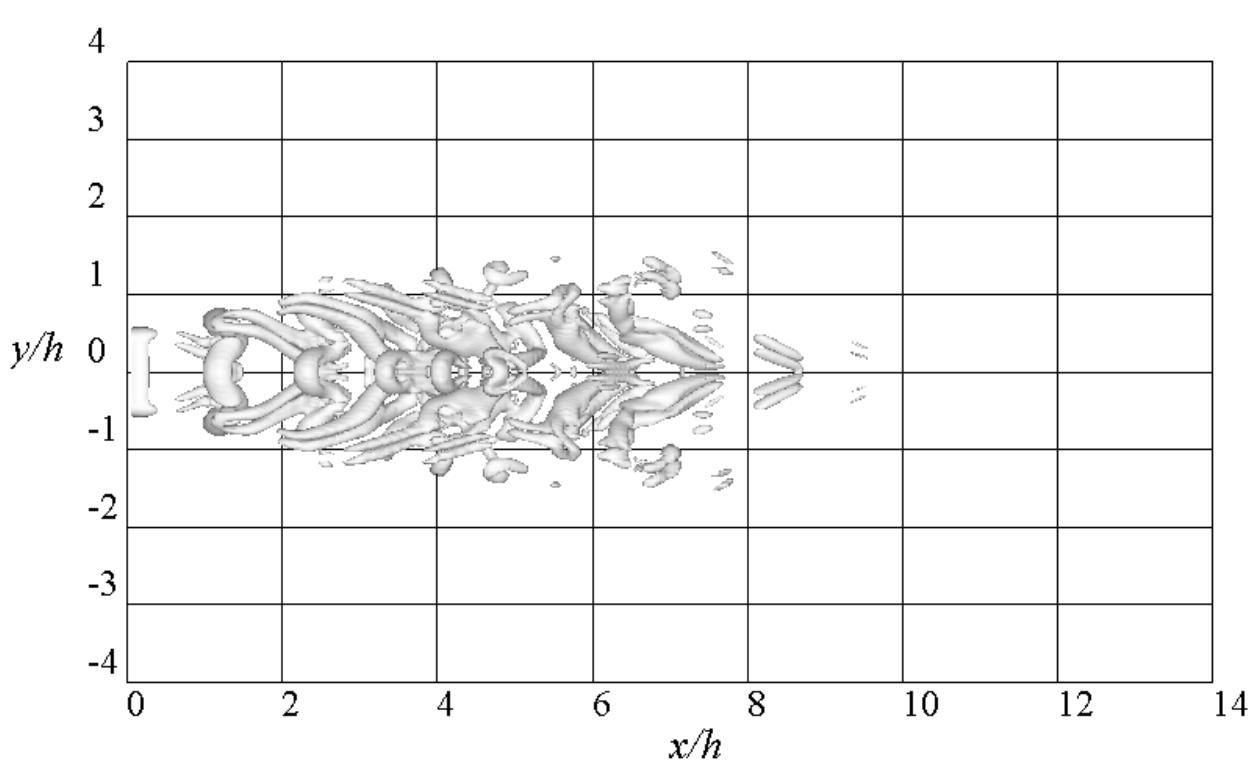} \\
(b) \\
\end{minipage}
\includegraphics[trim=0mm 5mm 0mm 0mm, clip, width=80mm]{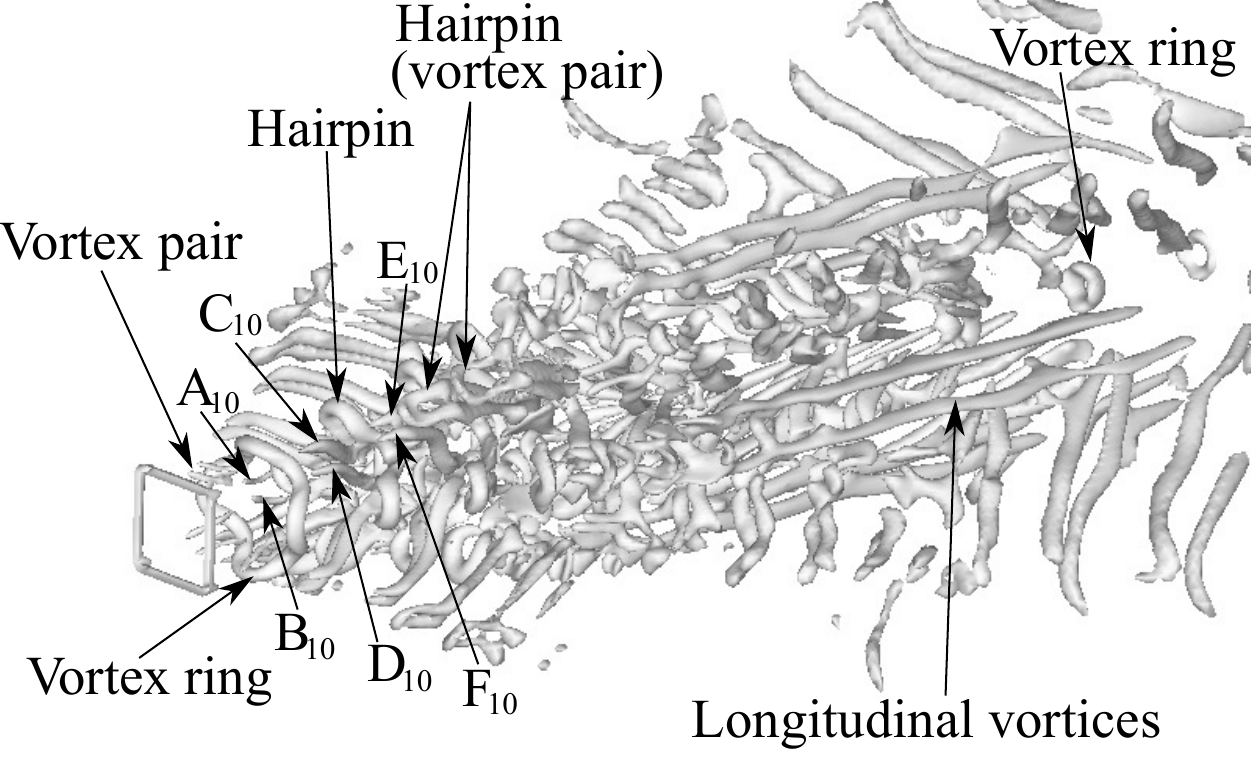}
\hspace{3mm}
\includegraphics[trim=0mm 0mm 0mm 0mm, clip, width=75mm]{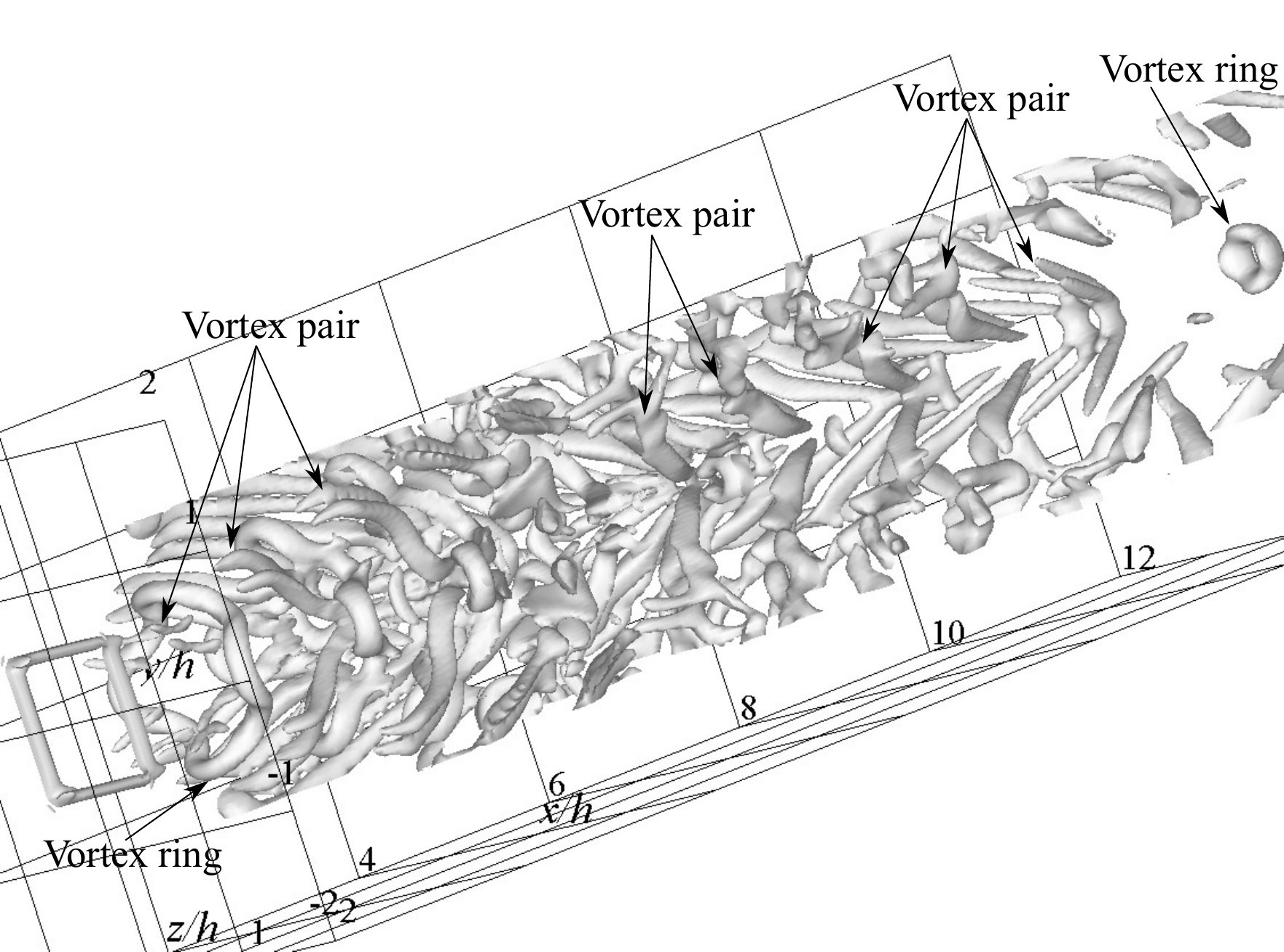} \\
(c) \\
\vspace*{-0.5\baselineskip}
\caption{Isosurface of curvature $\kappa$ of equipressure surface, and 
isosurface of second invariant $Q$ of velocity gradient tensor 
at $T = 0$ for $AR = 1.0$: 
Isosurface values of $\kappa$ and $Q$ are $-12/h$ 
and $0.5\overline{U}_\mathrm{max}^2/h^2$, respectively. 
(a) side view ($\kappa$), (b) side view ($Q$), 
and (c) perspective view ($\kappa$).}
\label{curv1}
\end{figure}
%------------------------------------------------------------------------------

%------------------------------------------------------------------------------
% Figure 6
%------------------------------------------------------------------------------
\begin{figure}[!t]
\centering
\begin{minipage}{0.48\linewidth}
\centering
\includegraphics[trim=0mm 2mm 0mm 0mm, clip, width=80mm]{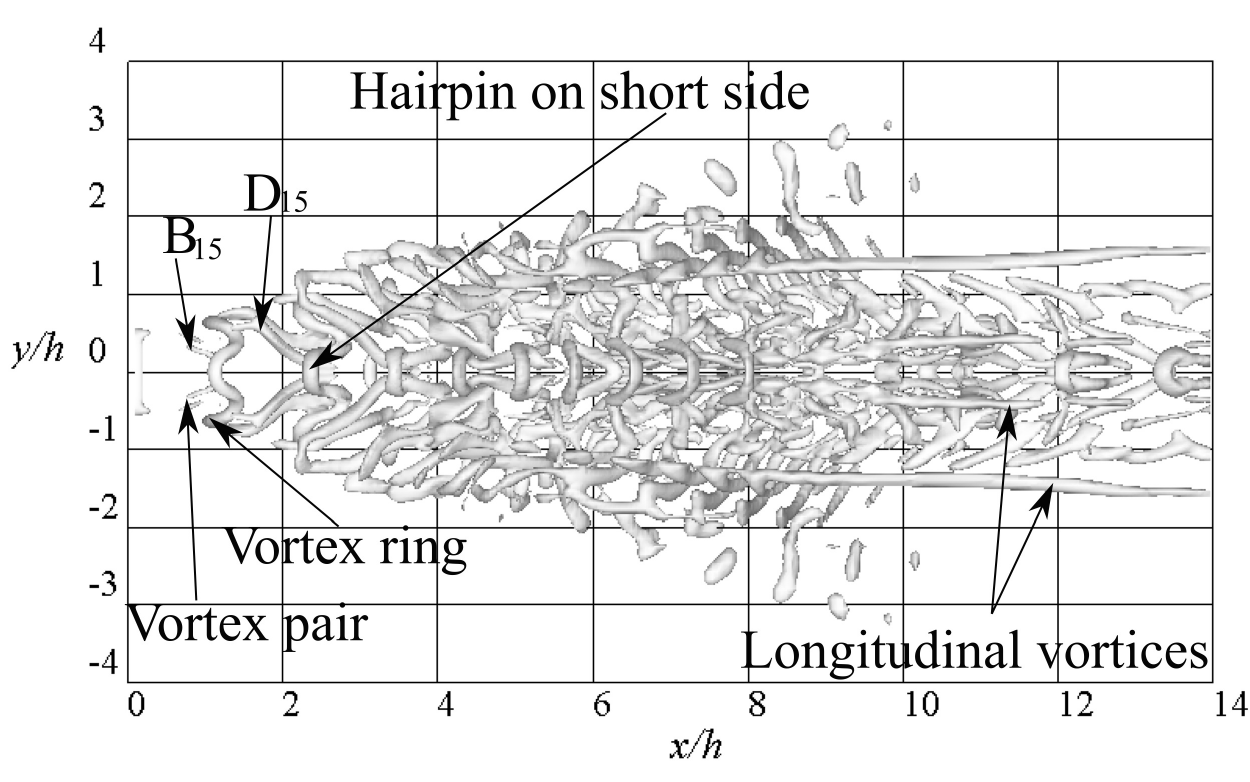} \\
(a) \\
\end{minipage}
\hspace{0.02\linewidth}
\begin{minipage}{0.48\linewidth}
\centering
\includegraphics[trim=0mm 5mm 0mm 0mm, clip, width=80mm]{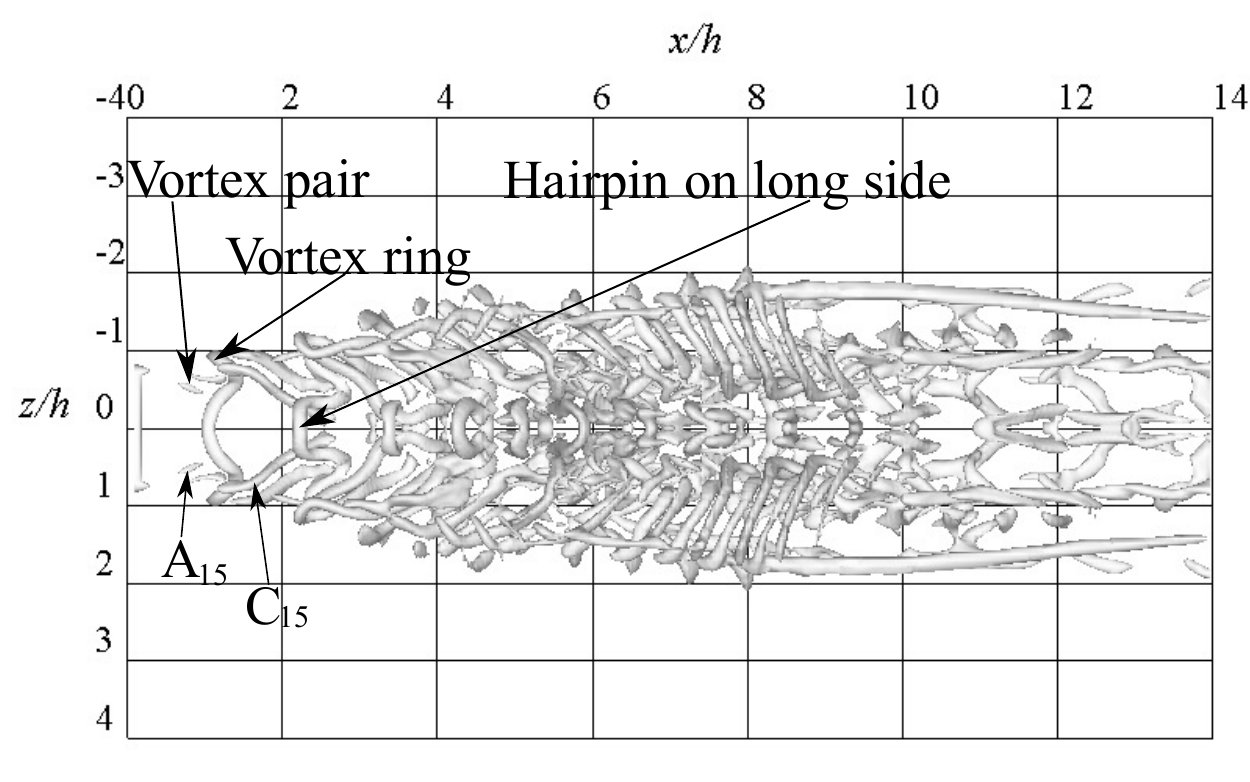} \\
(b) \\
\end{minipage}
\includegraphics[trim=0mm 0mm 0mm 0mm, clip, width=75mm]{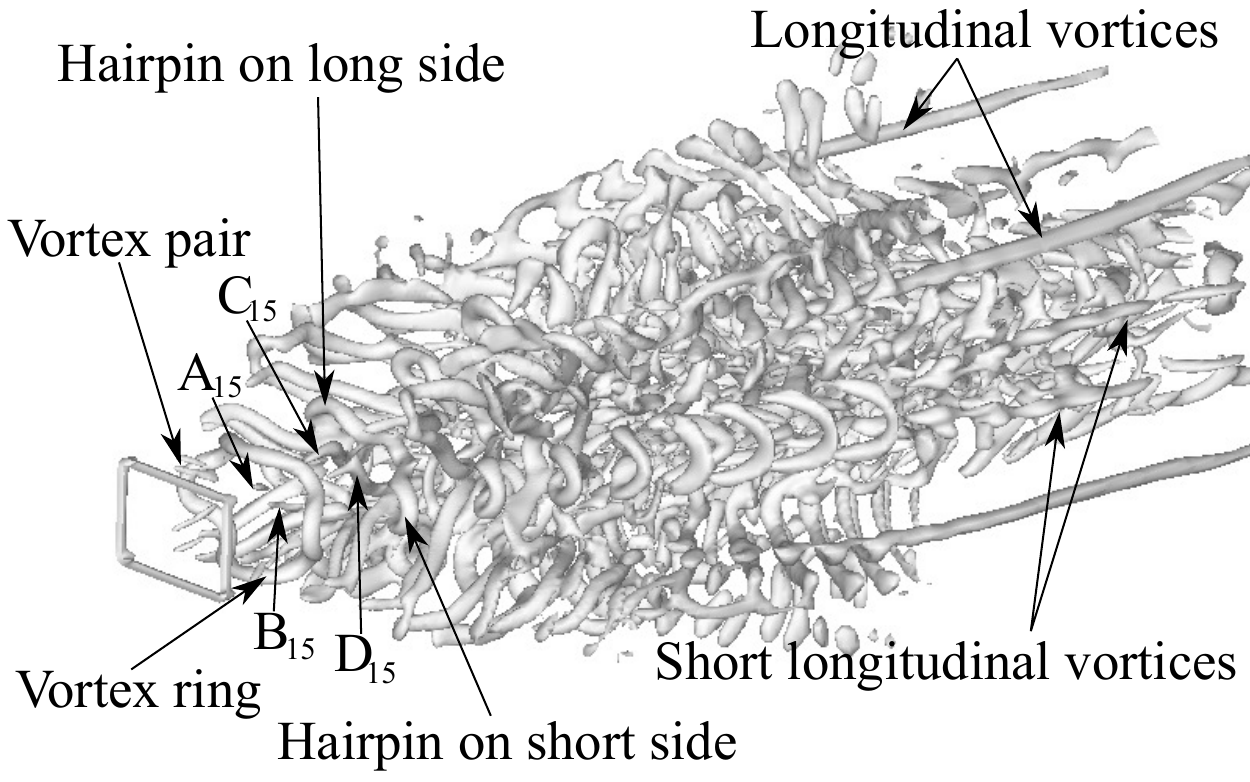}
\hspace{3mm}
\includegraphics[trim=0mm 0mm 0mm 0mm, clip, width=75mm]{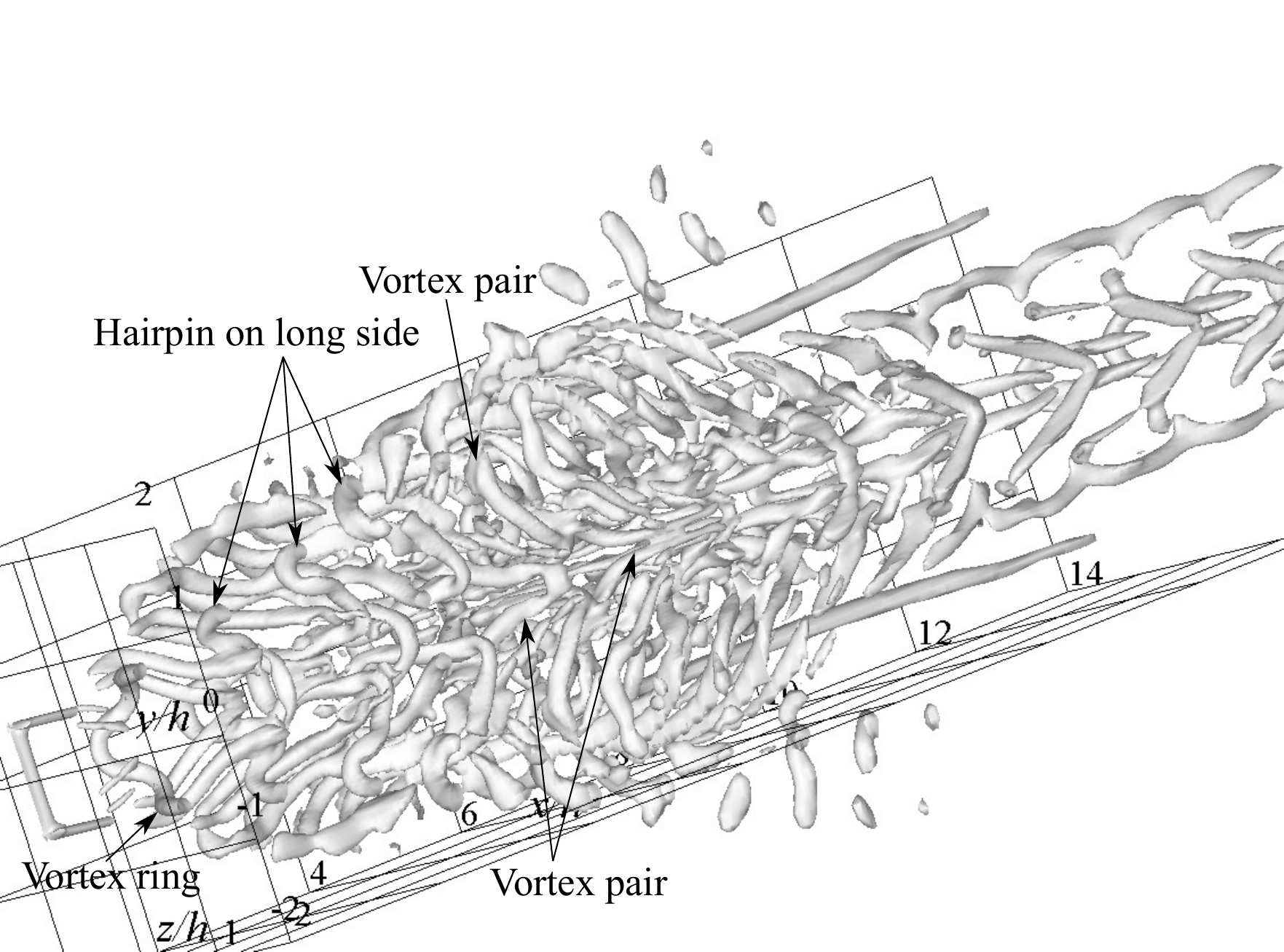} \\
(c)
\vspace*{-0.5\baselineskip}
\caption{Isosurface of curvature of equipressure surface at $T = 0$ for $AR = 1.5$: 
Isosurface value is $-12/h$. 
(a) side view, (b) top view, and (c) perspective view.}
\label{curv15}
\end{figure}
%------------------------------------------------------------------------------

%------------------------------------------------------------------------------
% Figure 7
%------------------------------------------------------------------------------
\begin{figure}[!t]
\centering
\begin{minipage}{0.48\linewidth}
\centering
\includegraphics[trim=0mm 0mm 0mm 4mm, clip, width=80mm]{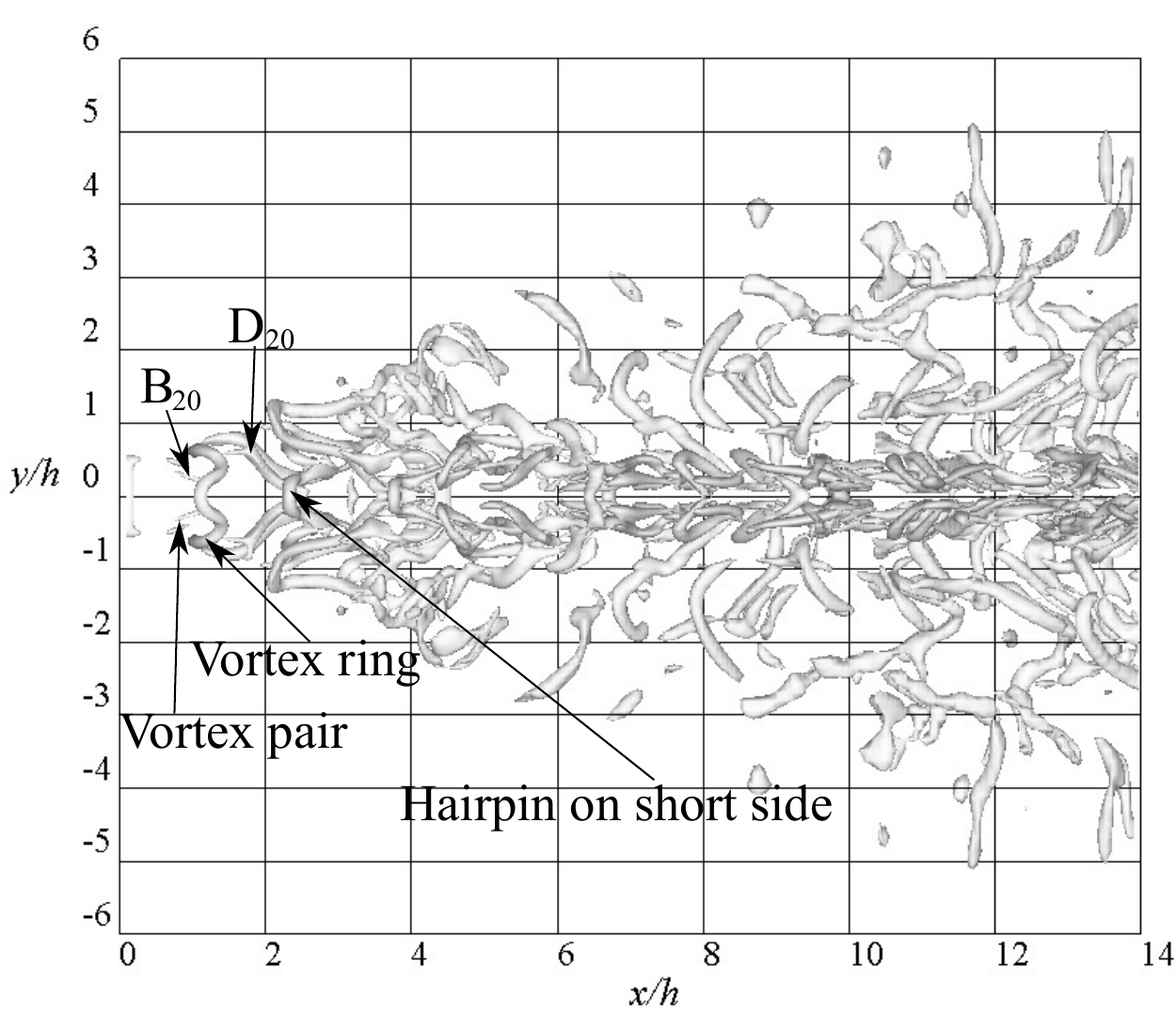} \\
(a) \\
\end{minipage}
\hspace{0.02\linewidth}
\begin{minipage}{0.48\linewidth}
\centering
\includegraphics[trim=0mm 6mm 0mm 0mm, clip, width=80mm]{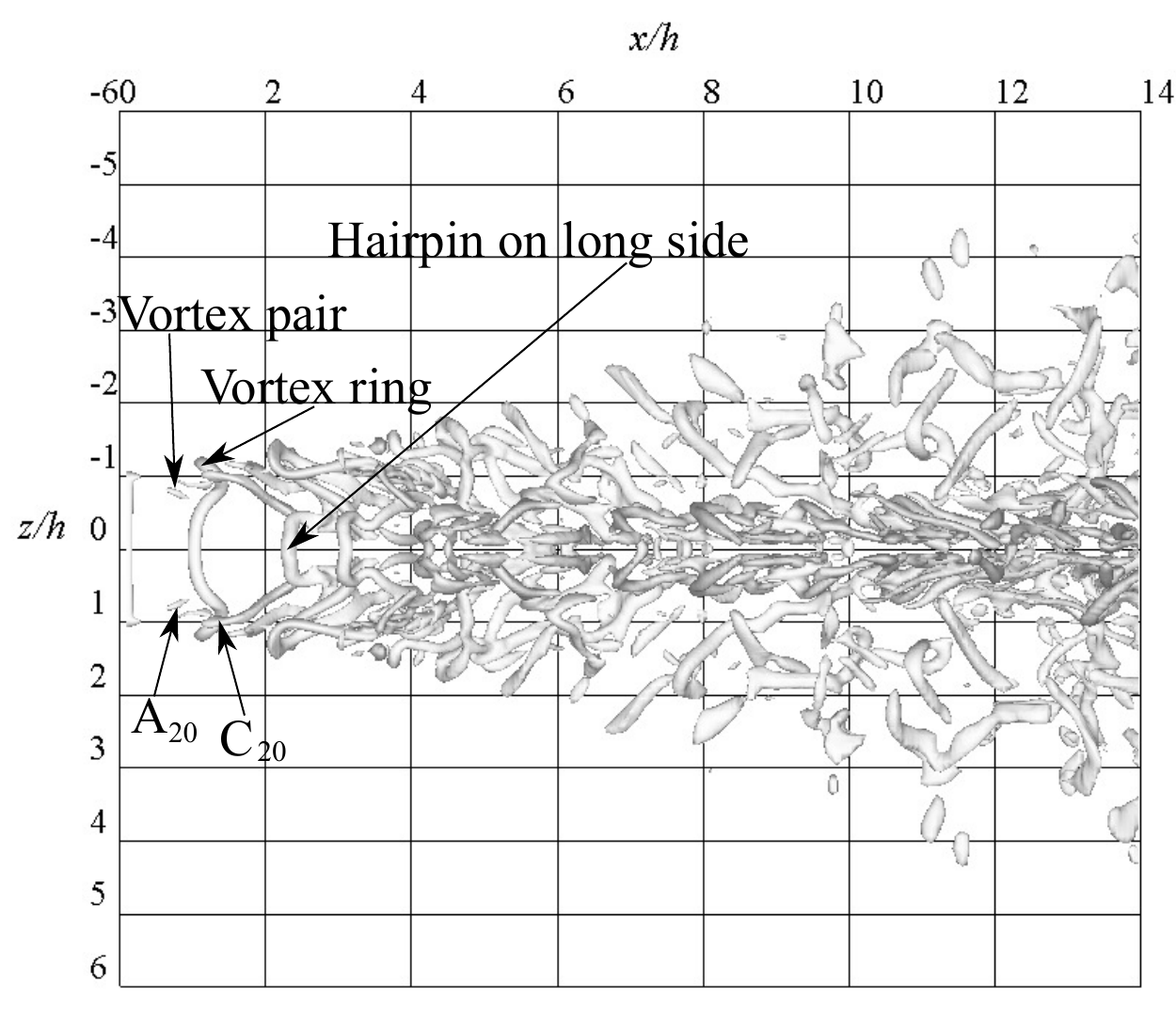} \\
(b) \\
\end{minipage}
\includegraphics[trim=0mm 0mm 0mm 0mm, clip, width=75mm]{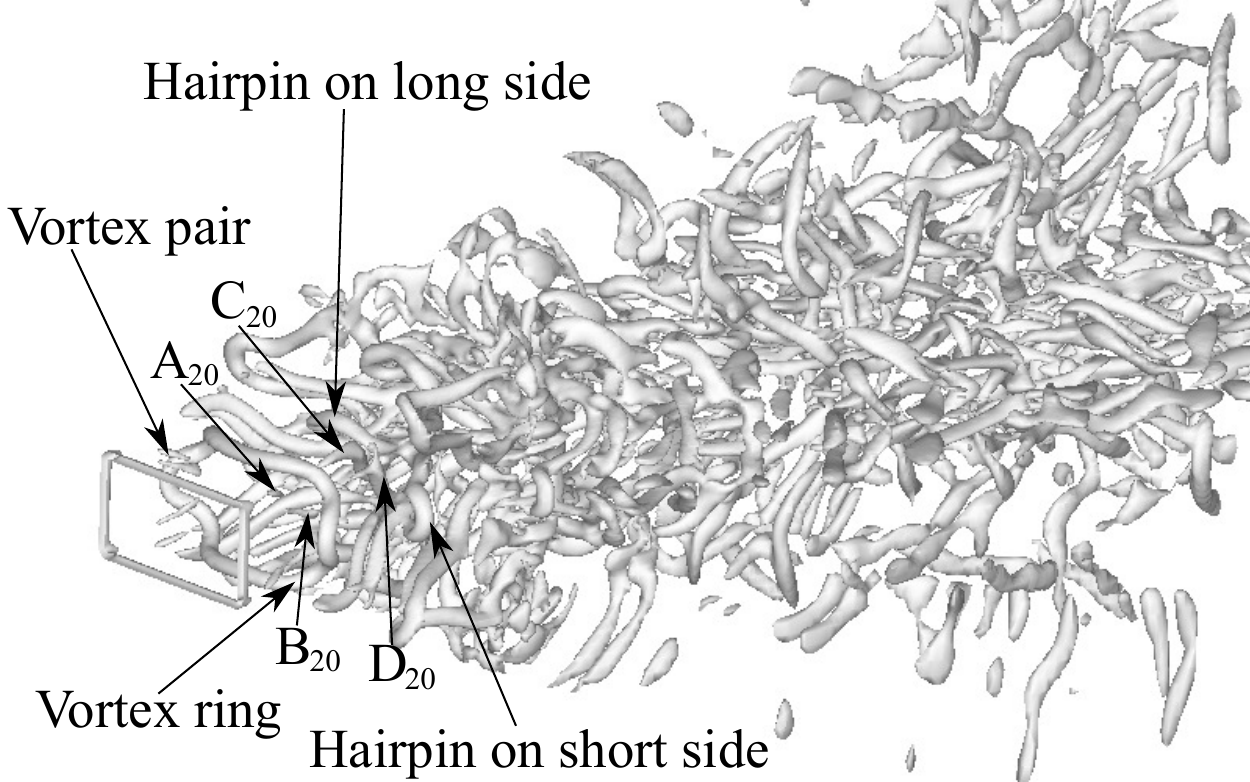}
\hspace{3mm}
\includegraphics[trim=0mm 0mm 0mm 0mm, clip, width=75mm]{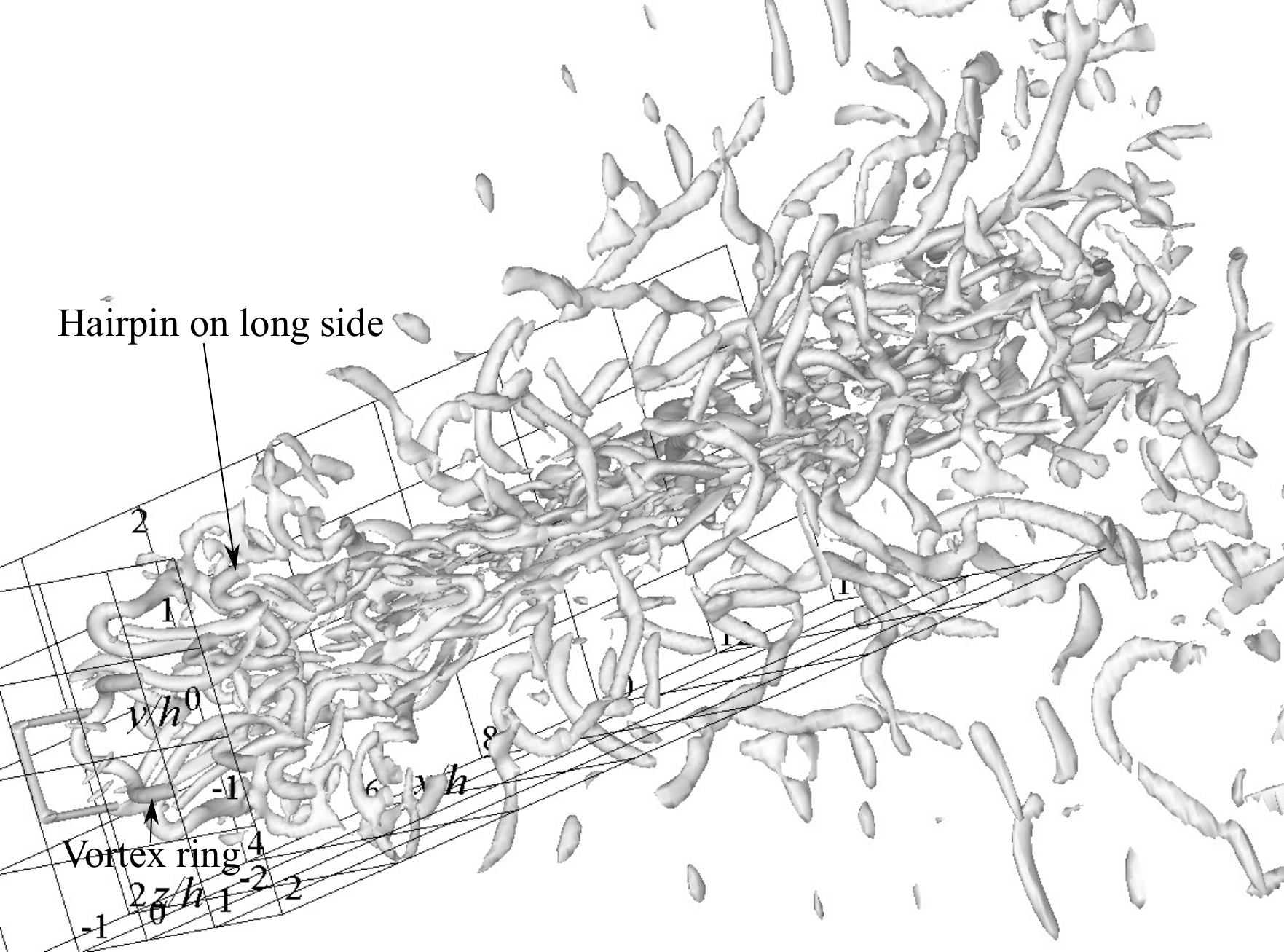} \\
(c)
\vspace*{-0.5\baselineskip}
\caption{Isosurface of curvature of equipressure surface at $T = 0$ for $AR = 2.0$: 
Isosurface value is $-12/h$. 
(a) side view, (b) top view, and (c) perspective view. 
Multi-media view 
(Movies of the time variation of vortex structures from $T = 0$ to $T = 4.75$ are provided).}
\label{curv2}
\end{figure}
%------------------------------------------------------------------------------

For $AR = 1.0$, it is found from Fig. \ref{curv1} (a) 
that the hairpin part of the vortex ring existing around $x/h = 2-3$ attenuates downstream 
and collapses around $x/h = 6$. 
Further downstream, two longitudinal vortices extending in the streamwise direction exist 
around $x/h = 7-12$ and $y/h = \pm 0.2$. 
The collapsed hairpin segment produces these vortices. 
As observed in Figs. \ref{curv1} (a) and (c), near $x/h = 1$, 
a vortex pair consisting of two vortices, A$_{10}$ and B$_{10}$, is symmetrical 
with respect to the diagonal direction of the nozzle. 
This vortex pair extends in the streamwise direction while maintaining symmetry 
as it moves downstream and grows like a vortex pair composed of two vortices, C$_{10}$ and D$_{10}$, at $x/h = 1-2$. 
This vortex pair is further elongated and grows like a vortex pair composed of the vortices, E$_{10}$ and F$_{10}$, 
at $x/h = 2-3$ downstream. 
In the right of Fig. \ref{curv1} (c), only vortex structures near the jet center are visualized. 
As observed from this figure, the vortex pair approaches the jet center. 
In addition, the vortex pair forms a hairpin vortex while spreading away from the jet center 
(see the left of Fig.\ref{curv1} (c)). 
That hairpin vortex collapses, and the remaining vortex pairs spread 
in the diagonal direction of the nozzle. 
After the collapse of the hairpin part of the vortex ring, 
we cannot confirm a clear vortex ring, 
but a vortex ring is regenerated downstream. 
It is considered that such regeneration of the vortex ring occurs 
because four vortex pairs (eight vortex tubes) generate a swirling flow toward the jet center. 
This regeneration is an intriguing phenomenon.

In the case of $AR = 1.5$ in Fig. \ref{curv15}, 
the vortex deformation processes downstream of the long side ($x$-$z$ plane) 
and short side ($x$-$y$ plane) of the nozzle are different. 
We can find from Figs. \ref{curv15} (a) and (b) 
that the scale of the hairpin part of the vortex ring around $x/h = 2-3$ 
on the long side is larger than that on the short side. 
The hairpin parts of this vortex ring exist up to around $x/h = 6$ and $x/h = 8$ 
on the long and short sides, respectively, and collapse downstream. 
For $AR = 1.5$, the vortices spread wider in the minor axis direction than in the major axis direction. 
As can be seen from Figs. \ref{curv15} (a), (b), and (c), 
in the vortex pair existing near $x/h = 1$, 
the scales of vortex A$_{15}$ on the long side and vortex B$_{15}$ on the short side are different. 
In addition, the vortex pair (composed of A$_{15}$ and B$_{15}$) exist 
asymmetrically with respect to the diagonal direction of the nozzle. 
This tendency becomes more conspicuous around $x/h = 1-2$, 
and the vortex C$_{15}$ elongates in the streamwise direction more than the vortex D$_{15}$. 
Around $x/h = 6-9$ downstream, vortex pairs extend in the major axis direction 
on the long side of the nozzle. 
On the short side, around $x/h = 9-12$ and $y/h = \pm 0.5$, 
two longitudinal vortices extending in the streamwise direction exist, 
similar to the result of $AR = 1.0$. 
The collapsed hairpin segment produces these vortices. 
In addition, long longitudinal vortices exist around $x/h = 8-14$ and $y/h = \pm 1.5$, 
and these vortices are also observed around $x/h = 8-14$ near $z/h = \pm 2.0$ on the long side. 
It is considered that this longitudinal vortex is generated by the interaction 
between the vortex pair and the surrounding fluid, 
but the detailed generation mechanism is unknown. 
The right side in Fig. \ref{curv15} (c) shows vortex structures existing at $z/h < 0$. 
The hairpin parts on the long and short sides break up from the vortex ring, 
and the remaining vortex ring transforms into four vortex pairs extending toward the jet center. 
The vortex pair on the short side is coupled near the jet center. 
Downstream, the regeneration of a vortex ring observed at $AR = 1.0$ does not occur.

For $AR = 2.0$ in Fig. \ref{curv2}, 
the trend of the flow field from $x/h = 0$ to 2 is similar to that at $AR = 1.5$. 
Around $x/h=5$, the hairpin part of a vortex ring collapses earlier than that for $AR = 1.5$. 
Around $x/h = 1$, the asymmetry of the vortex pair consisting of the vortices, 
A$_{20}$ and B$_{20}$, increases compared to $AR = 1.5$. 
Therefore, it can be seen that the asymmetry of the vortex pair consisting of the vortices, 
C$_{20}$ and D$_{20}$, existing downstream further increases. 
On the long side, we confirmed that after the hairpin part of the vortex ring collapsed, 
the remaining vortex pair combined, forming a hairpin part again. 
However, that hairpin part immediately collapses. 
The vortices for $AR = 2.0$ diffuse over a wider area of the major and minor axis directions 
compared to $AR = 1.0$ and 1.5. 
This tendency suggests that intensive interference between vortices occurs in the flow field. 
The trend of the jet widening with increasing the aspect ratio has also been observed 
in the previous study using a rectangular nozzle at high Reynolds numbers \citep{Grinstein_2001}. 
The right side of Fig. \ref{curv2} (c) shows vortex structures existing at $z/h < 0$. 
Due to the collapse of the vortex ring, 
vortex pairs elongated in the streamwise direction are observed near the jet center. 
Compared with the result of $AR = 1.5$, 
the regular arrangement of vortex pairs is not observed, 
and vortex structures are widely diffused.

As described above, we clarified the behavior of the vortex structure at $AR = 1$ 
in more detail than the existing research \citep{Gohil_et_al_2015}. 
In addition, we investigated the difference in the jet diffusion due to the aspect ratio 
for $AR > 1.0$ and obtained information to elucidate the mechanism of the axis switching.

%------------------------------------------------------------------------------
% Figure 8
%------------------------------------------------------------------------------
\begin{figure}[!t]
\centering
%
%%% x-y
\begin{minipage}{0.48\linewidth}
\centering
\vspace*{0.5\baselineskip}
\includegraphics[trim=0mm 0mm 0mm 0mm, clip, width=60mm]{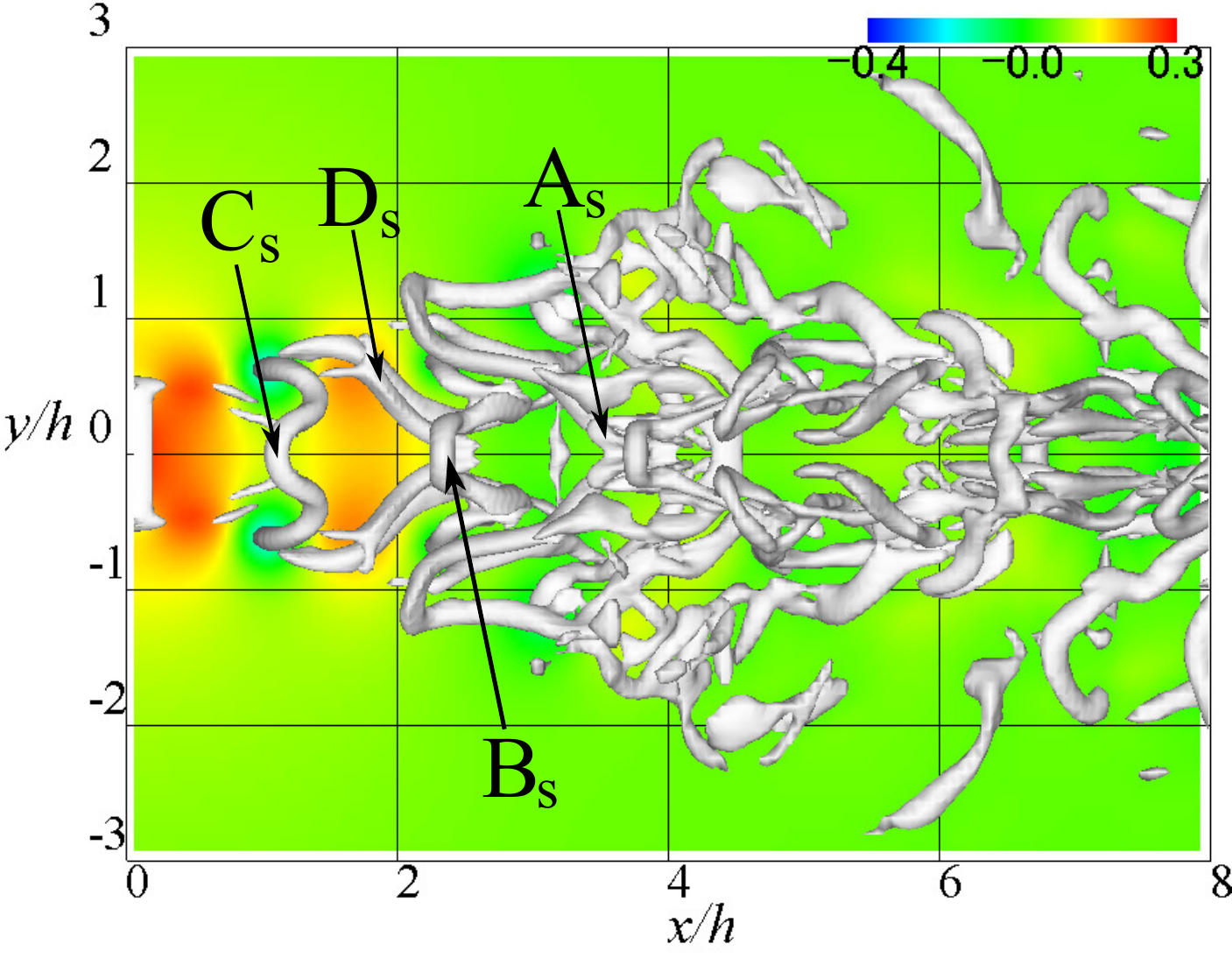} \\
\end{minipage}
%
%%% x-z
\hspace{0.02\linewidth}
\begin{minipage}{0.48\linewidth}
\centering
\includegraphics[trim=0mm 0mm 0mm 0mm, clip, width=60mm]{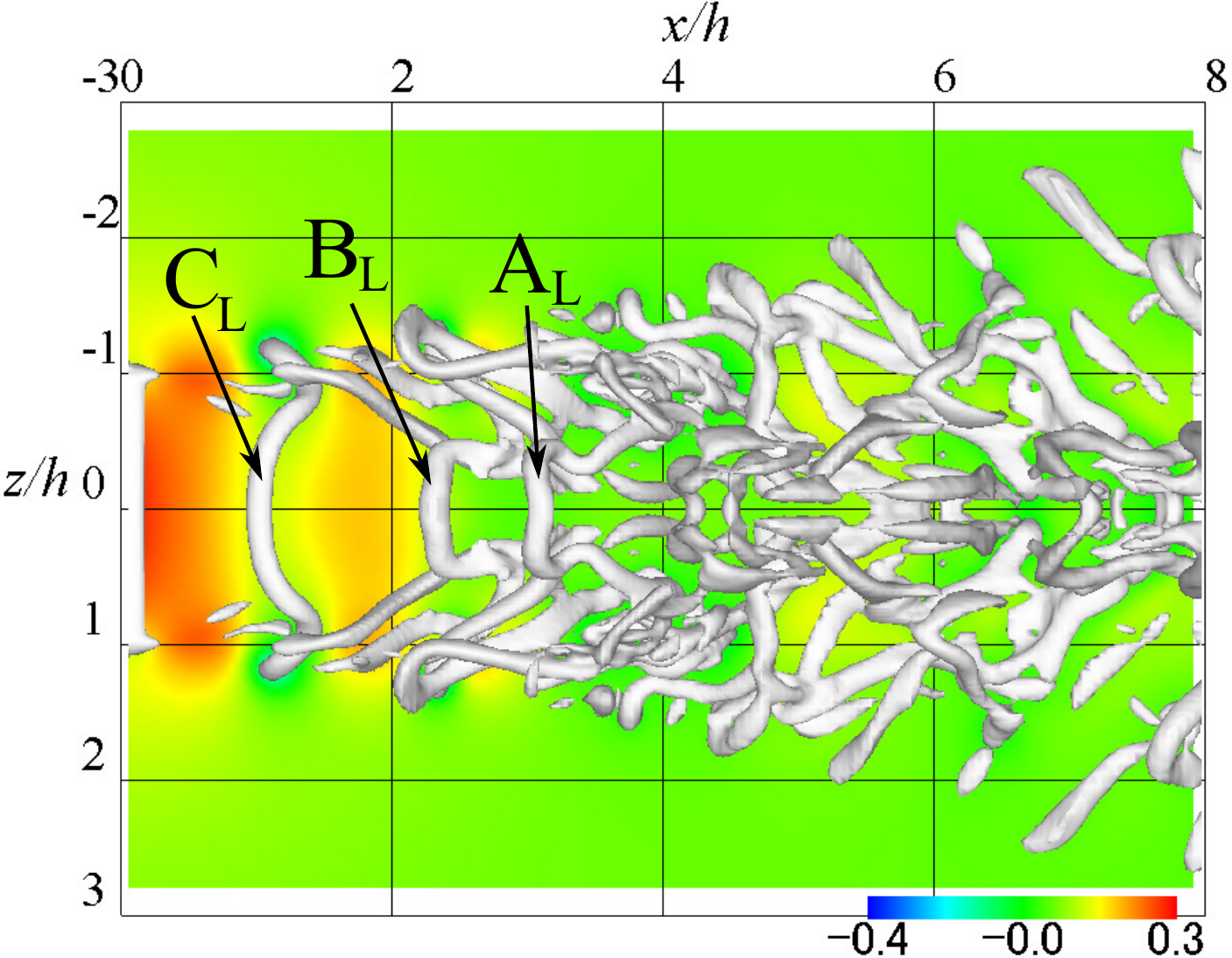} \\
\end{minipage} \\
(a) \\

%%% x-y
\begin{minipage}{0.48\linewidth}
\centering
\vspace*{0.5\baselineskip}
\includegraphics[trim=0mm 0mm 0mm 0mm, clip, width=60mm]{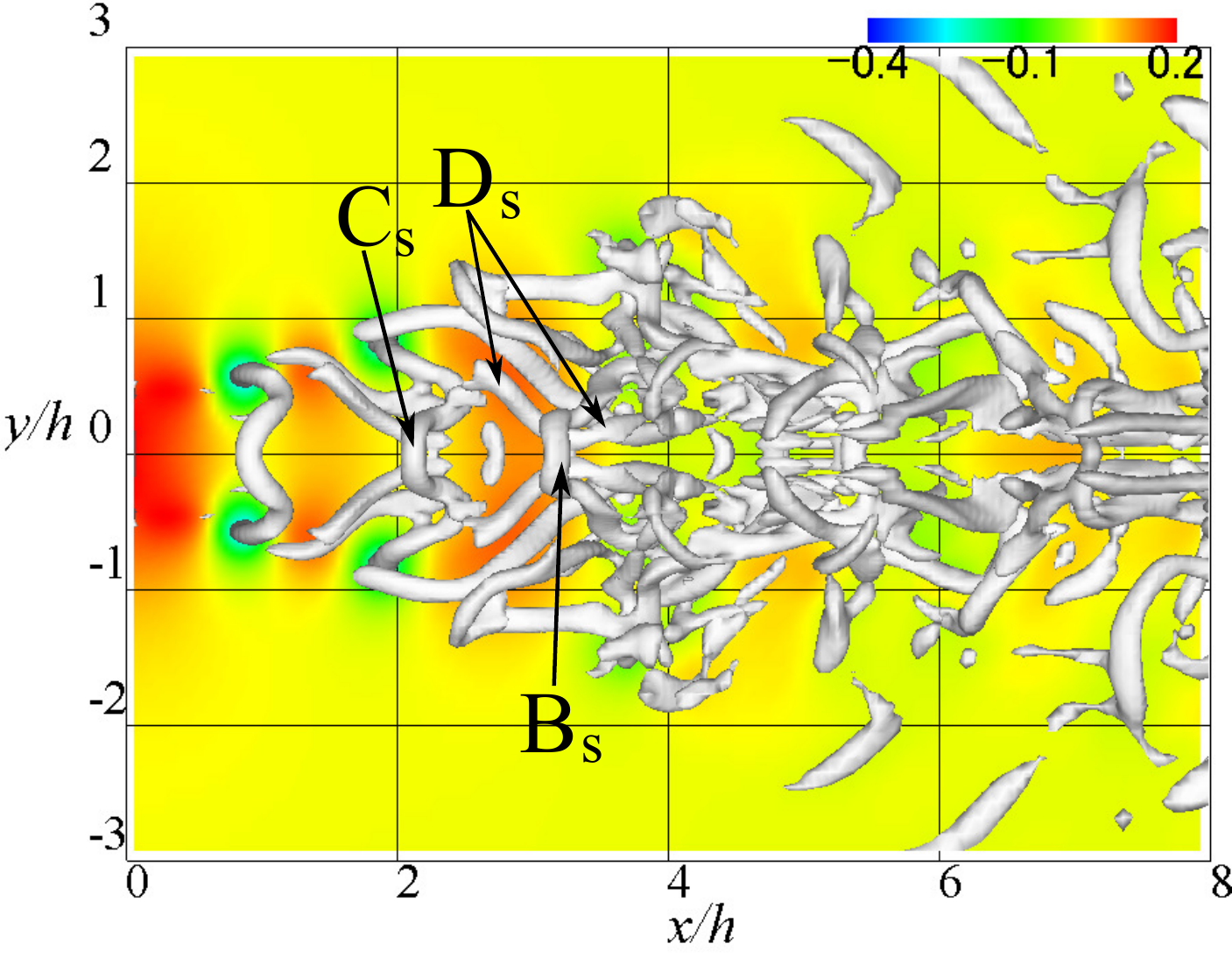} \\
\end{minipage}
%
%%% x-z
\hspace{0.02\linewidth}
\begin{minipage}{0.48\linewidth}
\centering
\includegraphics[trim=0mm 0mm 0mm 0mm, clip, width=60mm]{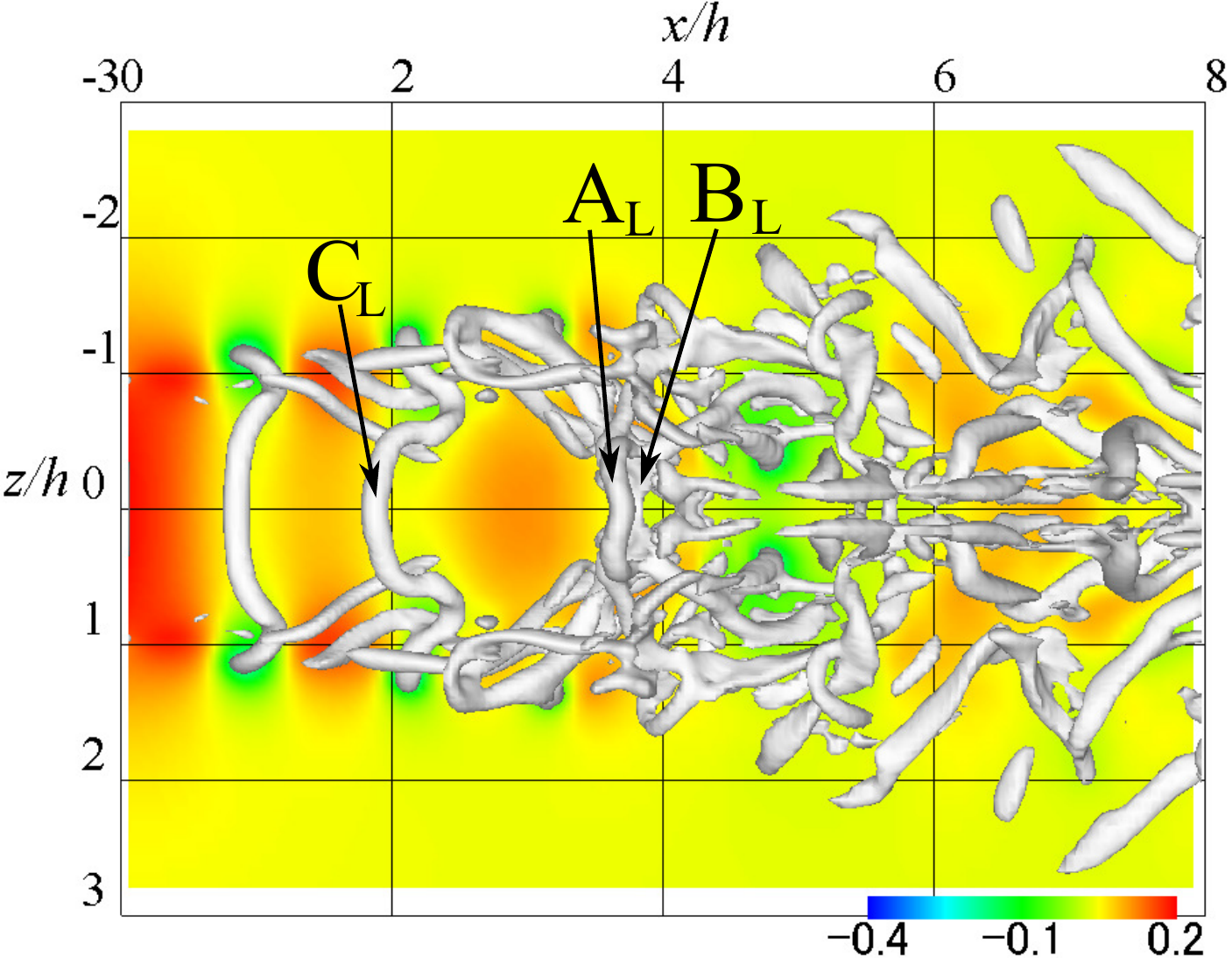} \\
\end{minipage}
(b) \\

%%% x-y
\begin{minipage}{0.48\linewidth}
\centering
\vspace*{0.5\baselineskip}
\includegraphics[trim=0mm 0mm 0mm 0mm, clip, width=60mm]{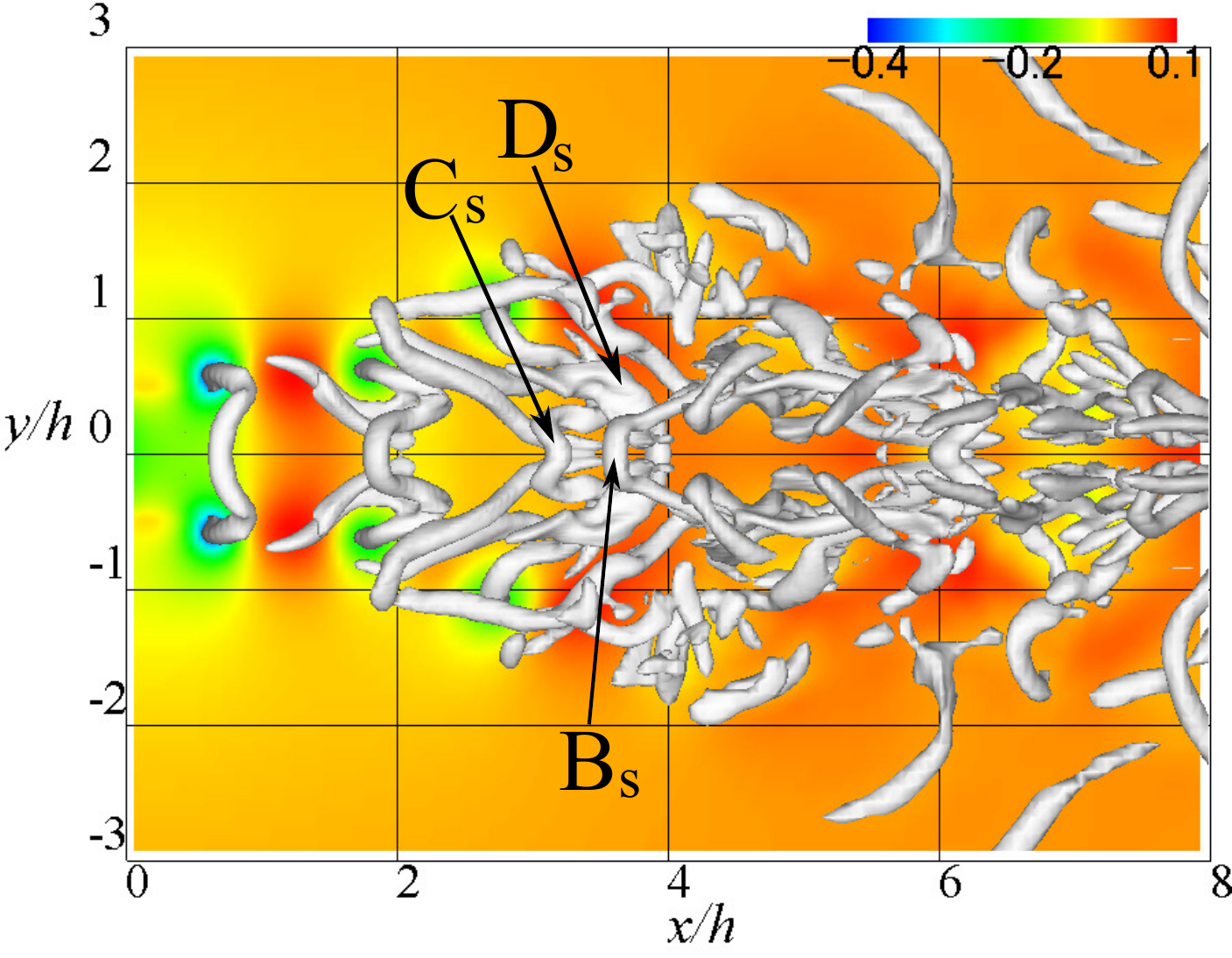} \\
\end{minipage}
%
%%% x-z
\hspace{0.02\linewidth}
\begin{minipage}{0.48\linewidth}
\centering
\includegraphics[trim=0mm 0mm 0mm 0mm, clip, width=60mm]{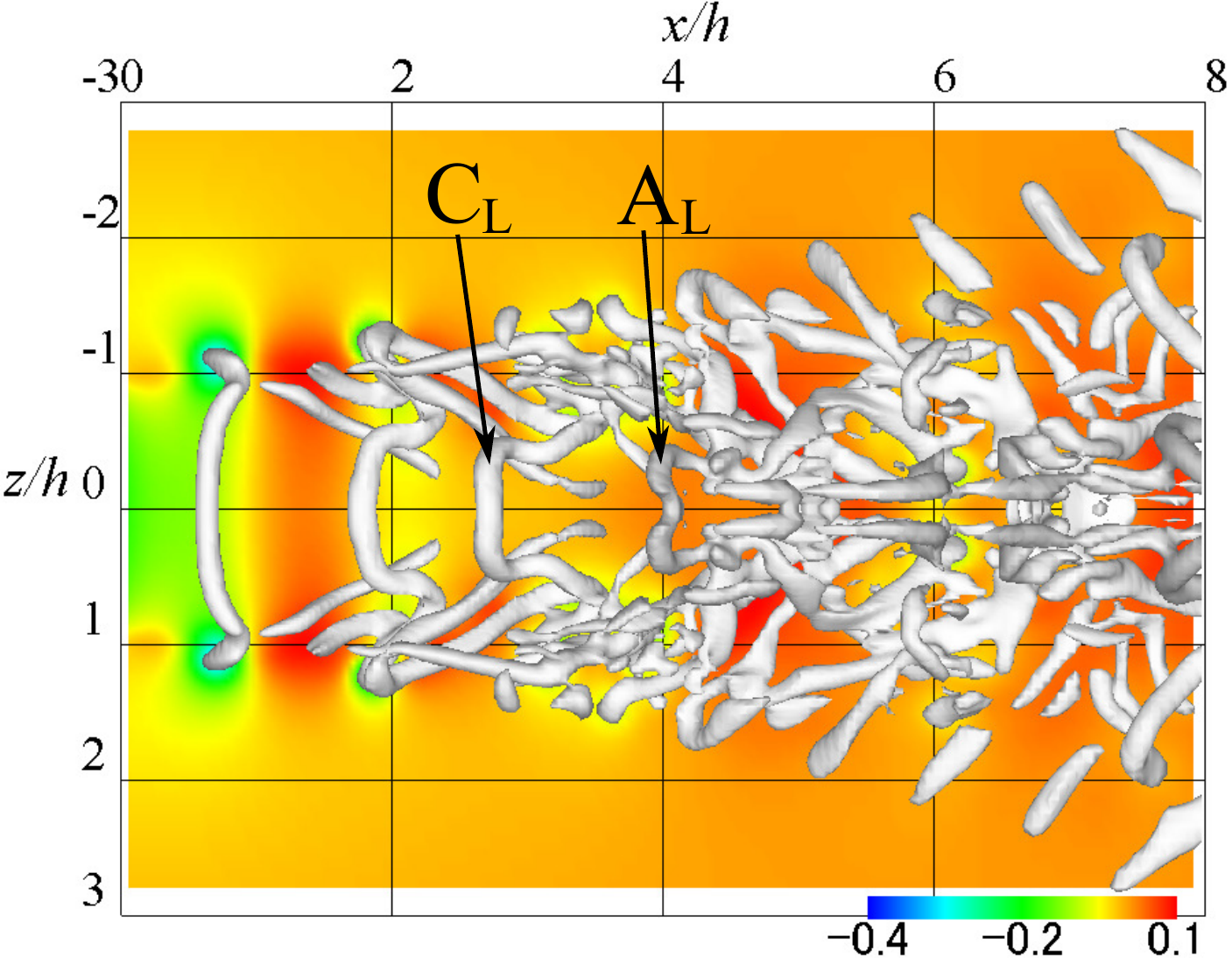} \\
\end{minipage}
(c) \\
\vspace*{-0.5\baselineskip}
\caption{Isosurface of curvature of equipressure surface, 
and pressure contour at $z/h = 0$ in $x$-$y$ plane and at $y/h = 0$ in $x$-$z$ plane for $AR = 2.0$: 
Isosurface value is $-12/h$. 
(a) $T = 0$, (b) $T = 2$, and (c) $T = 4$. 
Multi-media view 
(Movies of the time variation of vortex structures from $T = 0$ to $T = 4.75$ are provided).}
\label{curv2_time}
\end{figure}
%------------------------------------------------------------------------------

Next, to clarify the deformation process of the vortex ring generated 
under the condition $AR = 2.0$, 
where strong interaction between vortices is considered to occur, 
Fig. \ref{curv2_time} shows the time variations of the isosurface of the curvature of an equipressure surface. 
The left and right figures are the views from the short and long sides of the nozzle, respectively. 
The vortices are labeled A to D to distinguish them. 
The subscripts on the long and short sides are L and S, respectively. 
The contours of the pressure at $z/h = 0$ in the $x$-$y$ plane and $y/h = 0$ 
in the $x$-$z$ plane are also shown. 
As the fluid is ejected at $T = 0$, the pressure near the nozzle exit becomes high 
on the short and long sides. 
As a vortex ring flows downstream, 
the pressure around the center of the vortex ring decreases due to the accelerated flow 
caused by the rotation of the vortex ring. 
As a result, a pressure difference occurs between the upstream and center of the vortex ring. 
It is considered that interference between the upstream and downstream vortex rings 
occurs due to distortion and collapse of the vortex rings and this pressure difference.

On the long side, 
the vortices A$_\mathrm{L}$ and B$_\mathrm{L}$ are the hairpin parts of vortex rings, 
and vortex C$_\mathrm{L}$ is a vortex ring. 
At $T = 0$, the distance between the hairpin part B$_\mathrm{L}$ and vortex ring C$_\mathrm{L}$ is approximately $1.5h$. 
On the other hand, the distance between the hairpin parts A$_\mathrm{L}$ and B$_\mathrm{L}$ is approximately $1.0h$, 
and the hairpin parts of the vortex rings approach each other. 
At $T=2$, the hairpin part B$_\mathrm{L}$ collapses early due to the interaction with the hairpin part A$_\mathrm{L}$. 
At this time, the rotation of the hairpin part A$_\mathrm{L}$ also weakens. 
After $T = 0$, the hairpin part A$_\mathrm{L}$ diffuses in the minor axis direction, 
and the hairpin part B$_\mathrm{L}$ contracts. At $T = 2$, 
the hairpin part B$_\mathrm{L}$ of the vortex ring passes 
under the hairpin part A$_\mathrm{L}$ of the vortex ring and is pushed downstream. 
In other words, overtaking of the vortex ring occurs. 
The hairpin part A$_\mathrm{L}$ exists while maintaining its shape, 
but the hairpin part B$_\mathrm{L}$ has collapsed. 
A leapfrog phenomenon occurred at $AR = 2.0$ but was not observed at $AR = 1.0$ and 1.5. 
The previous study \citep{Husain&Hussain_1991,Grinstein_2001} reported this phenomenon. 
At $T = 4$, the rotation of the hairpin part A$_\mathrm{L}$ is weak, 
and the moving velocity downstream becomes slow, 
so the distance between the hairpin parts A$_\mathrm{L}$ and C$_\mathrm{L}$ becomes short.

On the short side, 
the vortices A$_\mathrm{S}$ and B$_\mathrm{S}$ are the hairpin parts of vortex rings, 
vortex C$_\mathrm{S}$ is a vortex ring, and vortex D$_\mathrm{S}$ is the vortex forming a rib structure. 
At $T = 0$, the hairpin part A$_\mathrm{S}$ collapses. 
At $T = 2$, the short side of the vortex ring C$_\mathrm{S}$ is deformed into a hairpin shape. 
The vortex D$_\mathrm{S}$ extends from the corner of the vortex ring C$_\mathrm{S}$ to the vicinity of the jet center, 
passing under the hairpin part B$_\mathrm{S}$ of the vortex ring. 
At this time, the hairpin part B$_\mathrm{S}$ deforms 
more than the vortex ring C$_\mathrm{S}$, 
so it is considered that the hairpin part B$_\mathrm{S}$ strongly interferes with the vortex D$_\mathrm{S}$. 
At $T = 4$, the hairpin part of the vortex ring C$_\mathrm{S}$ on the upstream side approaches 
the hairpin part B$_\mathrm{S}$, 
and the vortex D$_\mathrm{S}$ is deformed to expand in the minor axis direction. 
After $T = 4$, even on the short side, the vortex ring C$_\mathrm{S}$ goes 
under the hairpin part B$_\mathrm{S}$ 
while collapsing like the hairpin part A$_\mathrm{S}$ at $T = 0$, 
and the collapsed vortex structure flows out downstream. 
The hairpin part that existed downstream of the hairpin part A$_\mathrm{S}$ at $T = 0$ 
collapses at $T = 2$, and two hairpin parts disappear.

%------------------------------------------------------------------------------
% Figure 9
%------------------------------------------------------------------------------
\begin{figure}[!t]
\begin{minipage}{0.325\linewidth}
\centering
\includegraphics[trim=0mm 0mm 0mm 0mm, clip, width=65mm]{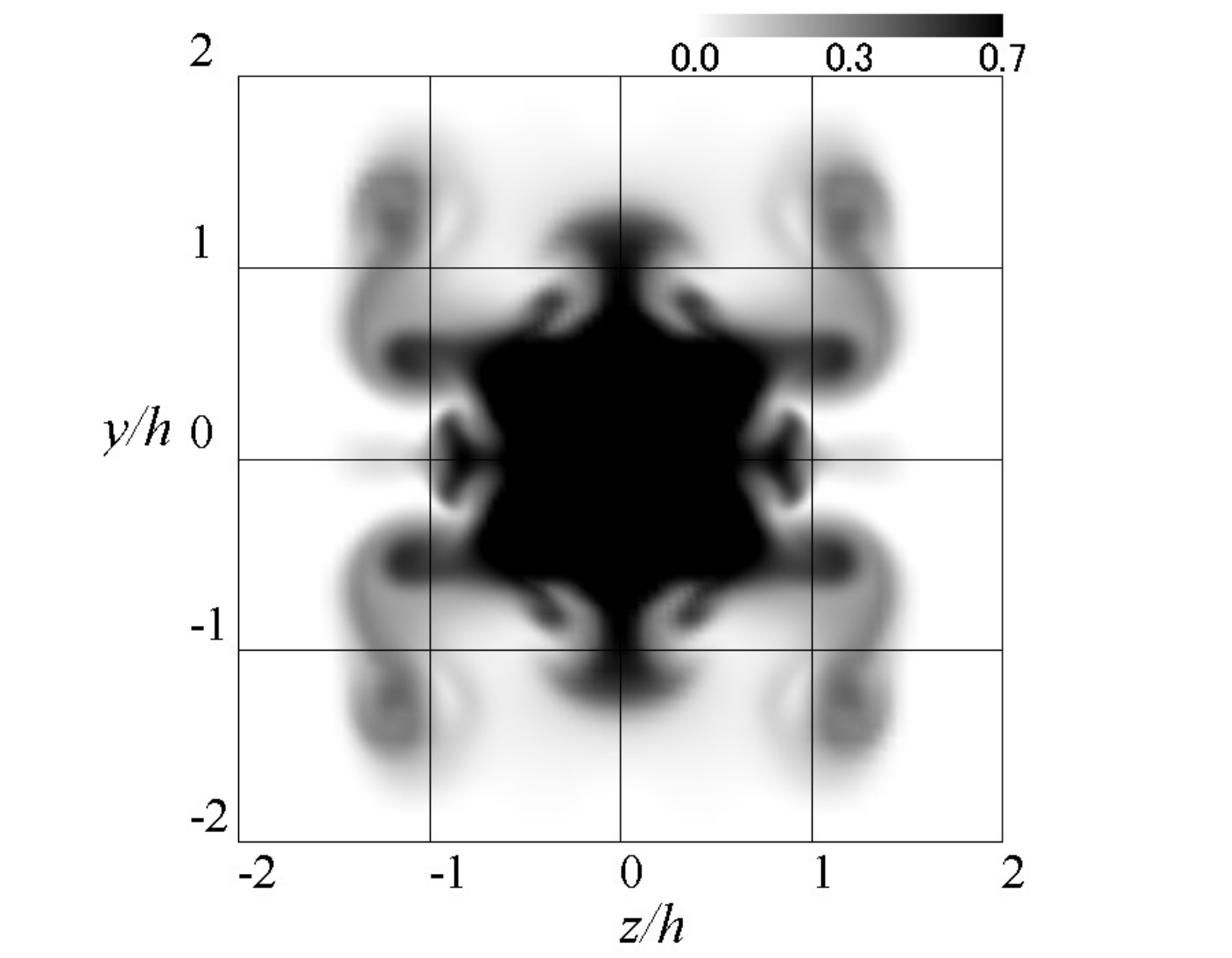} \\
(a)
\end{minipage}
\begin{minipage}{0.325\linewidth}
\centering
\includegraphics[trim=0mm 0mm 0mm 0mm, clip, width=65mm]{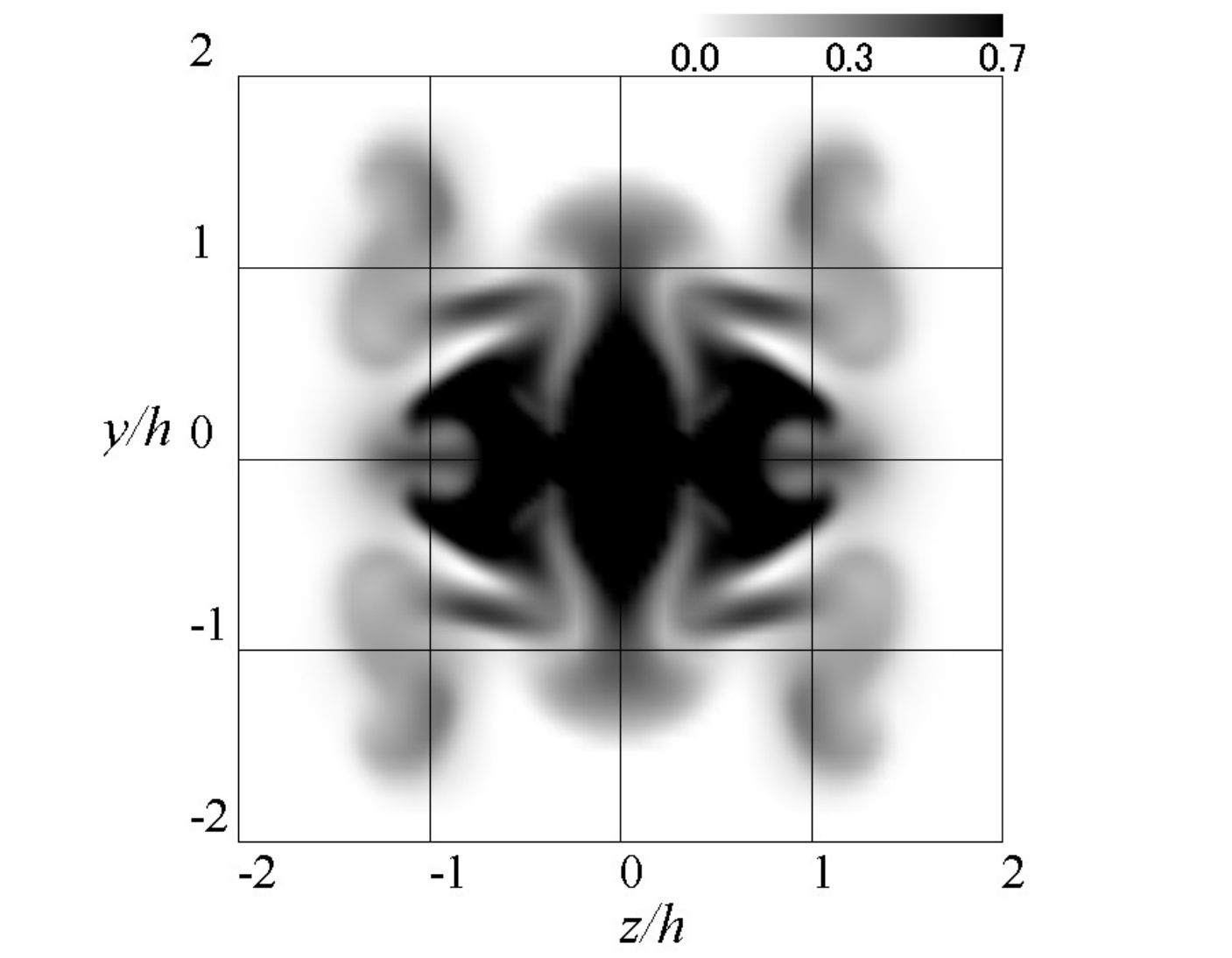} \\
(b)
\end{minipage}
\begin{minipage}{0.325\linewidth}
\centering
\includegraphics[trim=0mm 0mm 0mm 0mm, clip, width=65mm]{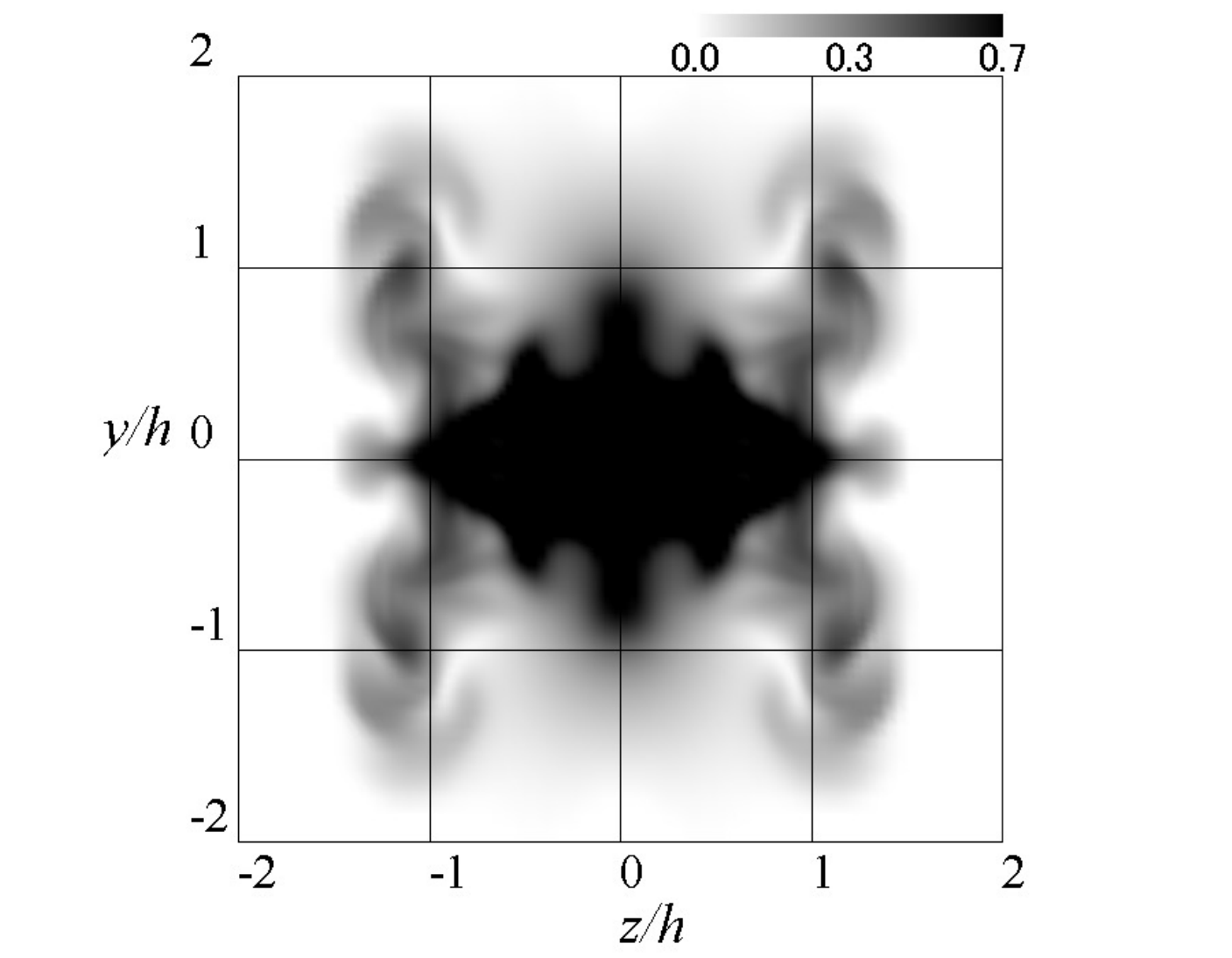} \\
(c)
\end{minipage}
\caption{Streamwise velocity contour at $x/h = 3.5$ for $AR = 2.0$: 
(a) $T = 0$, (b) $T = 2$, and (c) $T = 4$.}
\label{velocity_x/h=3.5}
\end{figure}
%------------------------------------------------------------------------------

On the long side, 
the tendency of the hairpin part A$_\mathrm{L}$ to remain without disappearing 
is different from that on the short side. 
Figure \ref{velocity_x/h=3.5} shows the streamwise velocity distribution 
in the $x/h=3.5$ plane for $AR=2.0$. 
On the long side, the velocity around the hairpin A$_\mathrm{L}$ near $y/h=1.5$ 
is fast at $T=0$ and slow at $T=2$. 
After the hairpin part B$_\mathrm{L}$ passes under the hairpin part A$_\mathrm{L}$, 
the induced velocity by the hairpin part A$_\mathrm{L}$ is attenuated. 
On the short side, the velocity around the hairpin part near $y/h=1.5$ 
is slow at $T=2$ and fast at $T=4$. 
At $T=4$, we confirmed that the flow is fast even at $x/h=3$ near the vortex ring C$_\mathrm{S}$. 
The vortex ring C$_\mathrm{S}$ passes under the hairpin part B$_\mathrm{S}$ at high speed and collapses. 
As the curvature of the vortex ring on the short side is large, 
the induced velocity also increases, 
and it is considered that the vortex ring collapses easily.

Like $AR = 2.0$, at $AR = 1.0$ and 1.5, 
the interference between upstream and downstream vortex rings 
and the pressure difference between the upstream and center of a vortex ring also occur. 
However, overtaking of a vortex ring does not occur. 
As the aspect ratio increases, 
the long side of a rectangular vortex ring becomes more susceptible 
to deformation, like a transverse vortex in shear flow. 
In addition, as the curvature of the hairpin section is also small, 
the increase in streamwise velocity due to the rotation of the vortex tube is suppressed, 
and the interference between the upstream and downstream vortex rings 
makes it easier for the downstream vortex ring to move in the minor axis direction. 
On the other hand, the upstream vortex ring is suppressed 
by the rotation of the downstream vortex ring and becomes smaller downstream. 
As a result, overtaking of the vortex ring is likely to occur. 
We will discuss the induced velocity by vortices later.

There is a difference of one period between the overtaking of vortex rings 
on the short and long sides. 
In the whole jet, the overtaking occurs at the same period as the pulsation of the flow. 
Although the mechanism by which this period difference occurs is unknown, 
we consider that a stable flow field is maintained when vortex rings alternately 
overtake on the short and long sides.

\begin{figure}[!t]
\begin{minipage}{0.48\linewidth}
\centering
\includegraphics[trim=0mm 0mm 0mm 0mm, clip, width=70mm]{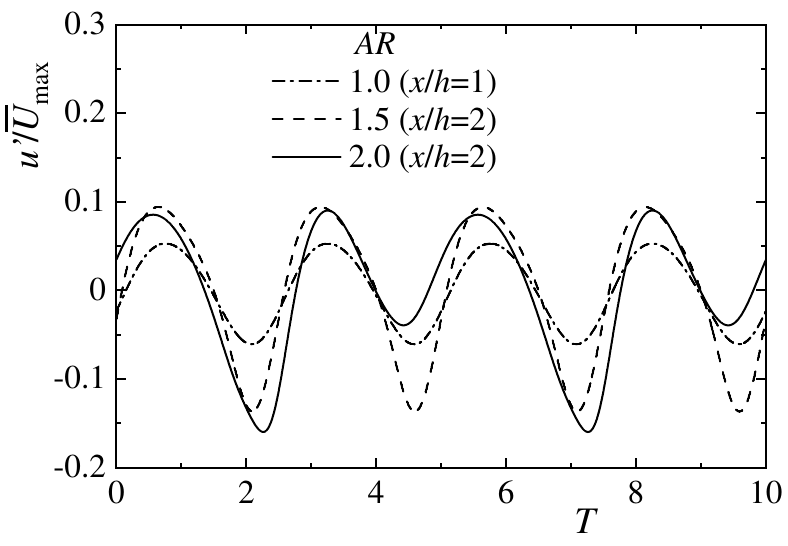} \\
\vspace*{-0.5\baselineskip}
(a)
\end{minipage}
\hspace{0.02\linewidth}
\begin{minipage}{0.48\linewidth}
\centering
\includegraphics[trim=0mm 0mm 0mm 0mm, clip, width=70mm]{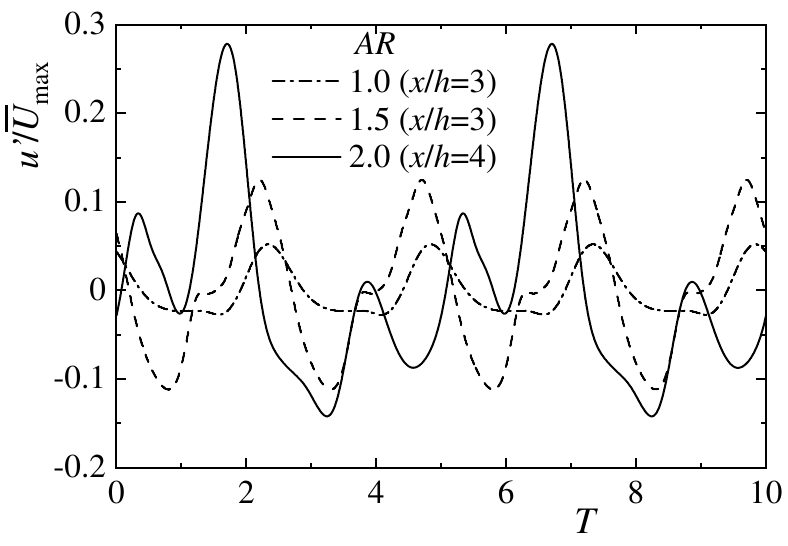} \\
\vspace*{-0.5\baselineskip}
(b)
\end{minipage}
\caption{Time variation of streamwise velocity fluctuation at $y/h = 1.0$ and $z/h = 0$: 
(a) upstream and (b) downstream.}
\label{vel_fluc}
\end{figure}

Figure \ref{vel_fluc} shows the time variation of streamwise velocity fluctuation 
at $y/h = 1.0$ and $z/h = 0$. 
The sampling positions on the upstream side for $AR = 1.0$, 1.5, and 2.0 are 
$x/h = 1$, 2, and 2, respectively, upstream where the axis switching occurs. 
The downstream sampling positions are $x/h = 3$, 3, and 4 near the axis switching 
for $AR = 1.0$, 1.5, and 2.0, respectively. 
On the upstream side, the velocity fluctuations for all $AR$ fluctuate at period $T = 2.5$. 
On the other hand, on the downstream side, 
the velocity fluctuations for $AR = 1.0$ and 1.5 fluctuate at period $T = 2.5$, 
while at $AR = 2.0$, they fluctuate at period $T = 5.0$. 
It is considered that vortex rings approach each other with a period 
longer than the period of vortex shedding. 
In the case of $AR = 2.0$, the vortex ring shed later overtakes the vortex ring shed earlier. 
Therefore, it is considered that intensive interference between vortex rings occurs, 
and the vortices spread over a wider area compared to $AR = 1.0$ and 1.5.

%------------------------------------------------------------------------------
% Figure 10
%------------------------------------------------------------------------------
%------------------------------------------------------------------------------

%------------------------------------------------------------------------------
% Figure 11
%------------------------------------------------------------------------------
\begin{figure}[!t]
\centering
%
%%% AR=1.0
\begin{minipage}{0.325\linewidth}
\centering
\includegraphics[trim=0mm 0mm 0mm 0mm, clip, width=55mm]{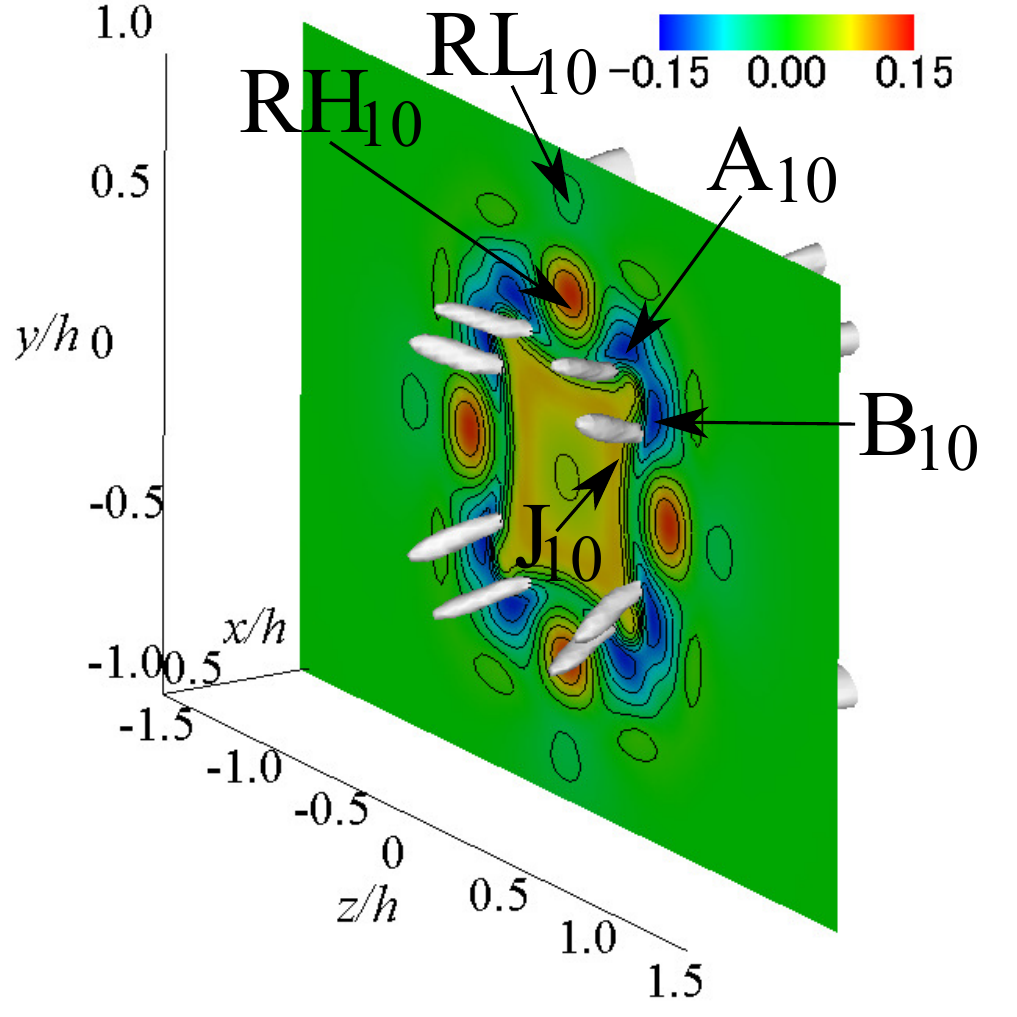} \\
\vspace*{-0.5\baselineskip}
(a)
\end{minipage}
%
%%% AR=1.5
\begin{minipage}{0.325\linewidth}
\centering
\includegraphics[trim=0mm 0mm 0mm 0mm, clip, width=55mm]{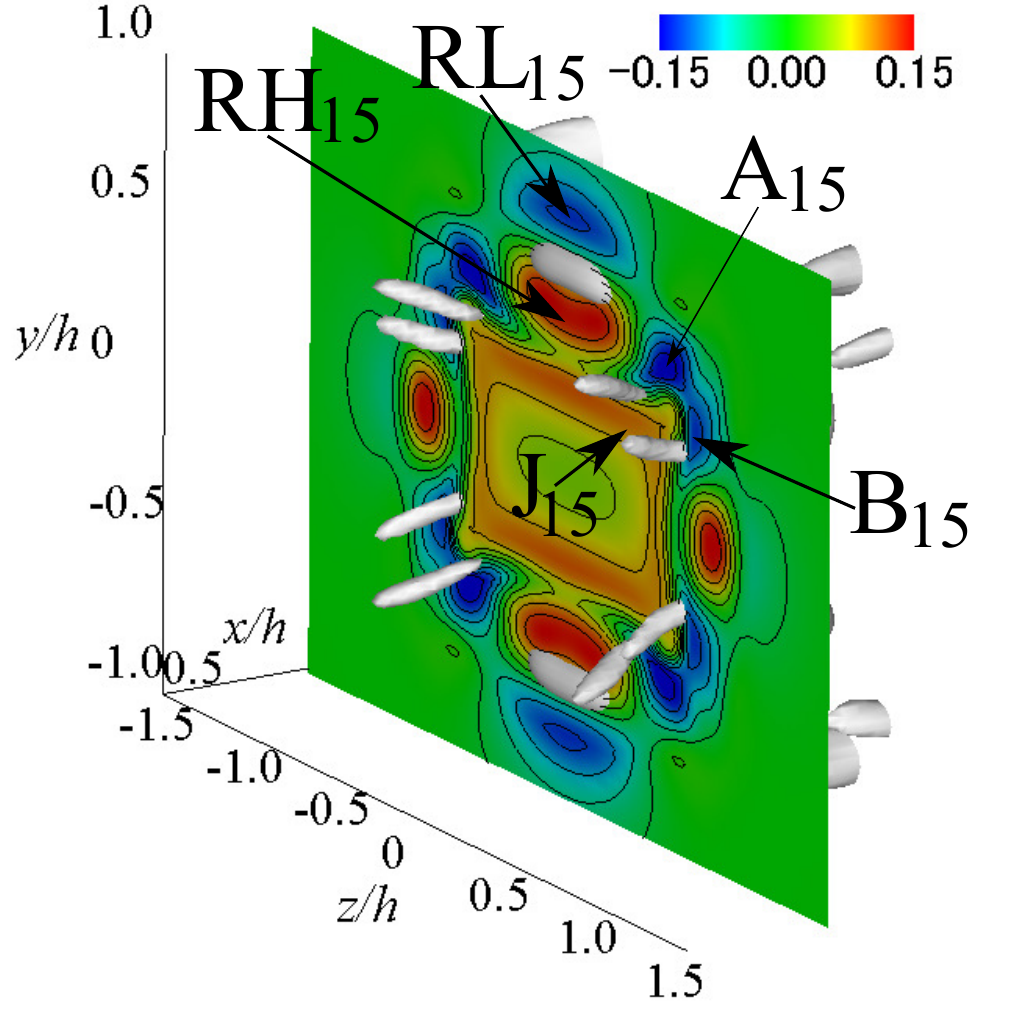} \\
\vspace*{-0.5\baselineskip}
(b)
\end{minipage}
%
%%% AR=2.0
\begin{minipage}{0.325\linewidth}
\centering
\includegraphics[trim=0mm 0mm 0mm 0mm, clip, width=55mm]{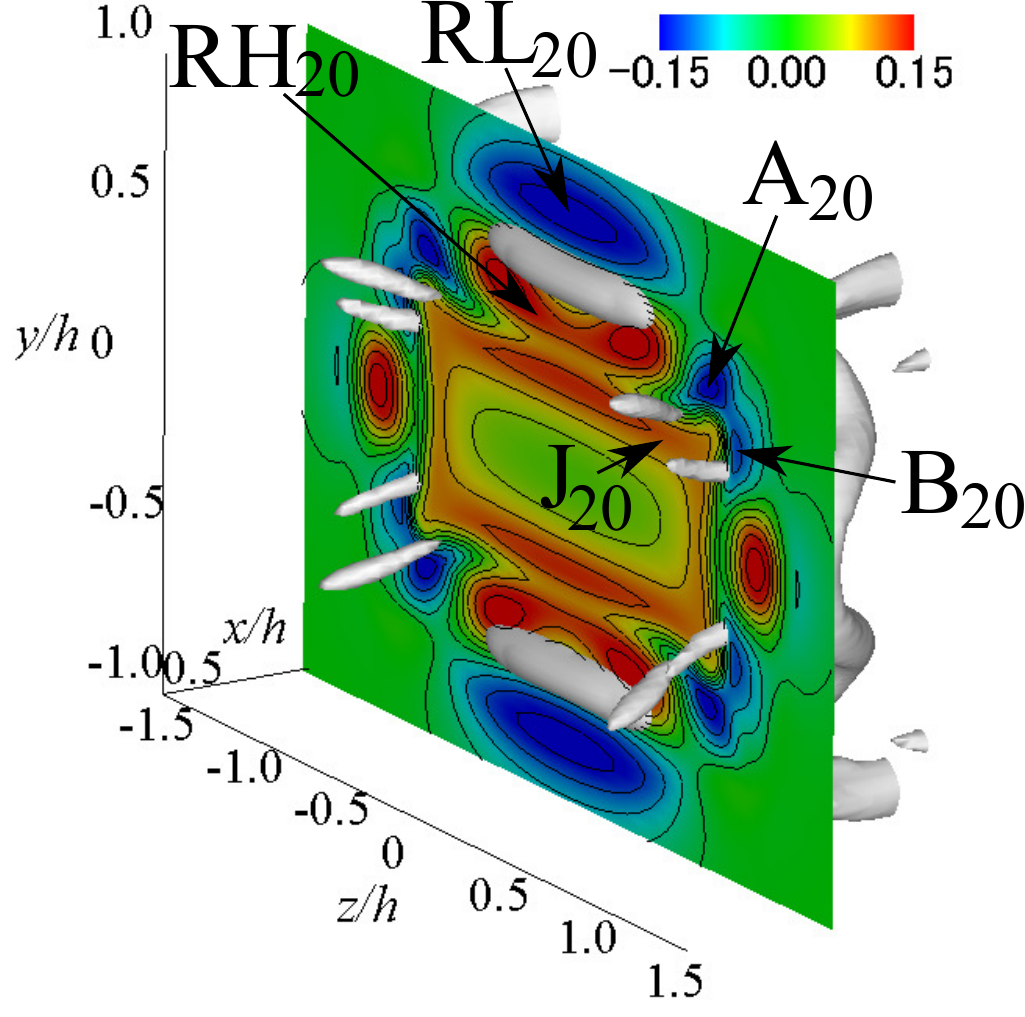} \\
\vspace*{-0.5\baselineskip}
(c)
\end{minipage}
\caption{Isosurface of curvature of equipressure surface and 
contours of streamwise velocity fluctuation in $y$-$z$ plane at $x/h = 1$ for $T = 0$: 
Isosurface value is $-12/h$. Contour interval is 0.03 from -0.15 to 0.15. 
(a) $AR = 1.0$, (b) $AR = 1.5$, and (c) $AR = 2.0$.}
\label{uf}
\end{figure}
%------------------------------------------------------------------------------

To clarify the flow induction by the vortex ring, 
the isosurface of the curvature calculated from the equipressure surface 
and the distribution of the streamwise velocity fluctuation in the $y$-$z$ plane at $x/h=1$ 
are shown in Fig. \ref{uf}. 
The regions of RH and RL indicate positive and negative fluctuations 
generated by the sides of the vortex ring, respectively. 
Areas A and B indicate negative fluctuations due to the corner of the vortex ring, 
and area J shows positive fluctuation due to the pulsating flow. 
Also, the fluctuations for $AR = 1.0$, 1.5, and 2.0 are distinguished 
by subscripts 10, 15, and 20, respectively. 
As the self-induced velocity of a vortex ring 
is inversely proportional to the radius of curvature \citep{Hama_1962}, 
the corner of the vortex ring with large curvature moves downstream 
faster than the side of the vortex ring. 
Therefore, the negative fluctuations A and B occur at $x/h = 1$ 
upstream from the corner of the vortex ring. 
A strong shear layer is generated between the negative fluctuations A and B 
and the positive fluctuation J. 
In the region where such an intensive shear layer exists, 
vortex pairs occur downstream near the corner of the nozzle. 
For $AR = 1.0$, the generated vortex pairs are symmetrical 
about the diagonal direction of the nozzle. 
For $AR = 1.5$ and 2.0, the short side of the vortex ring is shorter than the long side, 
so the radius of curvature on the short side becomes small. 
Therefore, a strong flow is induced on the short side, and the short side 
moves downstream faster than the long side. 
As a result, the generated vortex pairs become asymmetric, 
and the negative fluctuation B on the short side and the negative fluctuation A 
on the long side become asymmetric. 
Such the vortex pair generated from the corner of the rectangular vortex ring 
further distort the vortex ring itself and causes axis switching. 
\citet{Grinstein_2001} reported that at $AR = 1.0$, 
rib structures due to a vortex pair occur, 
but at $AR \ge 2$, a single rib appears. 
In this study, as the Reynolds number is low, 
the asymmetric vortex pair (ribs) is generated, but the rib on one side did not disappear.

%------------------------------------------------------------------------------
% Figure 12
%------------------------------------------------------------------------------
\begin{figure}[!t]
\centering
\begin{minipage}{0.48\linewidth}
\centering
\includegraphics[trim=0mm 0mm 0mm 0mm, clip, width=80mm]{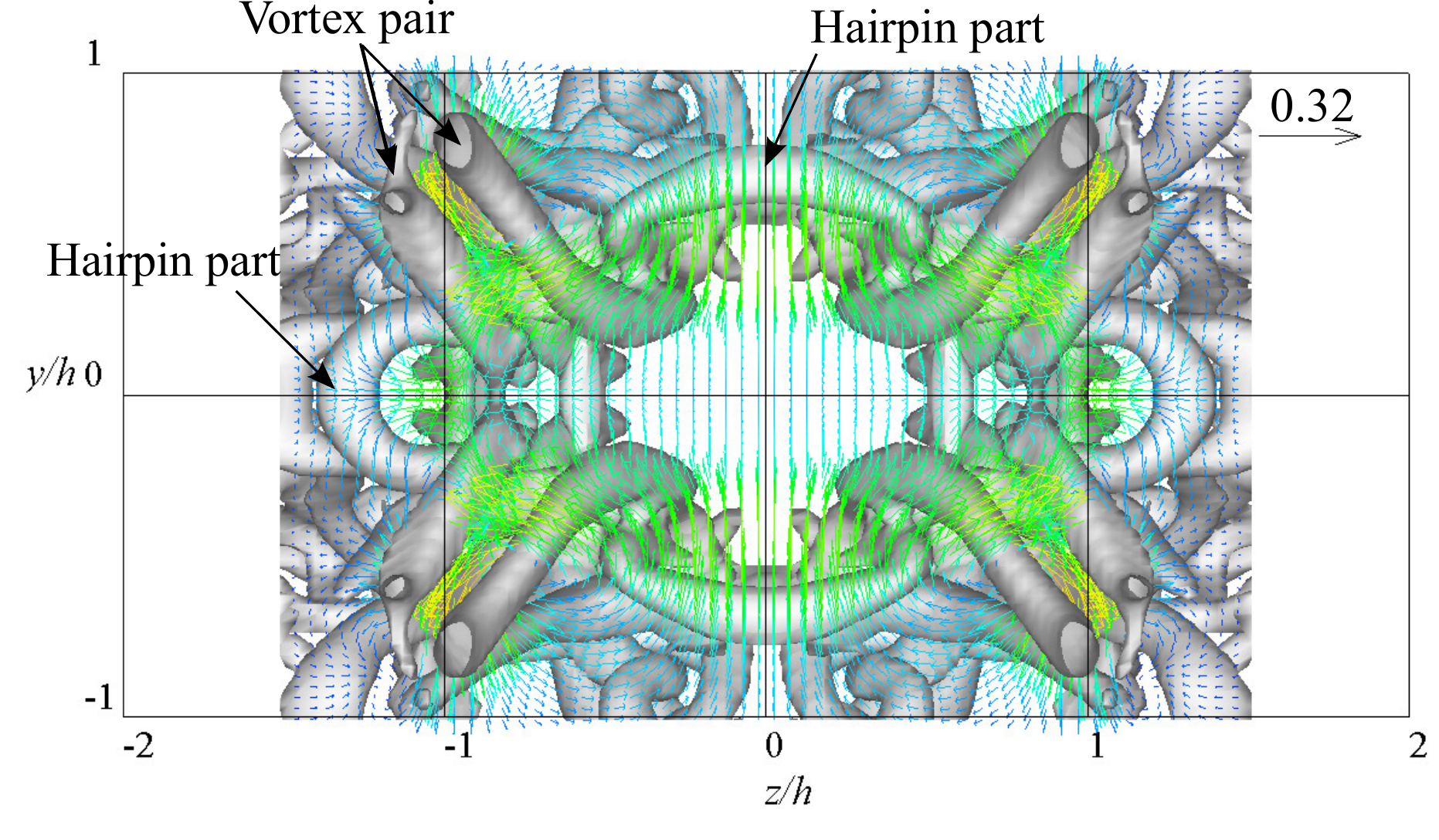} \\
(a)
\end{minipage}
\hspace{0.02\linewidth}
\begin{minipage}{0.48\linewidth}
\centering
\includegraphics[trim=0mm 0mm 0mm 0mm, clip, width=75mm]{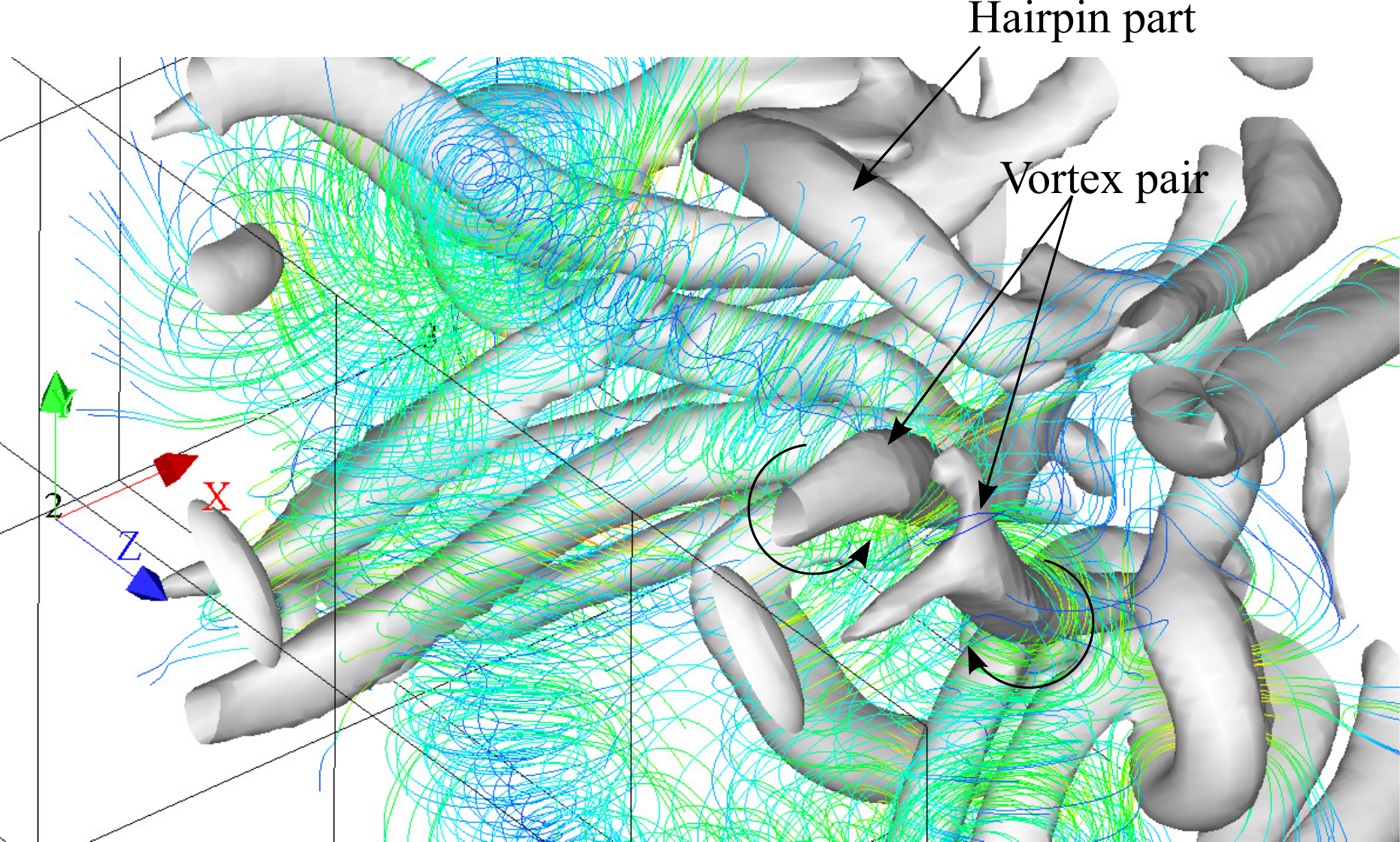} \\
(b)
\end{minipage}
\caption{Isosurface of curvature of equipressure surface and 
velocity fluctuation vectors at $T = 0$ for $AR = 2.0$ (left). 
Streamlines obtained from velocity fluctuation (right): 
Isosurface value is $-12/h$. 
(a) $y$-$z$ plane and (b) perspective view.}
\label{cur_vec_strm _2}
\end{figure}
%------------------------------------------------------------------------------

\begin{figure}[t]
\centering
\includegraphics[trim=0mm 0mm 0mm 0mm, clip, width=100mm]{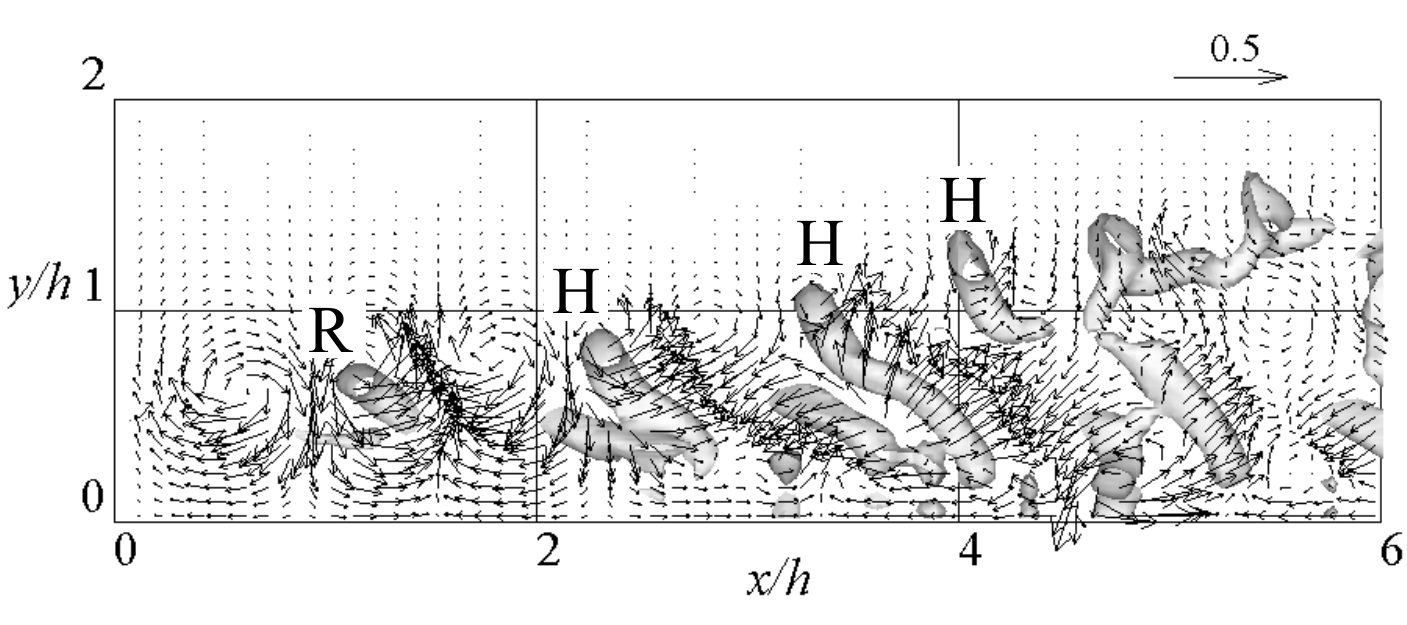} \\
\vspace*{-0.5\baselineskip}
\caption{Isosurface of curvature of equipressure surface and 
velocity fluctuation vectors in $x$-$y$ plane at $T = 0$ for $AR = 1.0$: 
Isosurface value is $-12/h$.}
\label{cur_vec_1}
\end{figure}

\begin{figure}[!t]
\centering
\includegraphics[trim=0mm 0mm 0mm 0mm, clip, width=100mm]{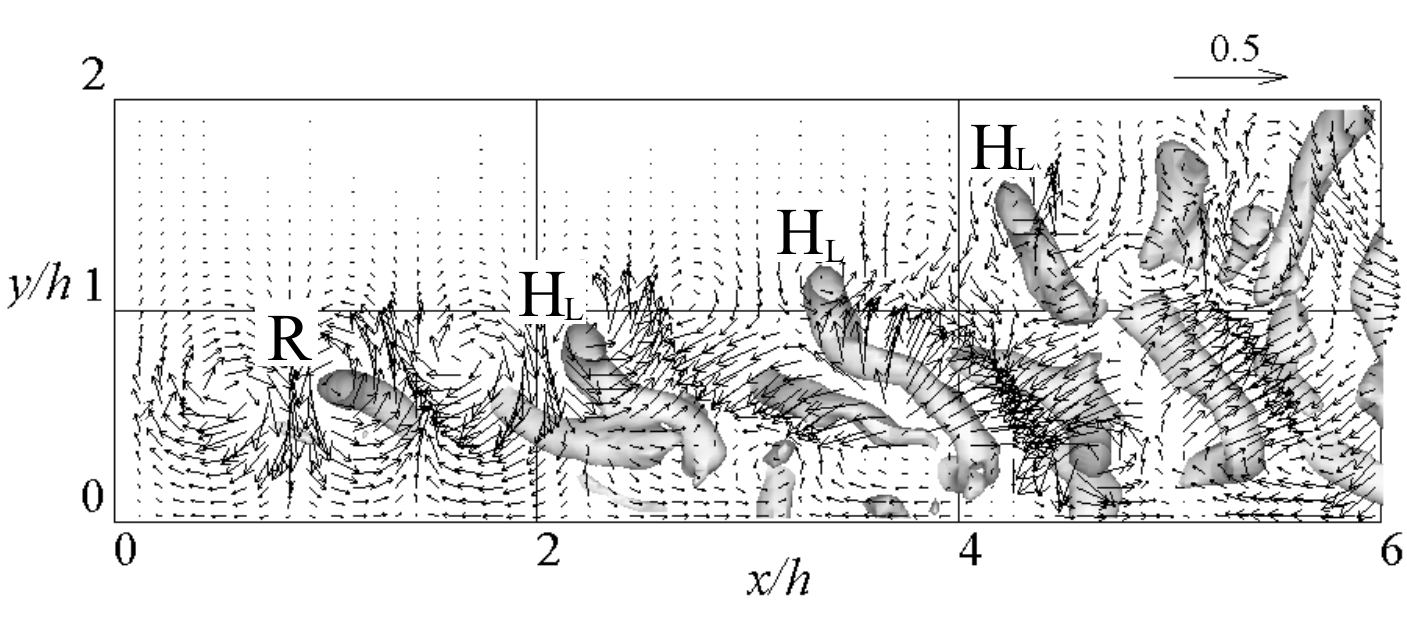} \\
\vspace*{-0.5\baselineskip}
(a) \\
\includegraphics[trim=0mm 0mm 0mm 0mm, clip, width=100mm]{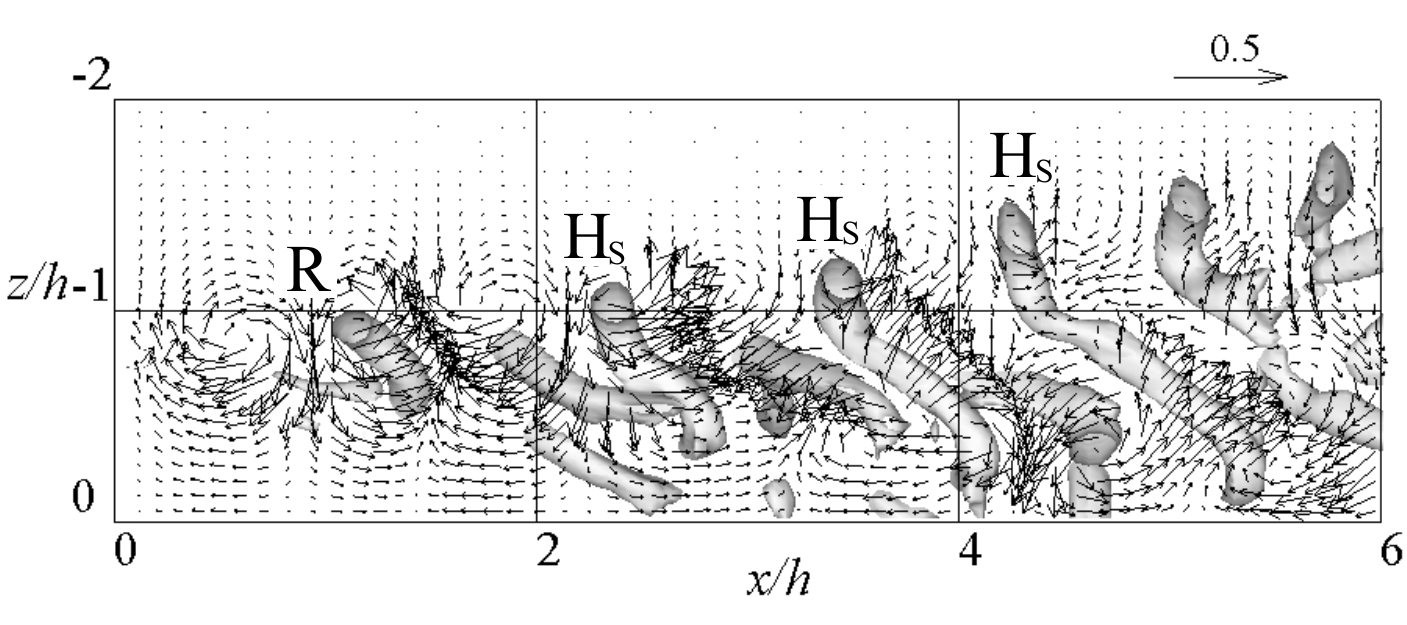} \\
\vspace*{-0.5\baselineskip}
(b)
\vspace*{-0.5\baselineskip}
\caption{Isosurface of curvature of equipressure surface and 
velocity fluctuation vectors at $T = 0$ for $AR = 1.5$: 
Isosurface value is $-12/h$. 
(a) $x$-$y$ plane and (b) $x$-$z$ plane.}
\label{cur_vec_15}
\end{figure}

To investigate the flow due to the vortex pairs shown in Fig. \ref{velocity_x/h=3.5}, 
Fig. \ref{cur_vec_strm _2} (a) shows vortex structures for $AR = 2.0$ 
and the velocity fluctuation vector in the $x/h=1.9$ plane as a representative example. 
Vortices upstream of $x/h = 1.9$ are not visualized. 
To easily understand the direction of velocity fluctuations, 
we represented the velocity fluctuation vectors as streamlines in Fig. \ref{cur_vec_strm _2} (b). 
Around $x/h=1.9$, the vortex ring does not collapse, 
and the hairpin parts on the long and short sides exist. 
As can be seen from the streamlines, 
a rotating flow occurs upstream of the hairpin part on the long side. 
This rotating flow is generated by the interference between the upstream and downstream hairpin parts. 
This flow is also shown in Fig. \ref{cur_vec_2} below. 
In addition, a vortex pair exists at the corner of the vortex ring. 
It can be seen that this vortex pair is a combination of two asymmetric vortices. 
As indicated by the arrows in Fig. \ref{cur_vec_strm _2} (b), 
the vortex pair exhibits counter-rotating flow, 
which generates the flow toward the upstream near the jet center axis. 
Such flow due to the vortex pair also occurs with other aspect ratios, 
as shown in Figs. \ref{cur_vec_1} and \ref{cur_vec_15}.

%------------------------------------------------------------------------------
% Figure 13
%------------------------------------------------------------------------------
%------------------------------------------------------------------------------

%------------------------------------------------------------------------------
% Figure 14
%------------------------------------------------------------------------------
%------------------------------------------------------------------------------

For each $AR$, we observe the differences in the behavior of the hairpin part of the vortex ring 
and the induced flow. 
Figures \ref{cur_vec_1} to \ref{cur_vec_2} show vortex structures, 
and the velocity fluctuation vectors in the $x$-$y$ plane at $z/h = 0$ 
and in the $x$-$z$ plane at $y/h = 0$. 
For $AR = 1$, only the $x$-$y$ sectional view is shown. 
A vortex ring is labeled as R, the hairpin portion of a vortex ring is labeled as H, 
and the long and short sides are distinguished with subscripts of L and S, respectively. 
In all conditions, the hairpin part of the vortex ring develops downstream 
away from the central axis of the jet. 
This is because a counterclockwise flow is induced around the hairpin part of the vortex ring, 
and an intensive flow is generated from the jet center to the outside. 
In addition, upstream of the hairpin part, 
the flow from the outside to the jet center occurs, entraining the surrounding fluid. 
As this downward flow and the rotation of the vortex pair interact, 
the flow toward the upstream occurs near the jet center at $x/h = 2$. 
It is considered that such flow around the hairpin part promotes the mixing 
between the ejected fluid and surrounding fluid.

In the case of $AR = 1.0$, 
the hairpin parts of the vortex rings existing around $x/h = 4-6$ 
move up to near $y/h = 1.5$. 
For $AR = 1.5$, the hairpin portions around $x/h = 4-6$ move up to near $y/h = 2$ on the long side. 
In the case of $AR = 2.0$, the hairpin part H$_\mathrm{L}$ of the vortex ring shed later approaches 
the hairpin part of the vortex ring existing near $x/h = 3$. 
The outward flow from the jet center due to the hairpin portion of the vortex ring 
on the upstream side pushes the hairpin portion on the downstream side outside. 
Therefore, the hairpin part of the vortex ring moves 
away from the central axis of the jet on the long side 
compared to the short side. 
As time passes, the vortex ring on the upstream side is sucked 
inside the vortex ring on the downstream side, and the vortex ring overtaking occurs. 
Due to such strong interaction between vortex rings, 
the vortices spread over a wider area compared to $AR = 1.0$ and 1.5.

%------------------------------------------------------------------------------
% Figure 15
%------------------------------------------------------------------------------
\begin{figure}[!t]
\centering
\includegraphics[trim=0mm 0mm 0mm 0mm, clip, width=100mm]{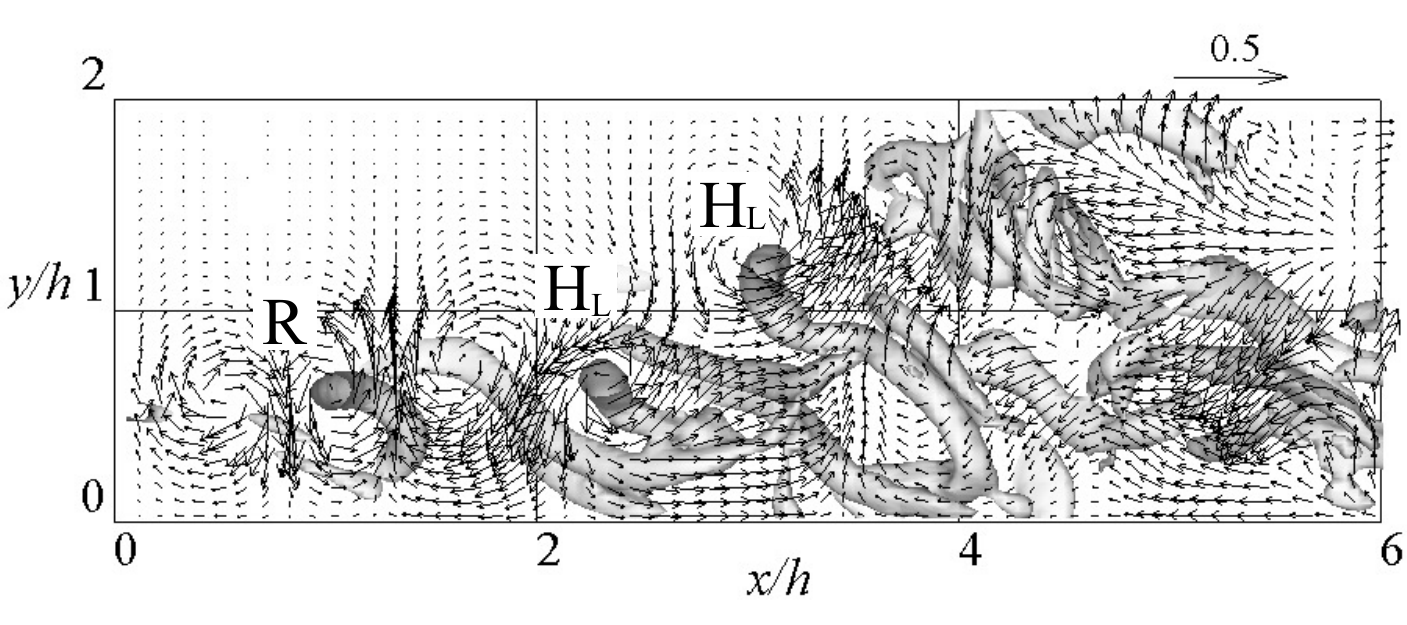} \\
\vspace*{-0.5\baselineskip}
(a) \\
\includegraphics[trim=0mm 0mm 0mm 0mm, clip, width=100mm]{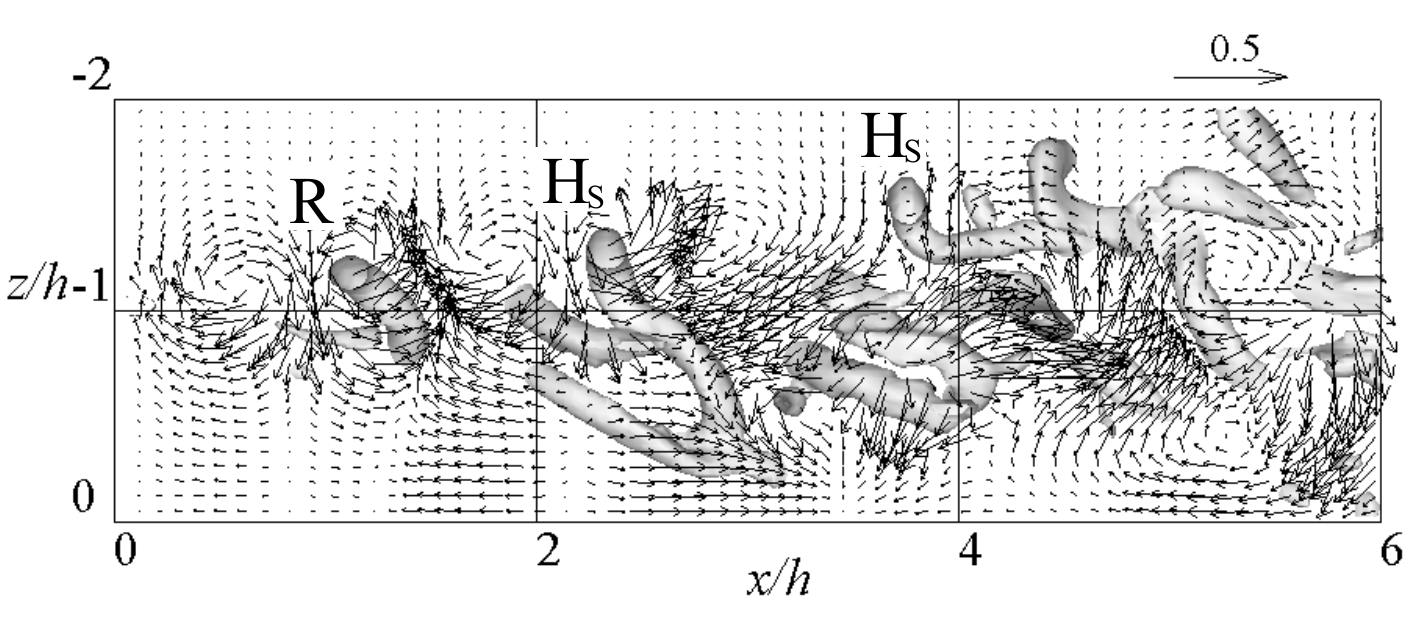} \\
\vspace*{-0.5\baselineskip}
(b)
\vspace*{-0.5\baselineskip}
\caption{Isosurface of curvature of equipressure surface and 
velocity fluctuation vectors at $T = 0$ for $AR = 2.0$: 
Isosurface value is $-12/h$. 
(a) $x$-$y$ plane and (b) $x$-$z$ plane.}
\label{cur_vec_2}
\end{figure}
%------------------------------------------------------------------------------

%++++++++++++++++++++++++++++++++++++++++++++++++++++++++++++++++++++++++++++++
\subsection{Axis switching of jet}
%++++++++++++++++++++++++++++++++++++++++++++++++++++++++++++++++++++++++++++++

\begin{figure}[!t]
\centering
%
%%% AR=1.0
\begin{minipage}{0.48\linewidth}
\centering
\includegraphics[trim=0mm 0mm 0mm 0mm, clip, width=60mm]{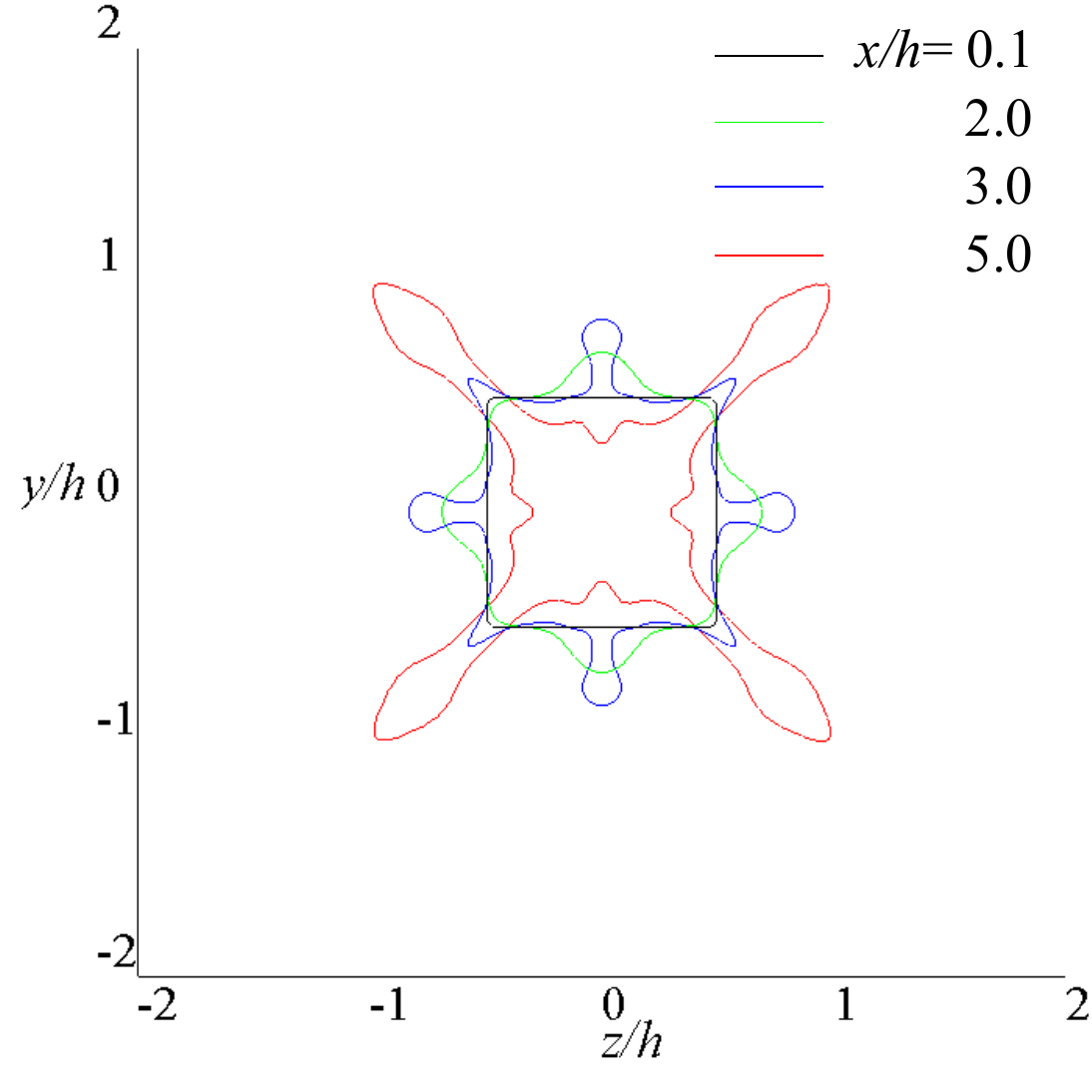} \\
(a) \\
\end{minipage}
%
%%% AR=1.5
\hspace{0.02\linewidth}
\begin{minipage}{0.48\linewidth}
\centering
\includegraphics[trim=0mm 0mm 0mm 0mm, clip, width=60mm]{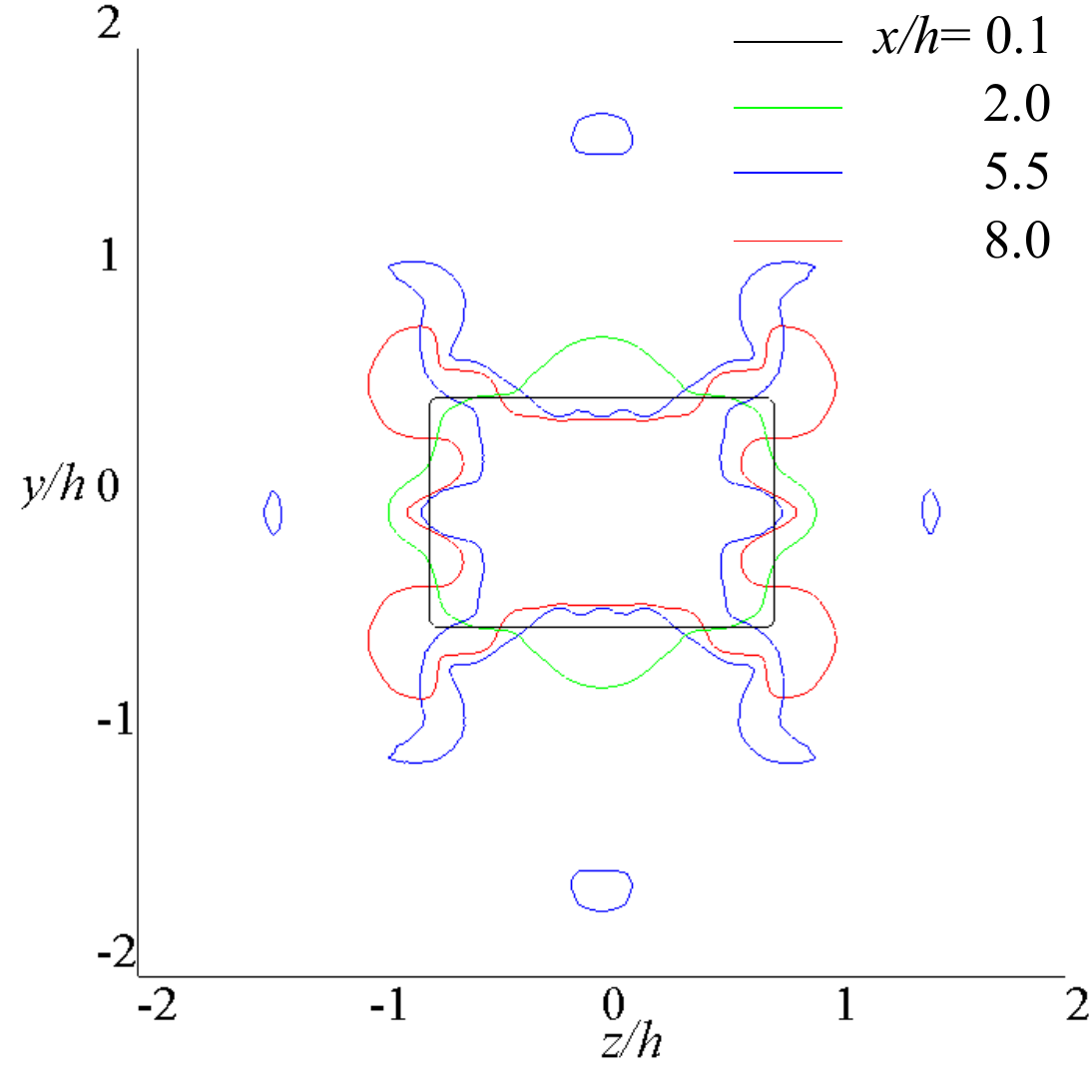} \\
(b) \\
\end{minipage}
%
%%% AR=2.0
\begin{minipage}{0.90\linewidth}
\centering
\includegraphics[trim=0mm 0mm 0mm 0mm, clip, width=60mm]{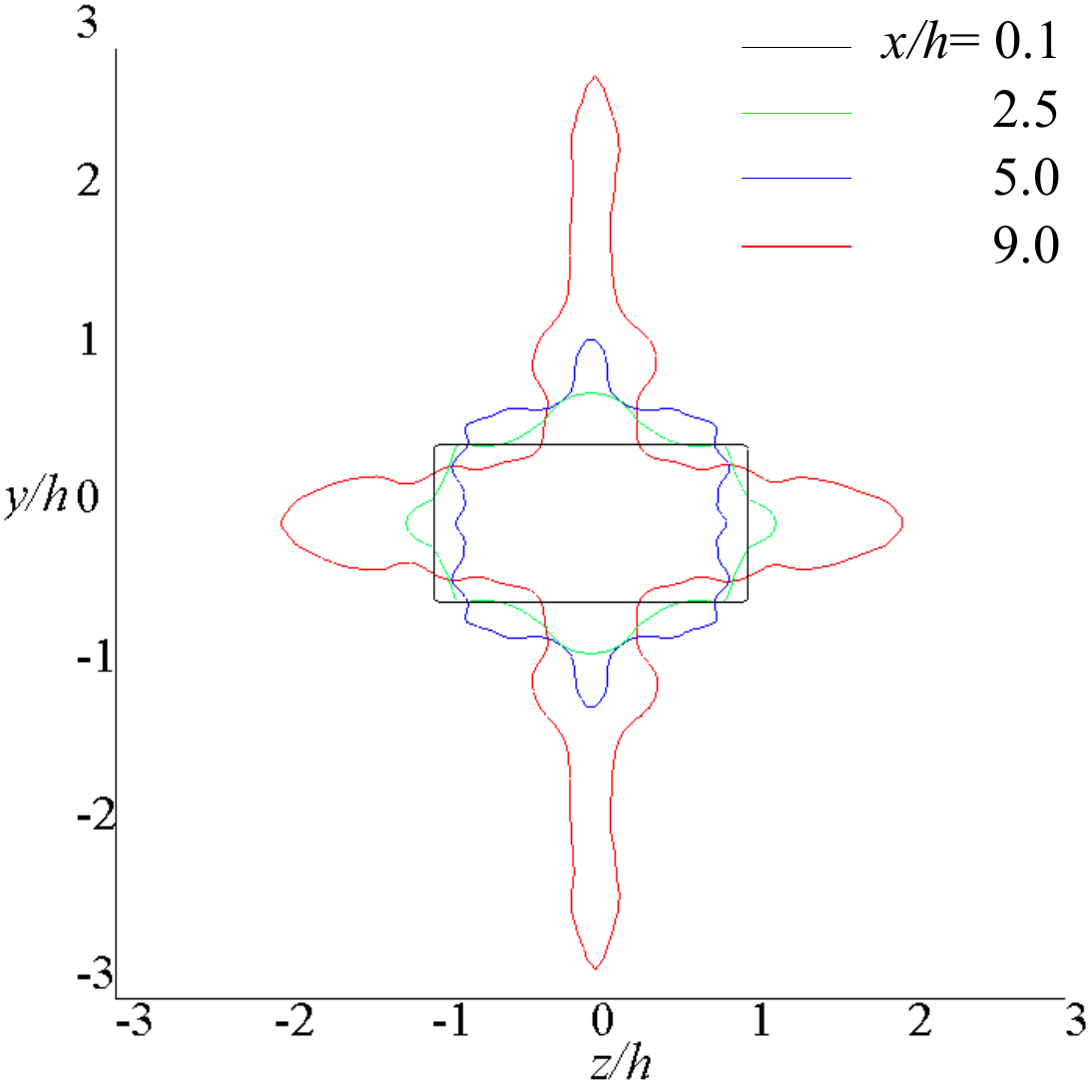} \\
(c)
\end{minipage}
\caption{Time-averaged contours of half-velocity-width in $y$-$z$ plane: 
(a) $AR = 1.0$, (b) $AR = 1.5$, and (c) $AR = 2.0$.}
\label{hvel}
\end{figure}

Figure \ref{hvel} shows the time-averaged distribution of the half-value 
$\overline{u_c}/2$ of the streamwise velocity on the central axis of the jet 
in the $y$-$z$ plane to confirm the axis switching of the jet. 
From this figure, we can see the shape of the cross-section of the jet. 
The switching (45-degree axis switching) of the axis and diagonal positions 
in the cross-section of the jet occurs at $x/h = 2.0$ for $AR = 1.0$ and 1.5 
and at $x/h = 2.5$ for $AR = 2.0$. 
In the case of $AR = 1.0$, at $x/h=3.0$ downstream, 
45-degree axis switching occurs again, 
and the axis and diagonal positions exist at almost the same position as the nozzle shape. 
At this position, the jet spreads in the opposite side and diagonal directions. 
The divergence of the jet in the diagonal direction is due to the vortex pairs 
that exist downstream near the nozzle corners. 
In addition, the high-velocity fluid of the jet is transported to the opposite side direction 
by the hairpin part of the vortex ring, 
and the jet becomes wider in the opposite side direction than in the $x/h = 0.1$ plane.

%------------------------------------------------------------------------------
% Figure 16
%------------------------------------------------------------------------------
%------------------------------------------------------------------------------

%------------------------------------------------------------------------------
% Figure 17
%------------------------------------------------------------------------------
\begin{figure}[!t]
\centering
\begin{minipage}{0.48\linewidth}
\centering
\includegraphics[trim=0mm 0mm 0mm 0mm, clip, width=80mm]{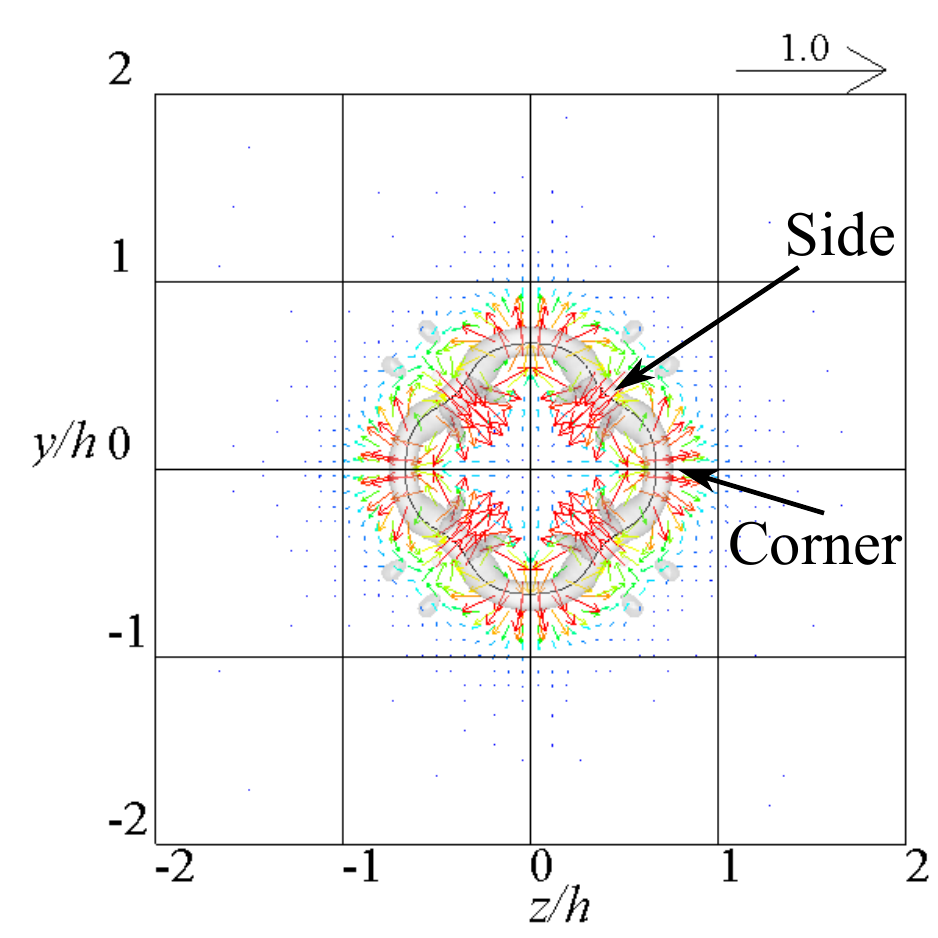} \\
(a) \\
\end{minipage}
\hspace{0.02\linewidth}
\begin{minipage}{0.48\linewidth}
\centering
\includegraphics[trim=0mm 0mm 0mm 0mm, clip, width=80mm]{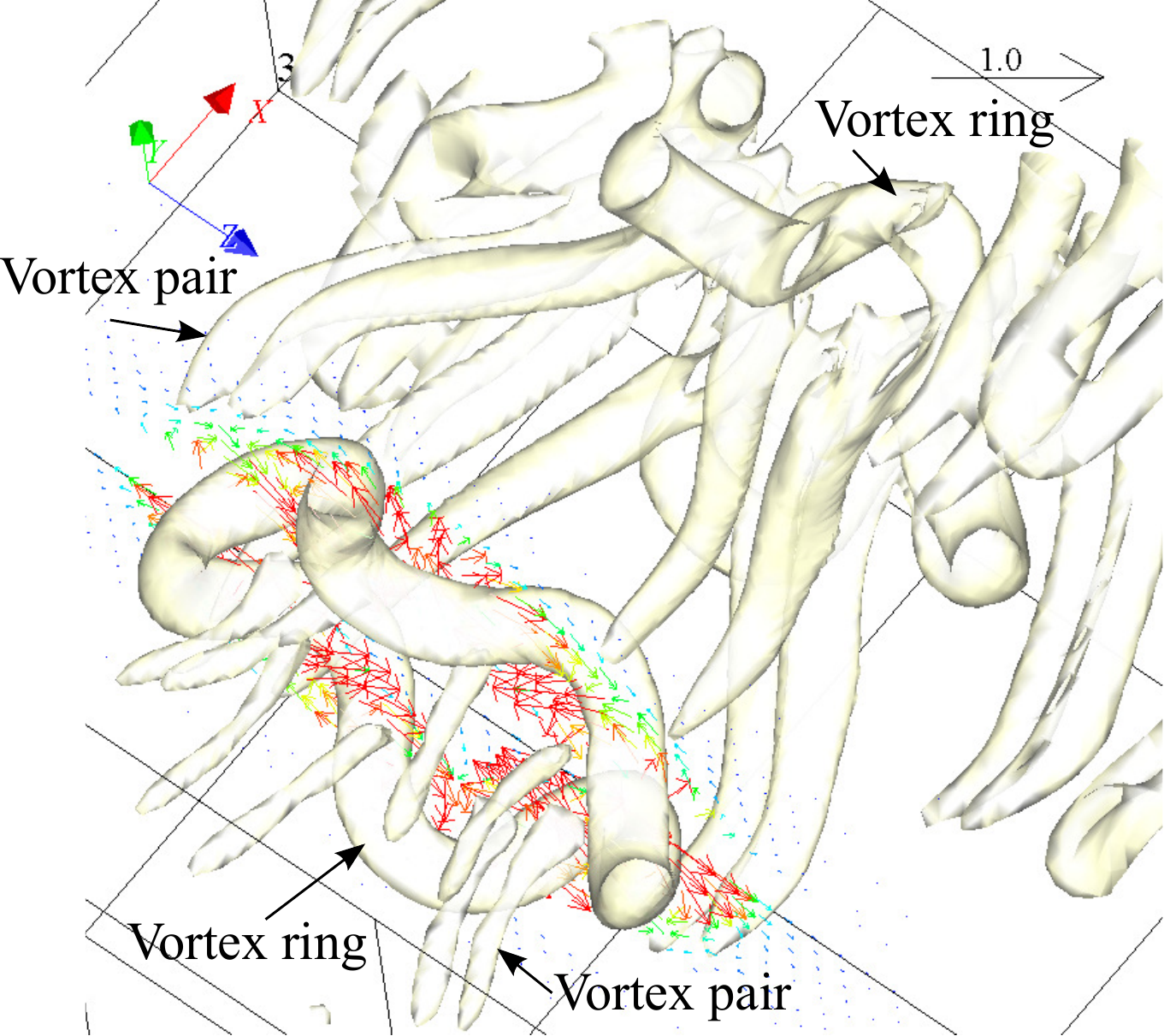} \\
(b)
\end{minipage}
\caption{Isosurface of curvature of equipressure surface, 
velocity fluctuation vectors and half-velocity-width 
in $y$-$z$ plane at $x/h = 1.3$ for $AR = 1.0$ and $T = 0$: 
Isosurface value is $-12/h$. 
(a) $y$-$z$ plane and (b) perspective view.}
\label{cur_vec_uc_1}
\end{figure}
%------------------------------------------------------------------------------

When $AR = 1.5$, $\overline{u_c}/2$ distributions exist at $x/h = 5.5$ 
around $y/h = \pm 1.8$ in the minor axis direction and around $z/h = \pm 1.4$ 
in the major axis direction. 
This is due to the hairpin part of the vortex ring. 
In the minor axis direction, 
as the hairpin part of the vortex ring develops further away from the jet center 
than in the major axis direction, 
the hairpin portion of the vortex ring transports the high-velocity fluid near the jet center outside. 
As a result, the jet becomes wider in the minor axis direction 
than in the major axis direction, 
and the switching (90-degree axis switching) of the major axis and minor axis 
in the cross-section of the jet occurs. 
In the diagonal direction, 
the elongated vortex pair generates a flow from the jet center toward the outside, 
which widens the jet. 
At $x/h = 8$ downstream, the distribution of $\overline{u_c}/2$ is narrower 
in the major and minor axis directions than at $x/h = 5.5$. 
This is because the hairpin part of the vortex ring collapses downstream, 
weakening flow induction.

For $AR = 2.0$, the jet becomes wider at $x/h = 5.0$ in the minor axis direction 
than in the major axis direction, 
and the 90-degree axis switching occurs. 
This result is due to the interaction induced near $x/h=3$ 
between the upstream and downstream vortex rings. 
Compared to the short side, 
the hairpin part of the vortex ring on the long side moves away from the jet center, 
and the vortex actively transports the high-velocity fluid outward from the jet center. 
Therefore, the jet becomes wider in the minor axis direction. 
At $x/h = 9$, unlike $AR = 1.5$, the interference between the vortex rings 
on the upstream and downstream sides diffuses vortices over a wide area. 
Thus, the jet spreads in the major and minor axis directions.

To clarify the occurrence of the 45-degree axis switching, 
Fig. \ref{cur_vec_uc_1} shows vortex structures, 
velocity fluctuation vectors at $x/h = 1.3$ in the $y$-$z$ plane, 
and distribution of the half-value $u_{\rm c}/2$ of the streamwise velocity 
on the central axis of the jet. 
In the figure, the side and corner of the vortex ring are labeled. 
Figure \ref{cur_vec_uc_1}(b) is a perspective view including downstream vortex structures. 
Here, only the result for $AR = 1.0$ is shown 
because the generation mechanism of axis switching was the same for all $AR$. 
As the corner with curvature moves downstream faster than the side of the vortex ring, 
the corner of the vortex ring moves toward the jet center on the downstream side. 
The positions of the sides and corners of the initial vortex ring are exchanged, 
and the vortex ring transforms into a new vortex ring. 
The labels represent new sides and corners. 
At this time, the side of the vortex ring exists downstream of $x/h = 1.3$, 
and the corner exists upstream. 
In the plane of $x/h = 1.3$ between the side and corner of the vortex ring, 
the side part of the vortex ring produces the flow toward the jet center, 
and the corner part generates outward flow. 
Therefore, as can be seen from the distribution of $u_{\rm c}/2$, 
the jet widens in the opposite side direction of the nozzle 
and shrinks in the diagonal direction of the nozzle, 
resulting in the 45-degree axis switching.

\begin{figure}[!t]
\centering
%
%%% AR=1.0
\begin{minipage}{0.48\linewidth}
\centering
\includegraphics[trim=0mm 0mm 0mm 0mm, clip, width=80mm]{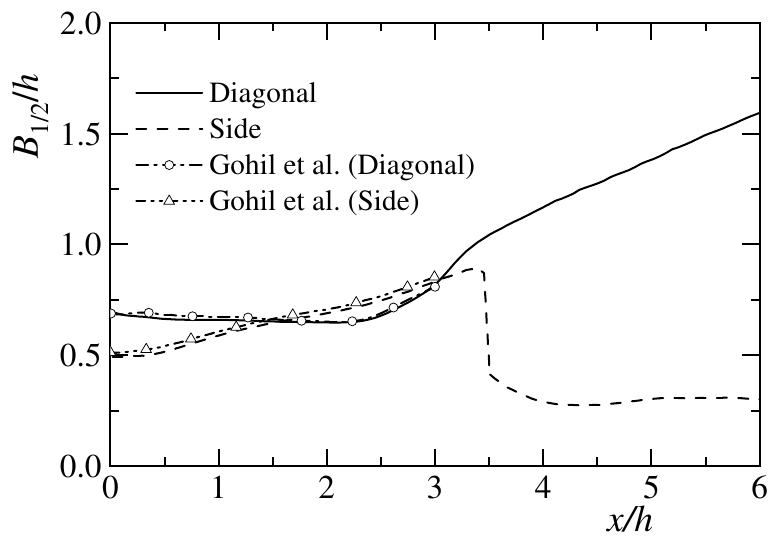} \\
\vspace*{-0.5\baselineskip}
(a)
\end{minipage}
%
%%% AR=1.5
\hspace{0.02\linewidth}
\begin{minipage}{0.48\linewidth}
\centering
\includegraphics[trim=0mm 0mm 0mm 0mm, clip, width=80mm]{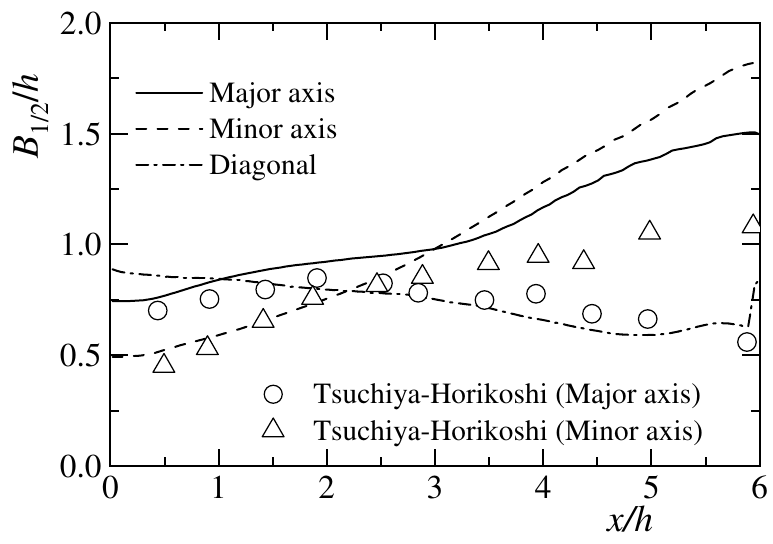} \\
\vspace*{-0.5\baselineskip}
(b)
\end{minipage}
%
%%% AR=2.0
\begin{minipage}{0.90\linewidth}
\centering
\includegraphics[trim=0mm 0mm 0mm 0mm, clip, width=80mm]{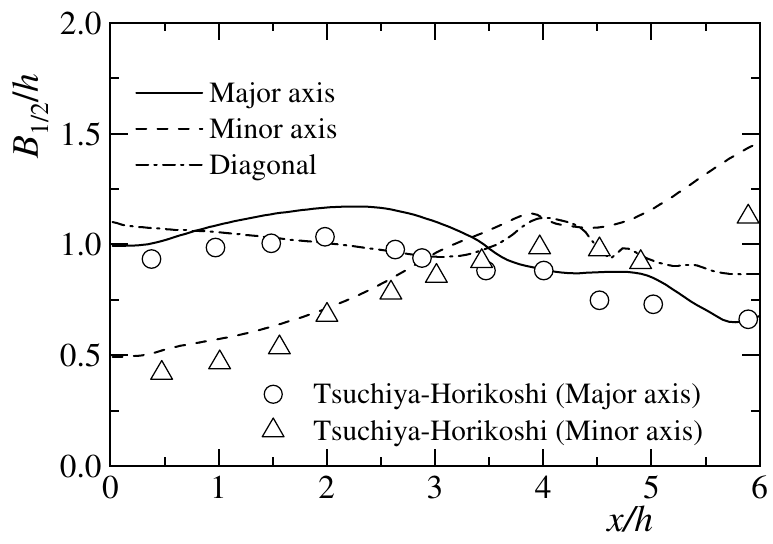} \\
\vspace*{-0.5\baselineskip}
(c)
\end{minipage}
\caption{Time-averaged distributions of half-velocity-width: 
(a) $AR = 1.0$, (b) $AR = 1.5$, and (c) $AR = 2.0$.}
\label{hw}
\end{figure}

Figure \ref{hw} shows the time-averaged distribution $B_{1/2}/h$ 
of the half-velocity-width 
to confirm the position where the flow axis switching occurs. 
As in the existing research \citep{Zaman_1996}, 
the half-width $B_{1/2}$ of the jet is defined as the distance 
between two points where the velocity is half the streamwise velocity, 
$u_c/2$, at the central axis. 
For $AR = 1.0$, the occurrence position of the 45-degree axis switching is defined as 
the point where two distributions in the opposite side and diagonal directions cross. 
For $AR > 1.0$, the occurrence position is defined as 
the intersection of the distributions in the minor axis and diagonal directions. 
In the case of $AR > 1.0$, the position where the 90-degree axis switching occurs 
is the intersection of the distributions in the major and minor axis directions. 
In the figure for $AR = 1.0$, the calculated results \citep{Gohil_et_al_2015} 
for the pulsating jet ejected from a square nozzle at $Re = 1000$ are compared. 
For $AR = 1.5$ and 2.0, the experimental results \citep{Tsuchiya&Horikoshi_1986} 
for free jets ejected from orifice nozzles at $Re = 15600-15800$ are also compared. 
For $AR = 1.0$, the present calculation result agrees well with the previous result \citep{Gohil_et_al_2015}, 
and 45-degree axis switching occurs at $x/h = 1.55$. 
At $x/h=3.0$ downstream, the half-velocity-width switches again, 
and 45-degree axis switching occurs. 
Further downstream, the half-velocity-width in the opposite side direction decreases sharply. 
This is because the hairpin part of the vortex ring that develops in the opposite side direction collapses, 
and the flow induced by that vortex weakens. 
When $AR = 1.5$ and 2.0, 45-degree axis switching occurs at $x/h = 2.2$ and $3.0$, respectively, 
and 90-degree axis switching occurs at $x/h = 3.0$ and $3.3$, respectively. 
As $AR$ increases, the position of axis switching moves downstream. 
This trend is similar to the results for free jets at high Reynolds numbers 
\citep{Krothapalli_et_al_1981,Tsuchiya&Horikoshi_1986,Straccia&Farnsworth_2021}.

%------------------------------------------------------------------------------
% Figure 18
%------------------------------------------------------------------------------
%------------------------------------------------------------------------------

%++++++++++++++++++++++++++++++++++++++++++++++++++++++++++++++++++++++++++++++
\subsection{Mean properties}
%++++++++++++++++++++++++++++++++++++++++++++++++++++++++++++++++++++++++++++++

\begin{figure}[!t]
\centering
\includegraphics[trim=0mm 0mm 0mm 0mm, clip, width=80mm]{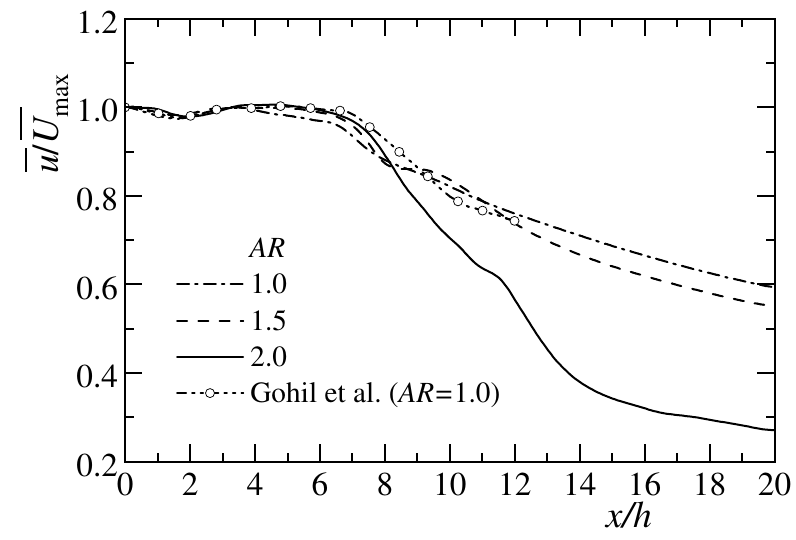} \\
\vspace*{-0.5\baselineskip}
\caption{Time-averaged distributions of streamwise velocity along centerline}
\label{streamc}
\end{figure}

Figure \ref{streamc} compares the time-averaged streamwise velocity distribution 
$\overline{u}$ on the central axis with the previous results of a pulsating jet 
for $Re = 1000$ and $AR = 1$ \citep{Gohil_et_al_2015}. 
The existing and present results agree well up to $x/h = 4$. 
In the downstream, the velocity decays earlier in our calculation results, 
but the trend is qualitatively consistent with the previous results. 
It is considered that this difference is due to the difference in the size of the computational domain. 
We used the large computational regions in the $y$- and $z$-directions. 
Therefore, compared with the case where the calculation area is narrow, 
the jet spreads more to the surroundings, 
so it is considered that the velocity decays earlier. 
There is a local minimum near $x/h = 2$ for all $AR$. 
This is because, as shown in Figs. \ref{cur_vec_1} to \ref{cur_vec_2}, 
the flow toward the upstream occurs near the jet center 
due to the interference between the hairpin part of the vortex ring and the vortex pair. 
At this time, a 45-degree axis switching occurs. 
It can be seen that the velocity decays earlier as $AR$ increases. 
This result indicates that the jet diffuses early and promotes mixing.

%------------------------------------------------------------------------------
% Figure 19
%------------------------------------------------------------------------------
%------------------------------------------------------------------------------

\begin{figure}[!t]
\begin{minipage}{0.48\linewidth}
\centering
\includegraphics[trim=0mm 0mm 0mm 0mm, clip, width=80mm]{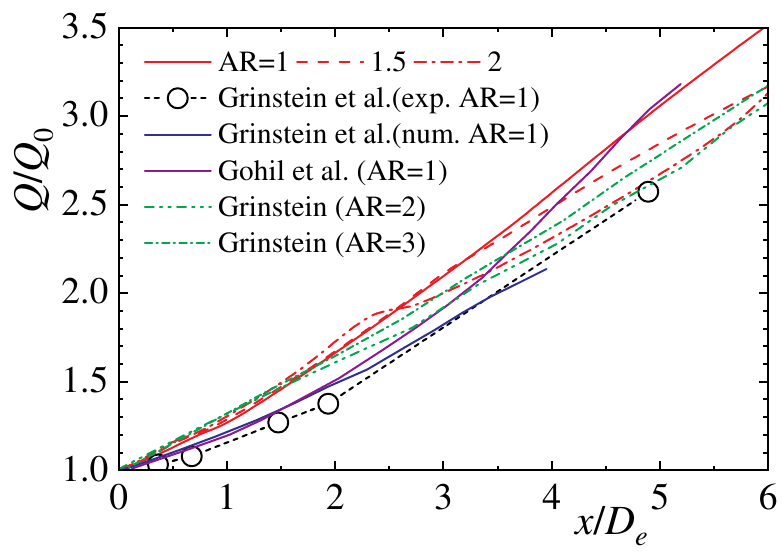} \\
\vspace*{-0.5\baselineskip}
(a)
\end{minipage}
\hspace{0.02\linewidth}
\begin{minipage}{0.48\linewidth}
\centering
\includegraphics[trim=0mm 0mm 0mm 0mm, clip, width=80mm]{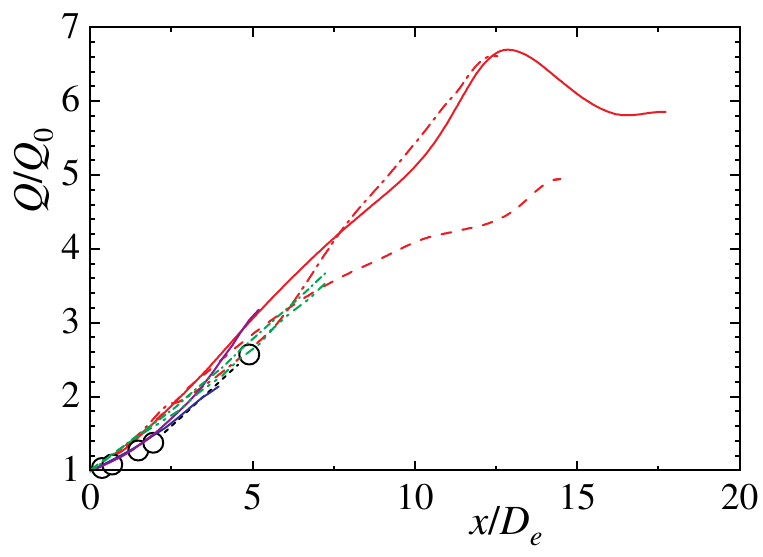} \\
\vspace*{-0.5\baselineskip}
(b)
\end{minipage}
\caption{Comparisons of jet entrainment with previous results 
\citep{Grinstein_et_al_1995,Grinstein_2001,Gohil_et_al_2015}: 
(a) near nozzle and (b) downstream.}
\label{entrainment}
\end{figure}

Figure \ref{entrainment} shows the distribution of entrainment 
in comparison with existing results \citep{Grinstein_et_al_1995,Grinstein_2001,Gohil_et_al_2015}. 
$Q$ is the mass flux (volumetric flux) obtained from the streamwise velocity, 
$Q_0$ is the mass flux at the nozzle exit, 
and $D_e$ is the circular-equivalent diameter. 
In Fig. \ref{entrainment}, the horizontal axis is set to $x/D_e$ 
to compare the results for different aspect ratios. 
Previous studies \citep{Grinstein_et_al_1995,Grinstein_2001,Gohil_et_al_2015} 
reported that entrainment with rectangular nozzles was more than that with circular nozzles. 
In Fig. \ref{entrainment}, only existing results using rectangular nozzles are shown. 
In addition, Fig. \ref{entrainment} (b) shows the downstream distribution. 
The results obtained by \citet{Grinstein_et_al_1995} are the experimental results 
for free jets at $Re_D = 4.2\times10^4$ 
and the calculated results for pulsating jets at $St_D = 0.48$ and $Re_D = 3.2\times10^3$. 
The results of \citet{Grinstein_2001} are the calculated results 
for $St_D = 0.48$ at a high Reynolds number of $Re_D > 8.5\times10^4$. 
The calculation conditions of \citet{Gohil_et_al_2015} are the same as in this study. 
The entrainment amount in this study is more than the previous results 
regarding free jet \cite{Grinstein_et_al_1995} 
and pulsating flow \cite{Grinstein_2001,Gohil_et_al_2015} 
and is different from the result \citep{Gohil_et_al_2015} 
under the same condition as this study. 
Similar to the results in Fig. \ref{streamc}, 
it is considered that the influence of the computational domain causes this difference. 
As $AR$ increases, the amount of entrainment decreases. 
This trend is consistent with the previous results \citep{Grinstein_2001,Jiang_et_al_2007}. 
\citet{Grinstein_2001} reported that entrainment decreased with increasing $AR$ 
and increased again at $AR = 3$, 
indicating that an optimal value for $AR$ exists. 
The trend for entrainment to decrease with increasing $AR$ also appeared in existing results 
for free jets at high Reynolds numbers \citep{Jiang_et_al_2007}. 
In this study, at $x/D_e > 7$, 
the entrainment for $AR = 2.0$ increases more than the results for $AR = 1.0$ and 1.5. 
This relationship between entrainment and aspect ratio agrees with the reported result 
by \citet{Grinstein_2001}.

%------------------------------------------------------------------------------
% Figure 20
%------------------------------------------------------------------------------
%------------------------------------------------------------------------------

\begin{figure}[!t]
\begin{minipage}{0.48\linewidth}
\centering
\includegraphics[trim=0mm 0mm 0mm 0mm, clip, width=80mm]{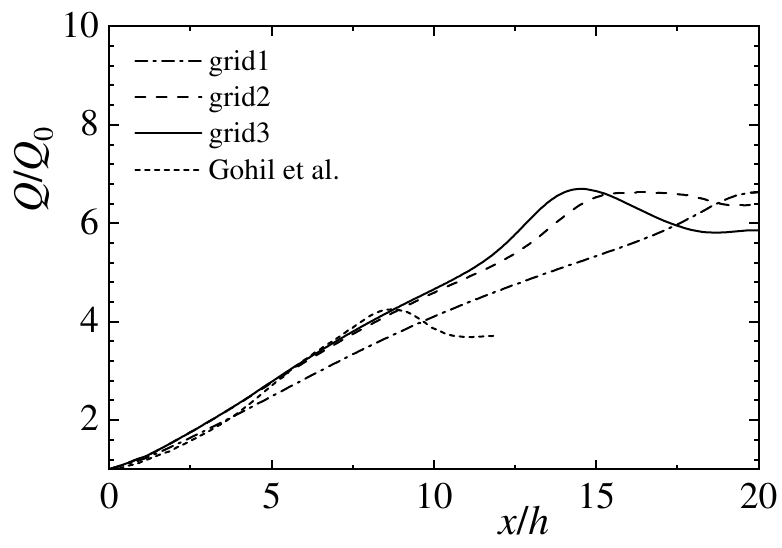} \\
\vspace*{-0.5\baselineskip}
(a)
\end{minipage}
\hspace{0.02\linewidth}
\begin{minipage}{0.48\linewidth}
\centering
\includegraphics[trim=0mm 0mm 0mm 0mm, clip, width=80mm]{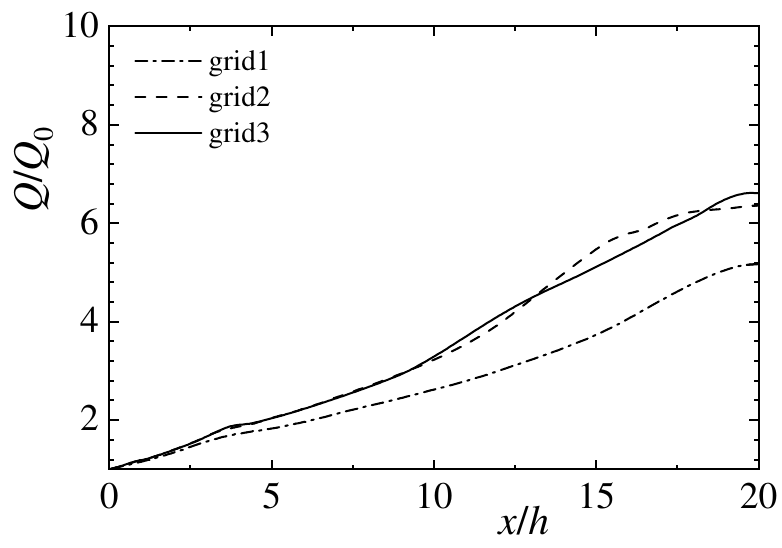} \\
\vspace*{-0.5\baselineskip}
(b)
\end{minipage}
\caption{Comparisons of jet entrainment obtained using three grids: 
(a) $AR = 1$ and (b) $AR = 2$.}
\label{entrainment_grid}
\end{figure}

Next, we examine the grid dependency on the calculation results. 
Figure \ref{entrainment_grid} shows the amount of entrainment for $AR = 1.0$ and 2.0. 
When $AR = 1.0$, the results of grid2 and grid3 are in relatively good agreement, 
but the results of grid1 and other grids differ. 
The present results agree with the previous one \citep{Gohil_et_al_2015} up to around $x/h = 8.5$, 
but the difference increases downstream. 
The distribution trends near the downstream boundary are similar 
between the grid2 and grid3 results and the previous results. 
Therefore, it is considered that this difference is due to the influence of the computational domain. 
For $AR = 2.0$, the results using grid2 and grid3 agree well, 
and the distribution of grid1 is different from those of other grids.

%------------------------------------------------------------------------------
% Figure 21
%------------------------------------------------------------------------------
%------------------------------------------------------------------------------

%------------------------------------------------------------------------------
% Figure 22
%------------------------------------------------------------------------------
\begin{figure}[!t]
\centering
\includegraphics[trim=0mm 0mm 0mm 0mm, clip, width=100mm]{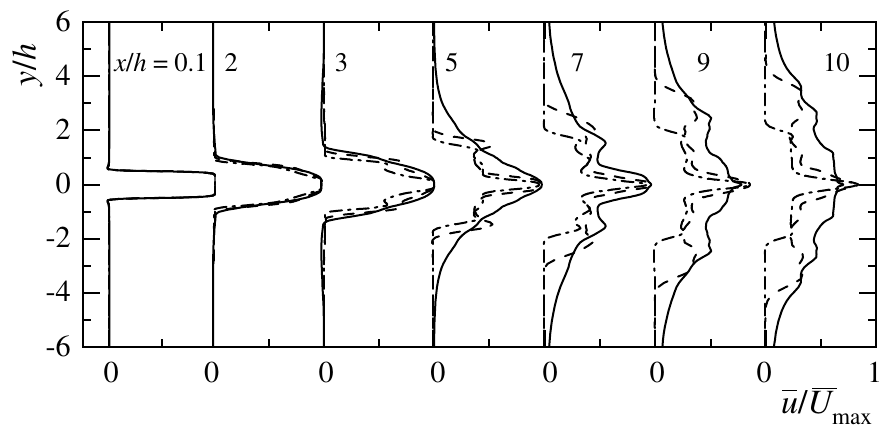} \\
\vspace*{-0.5\baselineskip}
(a) \\
\includegraphics[trim=0mm 0mm 0mm 0mm, clip, width=100mm]{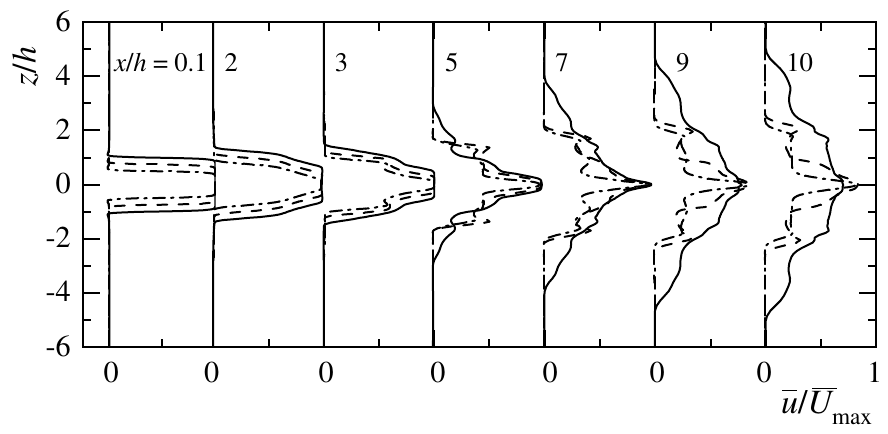} \\
\vspace*{-0.5\baselineskip}
(b)
\vspace*{-0.5\baselineskip}
\caption{Time-averaged distributions of streamwise velocity: 
- $\cdot$ -, $AR = 1.0$; - - -, $AR = 1.5$; ---, $AR = 2.0$: 
(a) $x$-$y$ plane and (b) $x$-$z$ plane.}
\label{stream}
\end{figure}
%------------------------------------------------------------------------------

Figure \ref{stream} shows the time-averaged streamwise velocity distribution 
$\overline{u}$ in the $x$-$y$ plane at $z/h = 0$ 
and the $x$-$z$ plane at $y/h = 0$ for each condition. 
From the distribution on the $x$-$y$ plane, 
it can be seen that the jets start to diffuse around $x/h = 0.1-2$ for all $AR$. 
Downstream from $x/h = 3$, when $AR = 1.5$, 
the jet diffuses in the minor axis direction compared to $AR = 1.0$ 
because of the 90-degree axis switching. 
At $x/h = 5$, when $AR = 1.0$ and 1.5, there are local maximums near $y/h = \pm 1.5$. 
This is due to the hairpin part of the vortex ring. 
In the case of $AR = 2.0$, the jet diffuses rapidly at $x/h = 5$. 
Further downstream at $x/h = 10$, 
the jet for $AR = 2.0$ is more diffuse than the jets for $AR = 1.0$ and 1.5, 
indicating that mixing is promoted.

In the $x$-$z$ plane, the jets diffuse to the same extent in the major axis direction 
around $x/h = 0.1-5$ for all $AR$. 
When $AR = 1.0$ and 1.5, the rotation of vortex rings and vortex pairs becomes weak downstream, 
and the mixing between the surrounding fluid and jet slows down. 
Therefore, at the downstream of $x/h = 7$, 
the jet diffusion at $AR = 1.0$ and 1.5 is weaker than at $AR = 2.0$. 
In the case of $AR = 2.0$, compared to $AR = 1.0$ and 1.5, 
the jet diffuses over a wide area up to $z/h = \pm 5$ 
because strong vortex structures exist downstream.

%++++++++++++++++++++++++++++++++++++++++++++++++++++++++++++++++++++++++++++++
\subsection{Turbulence properties}
%++++++++++++++++++++++++++++++++++++++++++++++++++++++++++++++++++++++++++++++

\begin{figure}[!t]
\begin{minipage}{0.48\linewidth}
\centering
\includegraphics[trim=0mm 0mm 0mm 0mm, clip, width=55mm]{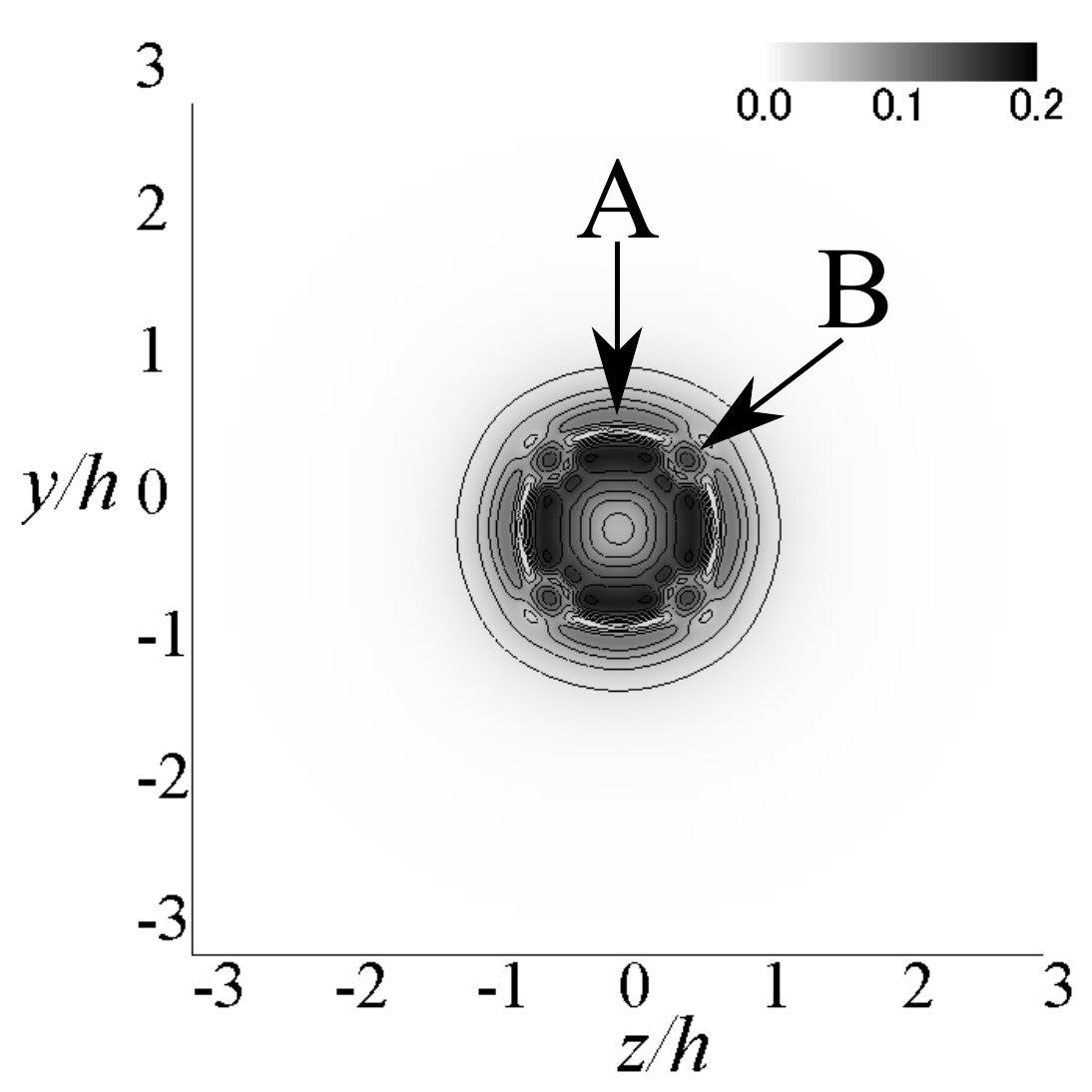} \\
\vspace*{-0.5\baselineskip}
(a)
\end{minipage}
\hspace{0.02\linewidth}
\begin{minipage}{0.48\linewidth}
\centering
\includegraphics[trim=0mm 0mm 0mm 0mm, clip, width=55mm]{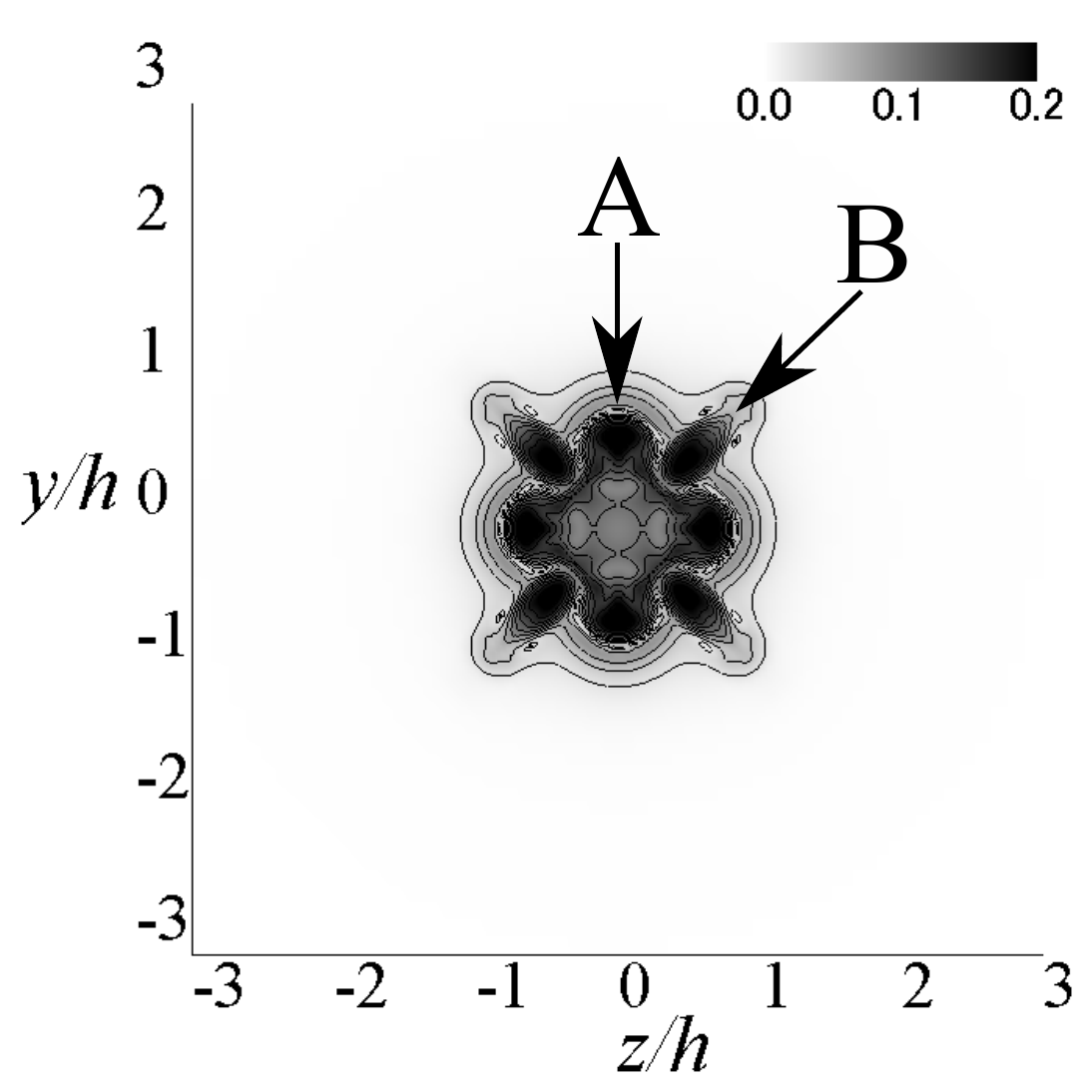} \\
\vspace*{-0.5\baselineskip}
(b)
\end{minipage}
\caption{Contours of turbulence intensity of streamwise velocity fluctuation 
in $y$-$z$ plane for $AR = 1.0$: 
Contour interval is 0.02 from 0 to 0.2. 
(a) $x/h = 1$ and (b) $x/h = 2$.}
\label{urms_ar1}
\end{figure}

\begin{figure}[!t]
\begin{minipage}{0.48\linewidth}
\centering
\includegraphics[trim=0mm 0mm 0mm 0mm, clip, width=55mm]{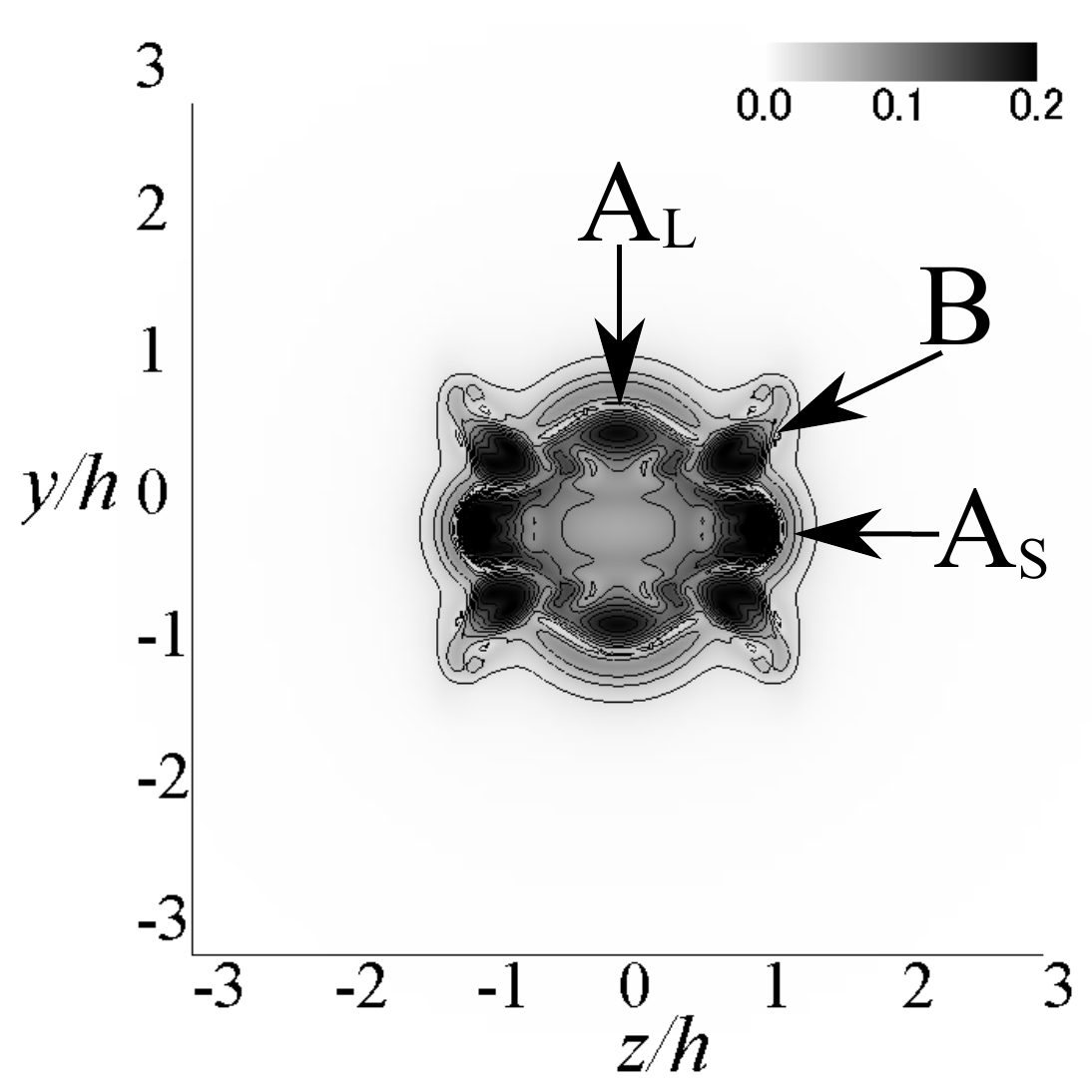} \\
\vspace*{-0.5\baselineskip}
(a)
\end{minipage}
\hspace{0.02\linewidth}
\begin{minipage}{0.48\linewidth}
\centering
\includegraphics[trim=0mm 0mm 0mm 0mm, clip, width=55mm]{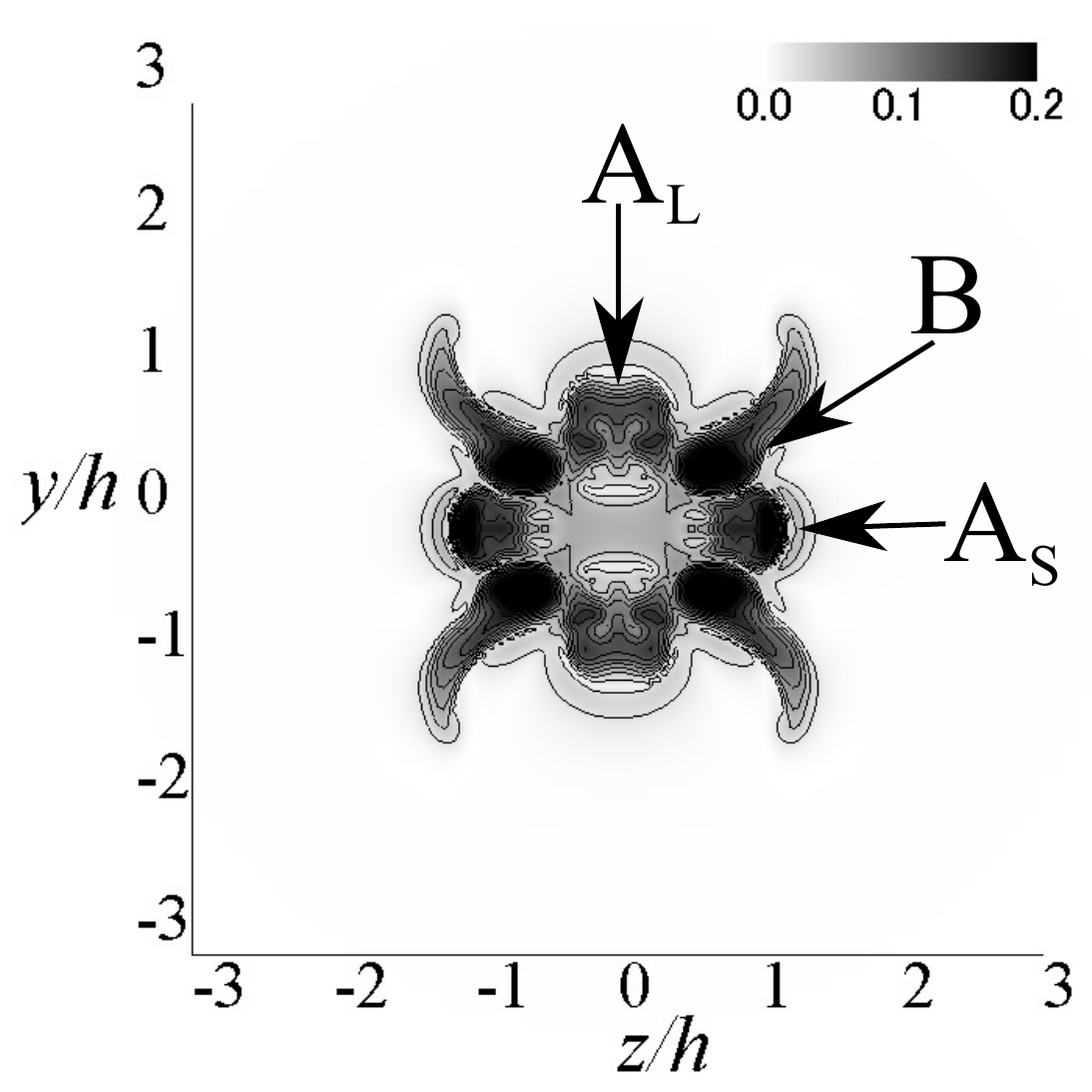} \\
\vspace*{-0.5\baselineskip}
(b)
\end{minipage}
\caption{Contours of turbulence intensity of streamwise velocity fluctuation 
in $y$-$z$ plane for $AR = 1.5$: 
Contour interval is 0.02 from 0 to 0.2. 
(a) $x/h = 2$ and (b) $x/h = 3$.}
\label{urms_ar15}
\end{figure}

\begin{figure}[!t]
\centering
\begin{minipage}{0.48\linewidth}
\centering
\includegraphics[trim=0mm 0mm 0mm 0mm, clip, width=55mm]{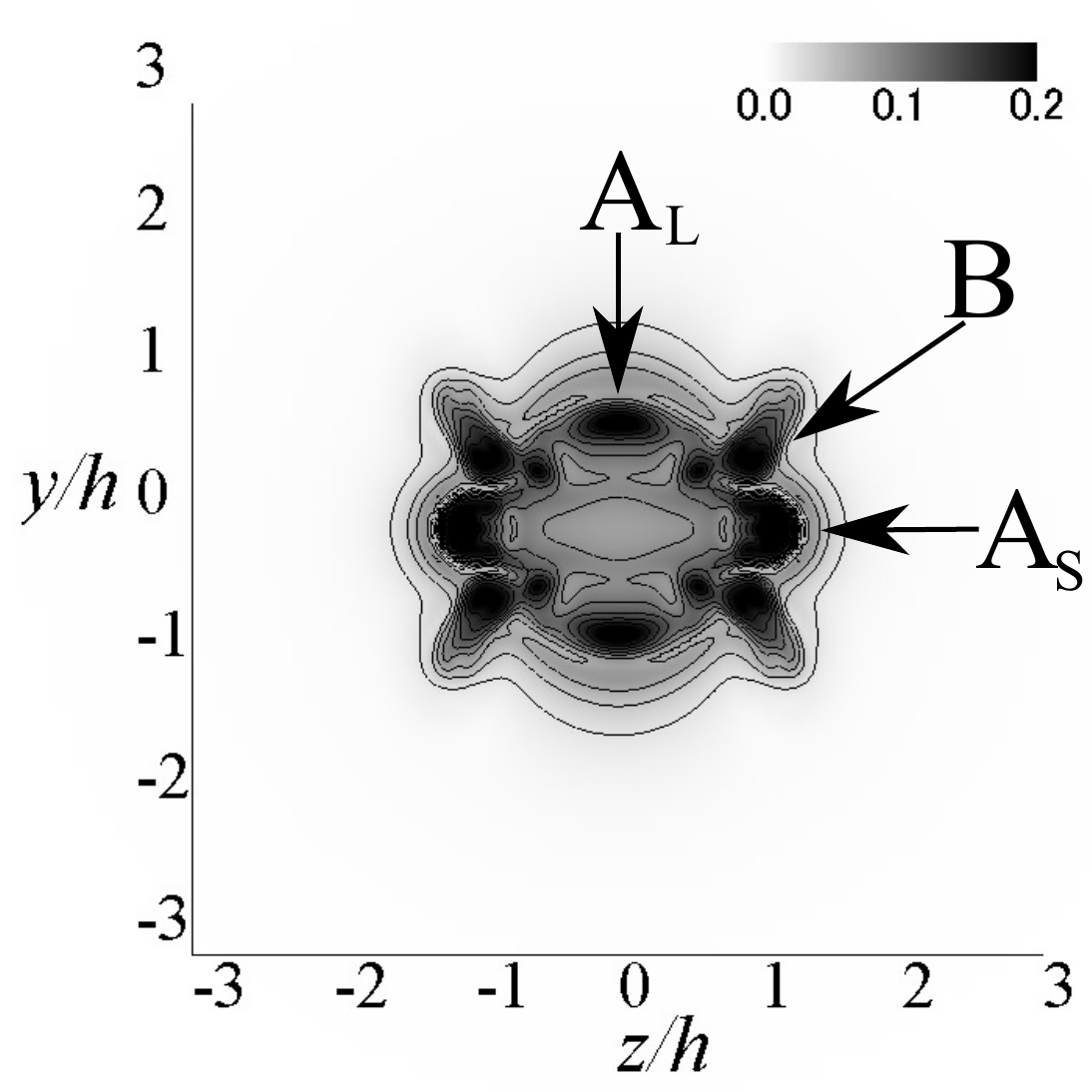} \\
\vspace*{-0.5\baselineskip}
(a)
\end{minipage}
\hspace{0.02\linewidth}
\begin{minipage}{0.48\linewidth}
\centering
\includegraphics[trim=0mm 0mm 0mm 0mm, clip, width=55mm]{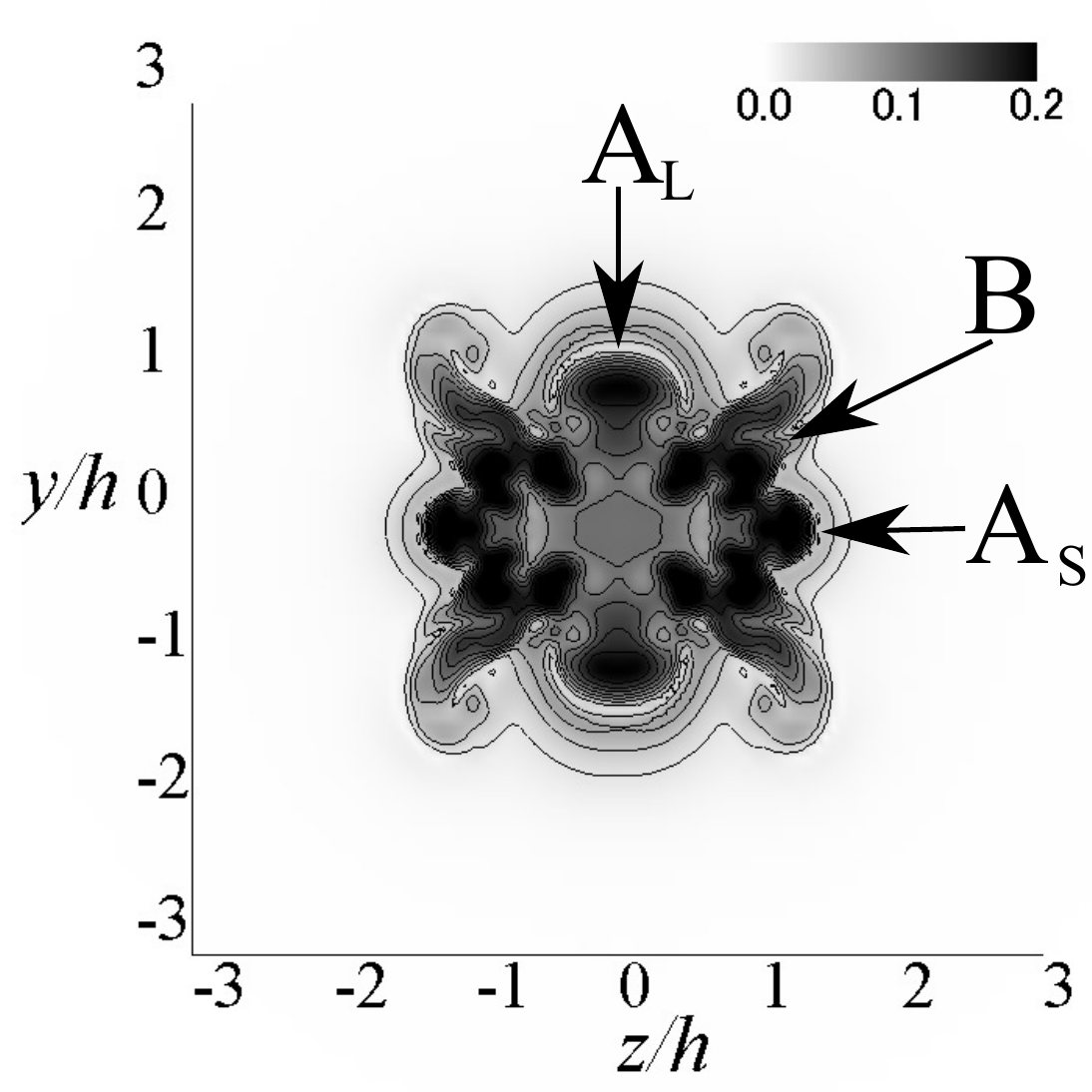} \\
\vspace*{-0.5\baselineskip}
(b)
\end{minipage}
\caption{Contours of turbulence intensity of streamwise velocity fluctuation 
in $y$-$z$ plane for $AR = 2.0$: 
Contour interval is 0.02 from 0 to 0.2. 
(a) $x/h = 2$ and (b) $x/h = 3$.}
\label{urms_ar2}
\end{figure}

Figures \ref{urms_ar1} to \ref{urms_ar2} show the distribution of the rms values 
of the streamwise velocity fluctuations in the $y-z$ plane. 
For $AR = 1.0$, the distributions are at $x/h = 1$ and 2, 
and for $AR = 1.5$ and 2.0, the distributions are at $x/h = 2$ and 3. 
Here, concerning individual regions where high turbulences appear, 
label A presents the turbulence region due to the side of a vortex ring, 
and label B presents the turbulence region due to the corner of a vortex ring and vortex pair. 
The turbulent regions A on the long and short sides are distinguished by subscripts L and S, respectively. 
In such a high turbulence region, mixing is promoted between the surrounding fluid and jet. 
In the $x/h = 1$ plane for $AR = 1.0$, 
the side and the corner of the vortex ring generate the high turbulence regions A and B, respectively, 
at the same position as the nozzle shape. 
Because the deformation of the vortex ring causes a 45-degree axis switching, 
the turbulence intensity levels in regions A and B at $x/h = 2$ are higher than at $x/h = 1$. 
This increase in turbulence suggests enhanced mixing. 
At this time, the high turbulence regions due to the side of the vortex ring, 
corner of the vortex ring, and vortex pair are symmetrical in the opposite side direction, 
and the positions of the turbulence regions A and B do not change downstream. 
When $AR = 1.5$ and 2.0, the turbulence in area A$_\mathrm{L}$ on the long side 
is lower than in area A$_\mathrm{S}$ on the short side at $x/h = 2$ and $x/h = 3$. 
For $AR = 1.5$ and 2.0, in the $x/h = 3$ plane where 90-degree axis switching occurs, 
a high turbulence area A$_\mathrm{L}$ due to the side of the vortex ring on the long side spreads 
in the minor axis direction. 
Therefore, the region where mixing is promoted becomes wide. 
The vortices on the long side diffuse widely, and the turbulence areas also spread. 
In addition, as the curvature of the hairpin part on the short side 
is higher than that on the long side, the rotational flow fluctuates more. 
Therefore, it is considered that a local increase in turbulence intensity occurs on the short side. 
For $AR = 2.0$, the vortex rings on the upstream and downstream sides interfere with each other, 
so the value of the turbulence area $A_\mathrm{L}$ on the long side becomes higher 
over a wide area than $AR = 1.5$. 
Therefore, active mixing may occur widely.

%------------------------------------------------------------------------------
% Figure 23
%------------------------------------------------------------------------------
%------------------------------------------------------------------------------

%------------------------------------------------------------------------------
% Figure 24
%------------------------------------------------------------------------------
%------------------------------------------------------------------------------

%------------------------------------------------------------------------------
% Figure 25
%------------------------------------------------------------------------------
%------------------------------------------------------------------------------

\begin{figure}[!t]
\centering
\includegraphics[trim=0mm 0mm 0mm 0mm, clip, width=80mm]{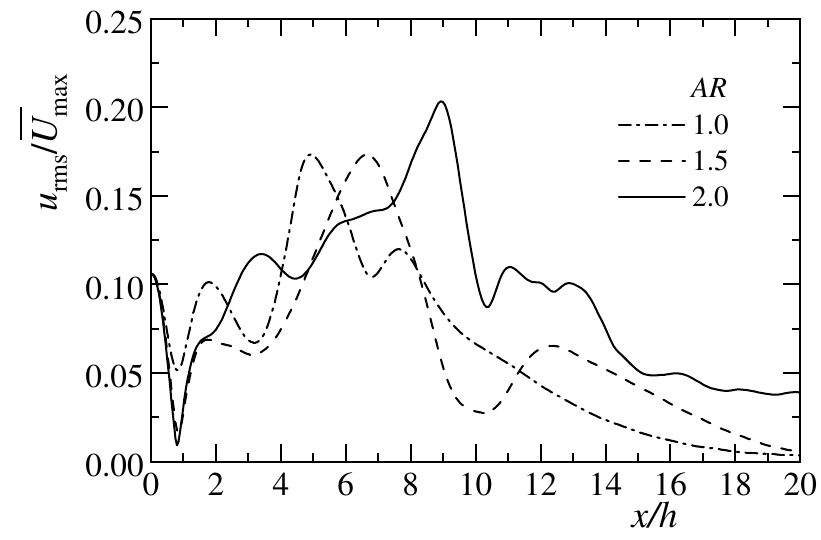} \\
\vspace*{-0.5\baselineskip}
\caption{Turbulence intensity distributions of 
streamwise velocity fluctuation along centerline}
\label{urmsc}
\end{figure}

Figure \ref{urmsc} shows the turbulence intensity $u_{\rm rms}$ 
of the streamwise velocity fluctuation on the jet center. 
As $AR$ increases, the maximum turbulence position moves downstream, 
and the maximum increases. 
For $AR = 1.0$, a local maximum occurs at $x/h = 2$. 
For $AR = 1.5$ and 2.0, the turbulence increases around $x/h = 2$. 
These increases in turbulence are due to the corner of the vortex ring 
approaching the jet center, causing a 45-degree axis switching. 
In the downstream, local maxima occur near $x/h = 5$, 7, and 9 for $AR = 1.0$, 1.5, and 2.0, respectively. 
The maxima are due to the elongated vortex pairs approaching the jet center. 
In the case of $AR = 1.0$, the vortex pair corresponds to the vortices generated 
near the corner of the vortex ring. 
For $AR = 1.5$ and 2.0, the vortex pairs correspond to the vortices 
generated by the collapse of vortex rings. 
Overall, the high turbulence forms at $AR = 2.0$.

%------------------------------------------------------------------------------
% Figure 26
%------------------------------------------------------------------------------
%------------------------------------------------------------------------------

%------------------------------------------------------------------------------
% Figure 27
%------------------------------------------------------------------------------
\begin{figure}[!t]
\centering
\includegraphics[trim=0mm 0mm 0mm 0mm, clip, width=100mm]{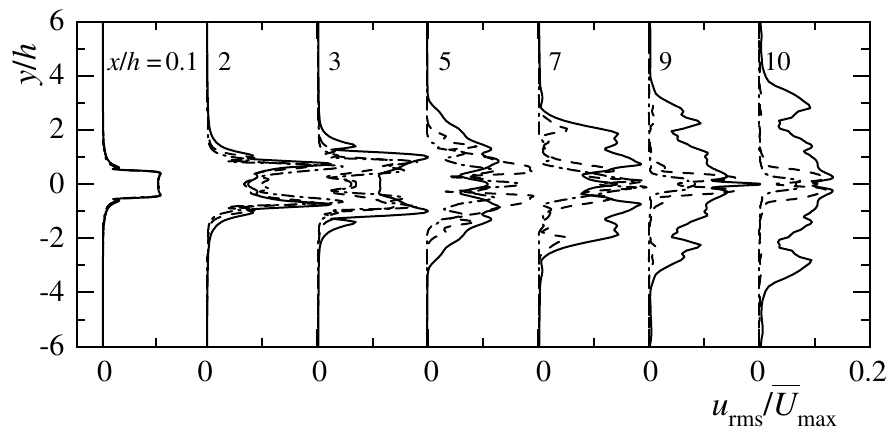} \\
\vspace*{-0.5\baselineskip}
(a) \\
\includegraphics[trim=0mm 0mm 0mm 0mm, clip, width=100mm]{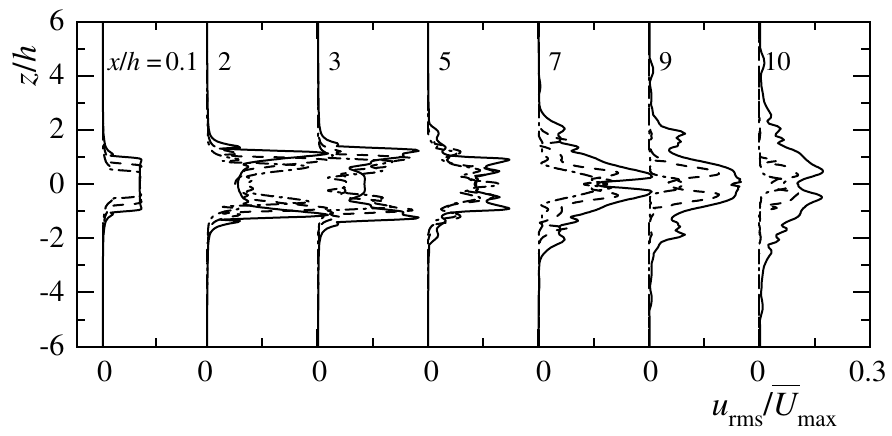} \\
\vspace*{-0.5\baselineskip}
(b)
\vspace*{-0.5\baselineskip}
\caption{Turbulence intensity distributions of streamwise velocity fluctuation: 
- $\cdot$ -, $AR = 1.0$; - - -, $AR = 1.5$; ---, $AR = 2.0$: 
(a) $x$-$y$ plane and (b) $x$-$z$ plane.}
\label{urms}
\end{figure}
%------------------------------------------------------------------------------

We confirm turbulence spread from the jet center to the surroundings for each $AR$. 
Figure \ref{urms} shows the turbulence intensities $u_{\rm rms}$ 
of streamwise velocity fluctuations in the $x$-$y$ plane at $z/h = 0$ 
and the $x$-$z$ plane at $y/h = 0$. 
The characteristics of the turbulence distribution are common 
in the $x$-$y$ and $x$-$z$ cross-sections. 
At $x/h = 0.1$ for all $AR$, turbulence due to pulsation occurs. 
At $x/h = 2$ downstream, the maximum value due to a vortex ring is observed. 
The maximum at $x/h = 3$ increases as $AR$ increases. 
This maximum is due to the hairpin part of the vortex ring. 
At $x/h = 3$, the turbulence for $AR = 2.0$ shows the highest value 
because the strong interference between the hairpin parts of the vortex rings 
on the upstream and downstream sides occurs. 
For $AR > 1.0$, a 90-degree axis switching occurs around $x/h = 3$, 
so at $x/h = 5$ in the $x-y$ plane, high turbulence occurs over a wide area. 
For $AR = 1.5$, the local maximum due to the hairpin part of the vortex ring is seen 
around $x/h = 5$ and $y/h = \pm 2$. 
At $x/h = 9$ downstream, because the hairpin part collapses, such a maximum does not occur. 
For $AR = 2.0$, intensive interference between upstream and downstream vortex rings occurs, 
and vortices diffuse over a wide area. 
Therefore, turbulence shows a high value in a wide area.

%------------------------------------------------------------------------------
% Figure 28
%------------------------------------------------------------------------------
\begin{figure}[!t]
\begin{minipage}{0.48\linewidth}
\centering
\includegraphics[trim=0mm 0mm 0mm 0mm, clip, width=80mm]{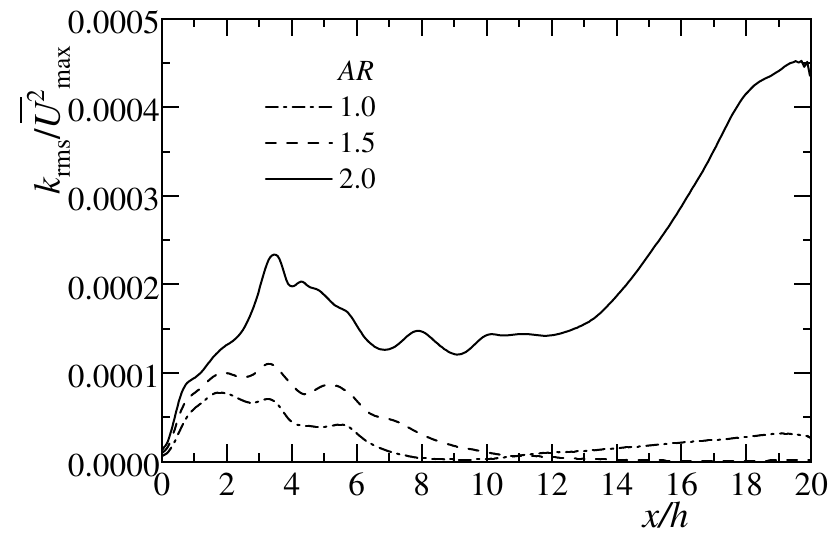} \\
\vspace*{-0.5\baselineskip}
(a)
\end{minipage}
\hspace{0.02\linewidth}
\begin{minipage}{0.48\linewidth}
\centering
\includegraphics[trim=0mm 0mm 0mm 0mm, clip, width=80mm]{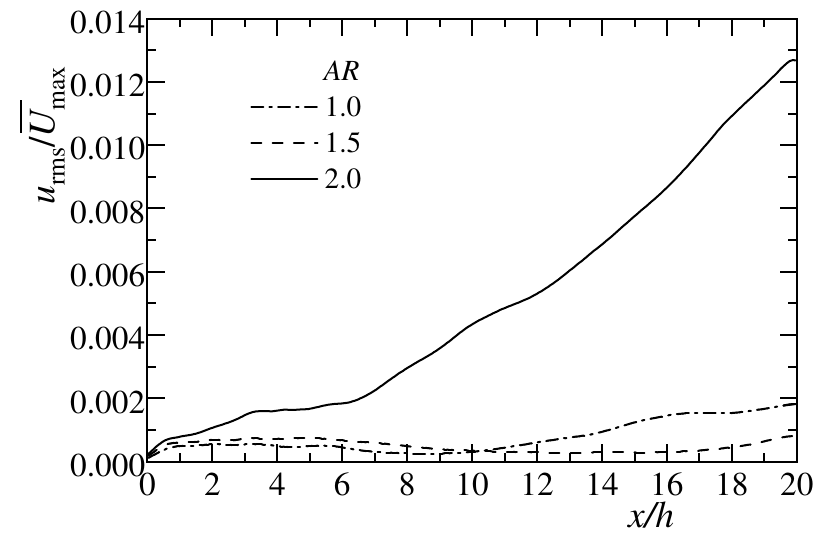} \\
\vspace*{-0.5\baselineskip}
(b)
\end{minipage}
\begin{minipage}{0.48\linewidth}
\centering
\includegraphics[trim=0mm 0mm 0mm 0mm, clip, width=80mm]{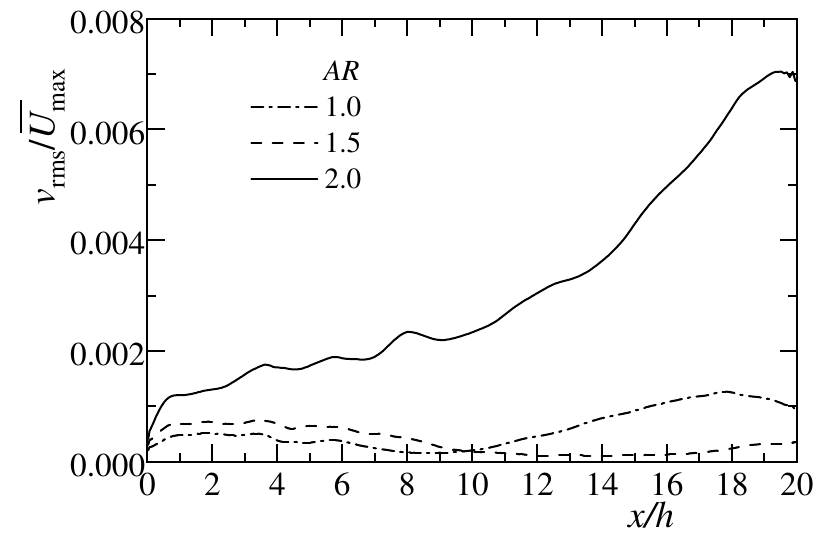} \\
\vspace*{-0.5\baselineskip}
(c)
\end{minipage}
\hspace{0.02\linewidth}
\begin{minipage}{0.48\linewidth}
\centering
\includegraphics[trim=0mm 0mm 0mm 0mm, clip, width=80mm]{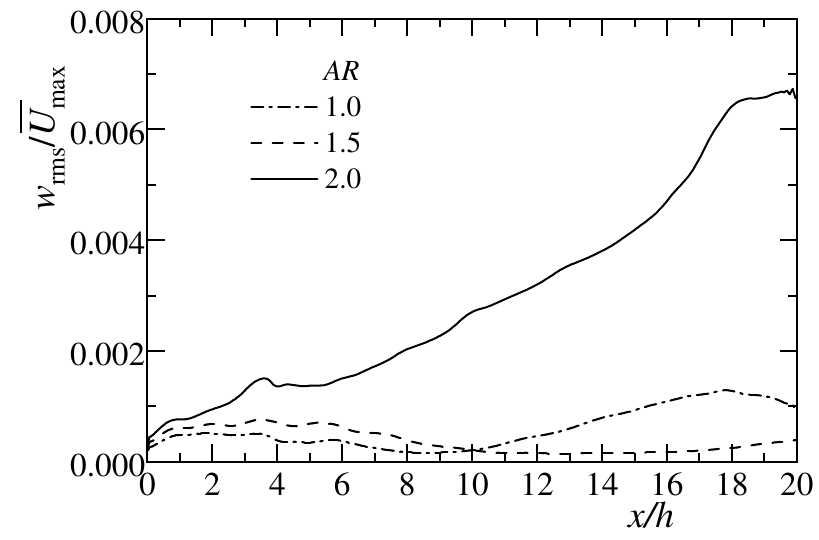} \\
\vspace*{-0.5\baselineskip}
(d)
\end{minipage}
\vspace*{-0.5\baselineskip}
\caption{Area- and time-averaged distributions in $y$-$z$ plane: 
(a) $k_\mathrm{rms}/\overline{U}_\mathrm{max}^{2}$, 
(b) $u_\mathrm{rms}/\overline{U}_\mathrm{max}$, 
(c) $v_\mathrm{rms}/\overline{U}_\mathrm{max}$, and 
(d) $w_\mathrm{rms}/\overline{U}_\mathrm{max}$.}
\label{rms_ave}
\end{figure}
%------------------------------------------------------------------------------

Next, to investigate turbulence spread, 
Fig. \ref{rms_ave} shows area-averaged turbulence distributions in the $y$-$z$ plane. 
$k_\mathrm{rms}$ is the time-averaged turbulence kinetic energy, 
and $u_\mathrm{rms}$, $v_\mathrm{rms}$, and $w_\mathrm{rms}$ are the turbulence intensities 
of velocity fluctuations in the $x$-, $y$-, and $z$-directions, respectively. 
All turbulence distributions are similar, 
but the effect of vortex structures appears well in the $k_\mathrm{rms}$ distribution. 
In the $k_\mathrm{rms}$ distribution, when $AR = 1.0$, 
the turbulence is maximum near $x/h = 2$ immediately after the 45-degree axis switching. 
Around $x/h = 6$, the turbulence decreases because the hairpin part of the vortex ring collapses. 
In the case of $AR = 1.5$, the turbulence shows a maximum near $x/h = 2$, 
which is the same as $AR = 1.0$. 
Around $x/h = 3$ downstream, the turbulence is maximum because a 90-degree axis switching occurs. 
Overall, the turbulence at $AR = 2.0$ is higher than $AR = 1.0$ and 1.5, 
and mixing is most likely to be promoted. 
This is because vortices spread over a wide area due to the intensive interference 
between the vortex rings on the upstream and downstream sides. 
In addition, the turbulence has a maximum near $x/h = 3$ 
where the 90-degree axis switching occurs. 
For $AR = 1.0$ and 1.5, the turbulence attenuates downstream, 
but for $AR = 2.0$, the turbulence increases downstream from $x/h = 12$. 
Even with $AR = 2.0$, the turbulence attenuates near the jet center. 
However, the turbulence is generated over a wide area as vortices move 
away from the jet center, resulting in an average increase in turbulence.

%------------------------------------------------------------------------------
% Figure 29
%------------------------------------------------------------------------------
\begin{figure}[!t]
\begin{minipage}{0.48\linewidth}
\centering
\includegraphics[trim=0mm 0mm 0mm 0mm, clip, width=80mm]{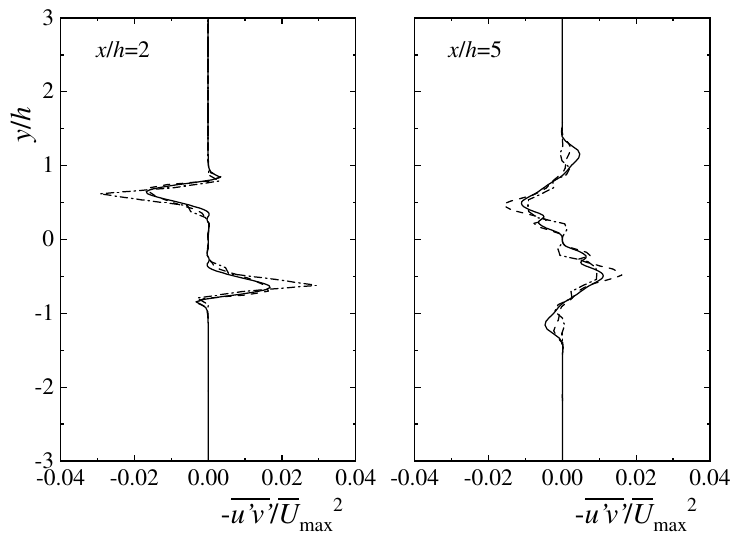} \\
\vspace*{-0.5\baselineskip}
(a)
\end{minipage}
\hspace{0.02\linewidth}
\begin{minipage}{0.48\linewidth}
\centering
\includegraphics[trim=0mm 0mm 0mm 0mm, clip, width=80mm]{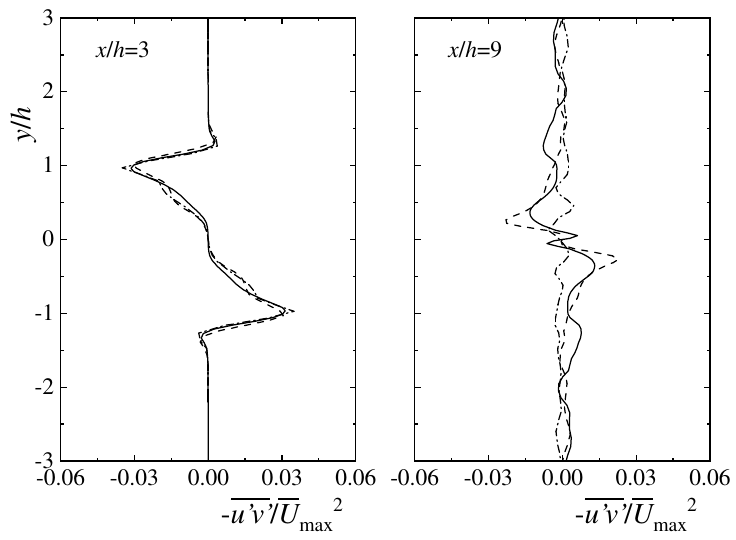} \\
\vspace*{-0.5\baselineskip}
(b)
\end{minipage}
\caption{Distributions of Reynolds shear stress at $z/h = 0$: 
- $\cdot$ -, grid1; - - -, grid2; \mbox{---, grid3}: 
(a) $AR = 1$ and (b) $AR = 2$.}
\label{correlation_grid}
\end{figure}
%------------------------------------------------------------------------------

%------------------------------------------------------------------------------
% Figure 30
%------------------------------------------------------------------------------
\begin{figure}[!t]
\begin{minipage}{0.48\linewidth}
\centering
\includegraphics[trim=0mm 0mm 0mm 0mm, clip, width=80mm]{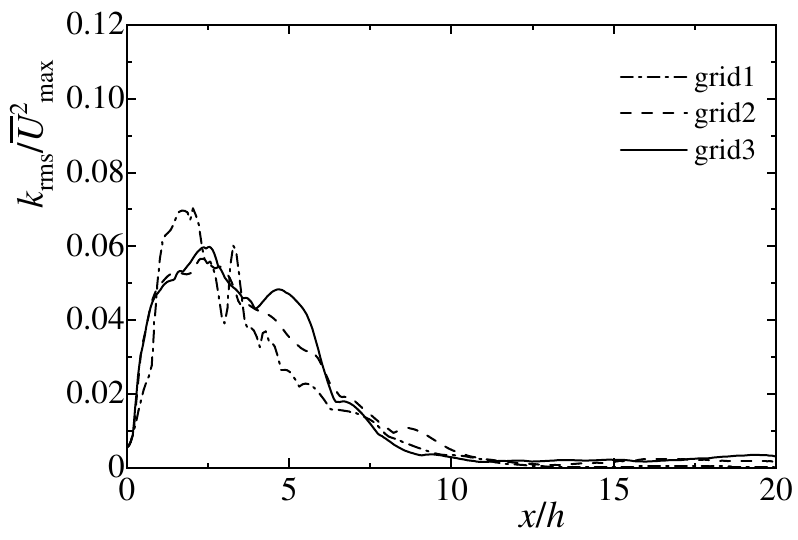} \\
\vspace*{-0.5\baselineskip}
(a)
\end{minipage}
\hspace{0.02\linewidth}
\begin{minipage}{0.48\linewidth}
\centering
\includegraphics[trim=0mm 0mm 0mm 0mm, clip, width=80mm]{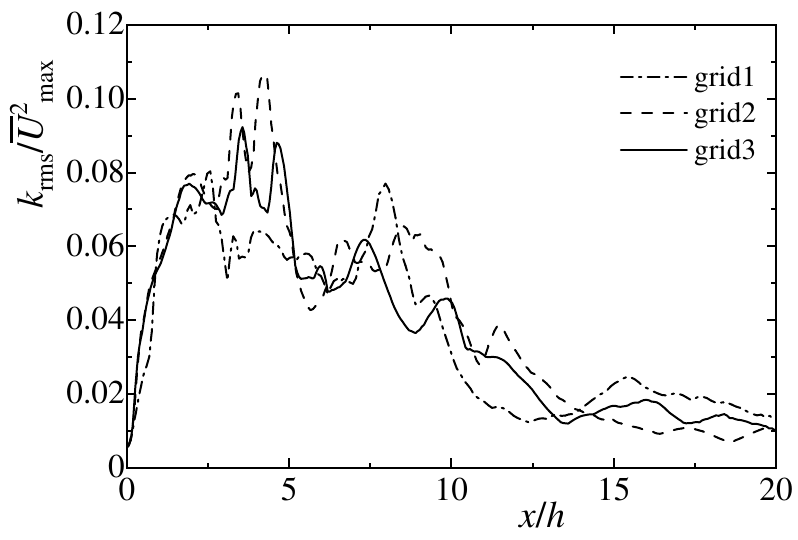} \\
\vspace*{-0.5\baselineskip}
(b)
\end{minipage}
\caption{Distributions of maximum turbulence kinetic energy: 
- $\cdot$ -, grid1; - - -, grid2; \mbox{---, grid3}: 
(a) $AR = 1$ and (b) $AR = 2$.}
\label{kmax_grid}
\end{figure}
%------------------------------------------------------------------------------

%------------------------------------------------------------------------------
% Figure 31
%------------------------------------------------------------------------------
\begin{figure}[!t]
\begin{minipage}{0.48\linewidth}
\centering
\includegraphics[trim=0mm 0mm 0mm 0mm, clip, width=80mm]{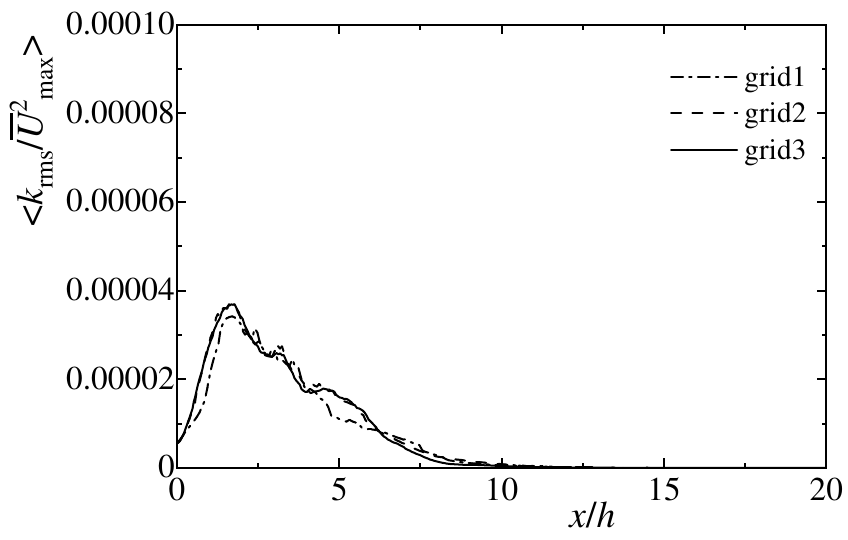} \\
\vspace*{-0.5\baselineskip}
(a)
\end{minipage}
\hspace{0.02\linewidth}
\begin{minipage}{0.48\linewidth}
\centering
\includegraphics[trim=0mm 0mm 0mm 0mm, clip, width=80mm]{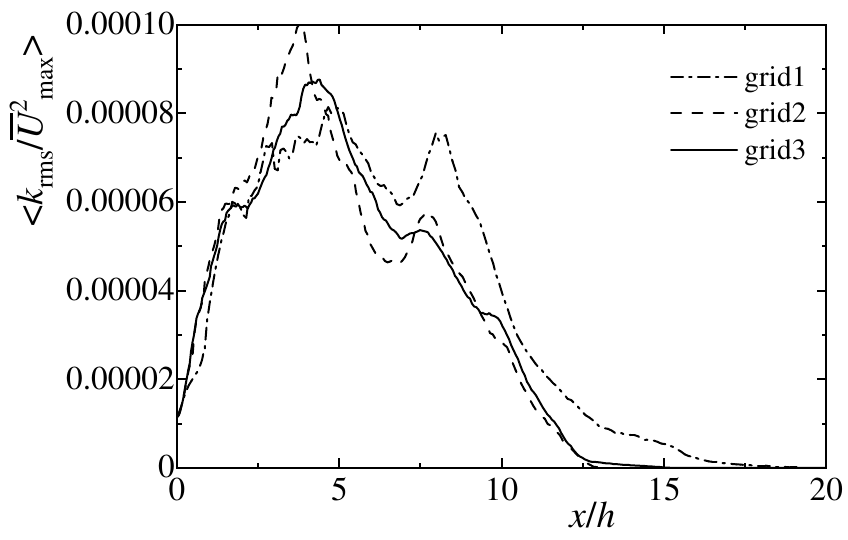} \\
\vspace*{-0.5\baselineskip}
(b)
\end{minipage}
\caption{Distributions of average turbulence kinetic energy: 
- $\cdot$ -, grid1; - - -, grid2; \mbox{---, grid3}: 
(a) $AR = 1$ and (b) $AR = 2$.}
\label{kave_grid}
\end{figure}
%------------------------------------------------------------------------------

Finally, we consider the grid dependency on turbulence properties. 
Figure \ref{correlation_grid} shows the distribution of Reynolds stress 
at the position where the $u_\mathrm{rms}$ distribution shown in Fig. \ref{urmsc} is maximum. 
At $x/h = 2$ for $AR = 1.0$ and $x/h = 5$ for $AR = 2.0$, 
the results of grid2 and grid3 are consistent, 
but the results of grid1 and other grids differ. 
At $x/h = 5$ for $AR = 1.0$, there is a difference among the three grids. 
At $x/h = 9$ for $AR = 2.0$, a difference between the results of grid2 and grid3 
near the jet center occurs. 
Figures \ref{kmax_grid} and \ref{kave_grid} show the maximum and average values 
of the turbulence kinetic energies, respectively. 
To confirm the grid dependency on the turbulence characteristics 
around the high-velocity area, 
we calculated the average value of the turbulence kinetic energy 
in the region of velocities higher than $\overline{U}_\mathrm{max}/2$. 
The average value indicates the turbulence kinetic energy near the jet center. 
The maximum value decreases because the jet diffuses downstream. 
In addition, the average value also decreases near the jet center. 
On the other hand, as shown in Fig. \ref{rms_ave} (a), 
the turbulence kinetic energy averaged over the $y$-$z$ plane increased downstream. 
These two results indicate that the mixing region spreads to the surroundings. 
Although the results vary depending on the number of grid points, 
the results using grid2 and grid3 for $AR = 1.0$ are relatively consistent. 
For $AR = 2.0$, there is some difference among the grids in the average value distribution, 
but the number of grid points causes a difference in the maximum value distribution. 
In the case of $AR = 2.0$, turbulence characteristics are likely to be affected by grids 
because vortex structures diffuse over a wide area. 
Although there are quantitative differences, the qualitative trend is the same. 
Therefore, we believe that the influence of grid resolution is small 
in discussing the phenomenon of axis switching.

%##############################################################################
\section{Conclusion}
%##############################################################################

This study performed a numerical analysis of the axis switching 
of a pulsating jet ejected from a rectangular nozzle at a low Reynolds number. 
We clarified the formation of vortex structures in the jet, 
the vortex deformation process, the interaction between vortices, 
and the generated local turbulence. 
The findings obtained are summarized below.

%------------------------------------------------------------------------------
\begin{enumerate}[(1)]
\item 
At all aspect ratios, 
a rectangular vortex ring similar to the shape of the nozzle cross-section is periodically shed downstream, 
and the side of the vortex ring deforms into a hairpin shape downstream. 
A vortex pair is generated inside the vortex ring downstream of the nozzle corner, 
and further downstream, the vortex pair extending in the streamwise direction forms a rib structure. 
When the aspect ratio is $AR = 1.0$, the vortex pair consists of symmetrical vortices, 
while as $AR$ increases, the asymmetry of the vortex pair enlarges. 
At $AR = 1.0$, regeneration of a vortex ring occurs downstream due to vortex pairs 
existing near the jet center. 
For $AR>1.0$, the hairpin part of the vortex ring on the long side of the nozzle develops 
in the direction away from the jet center compared to the short side at the nozzle. 
As $AR$ increases, intensive interaction between the vortex rings 
on the upstream and downstream sides occurs periodically. 
For $AR = 2.0$, interference between the vortex rings occurs, 
and alternately on the long and short sides of the nozzle, 
an upstream vortex ring overtakes a downstream vortex ring.

\item 
Regardless of $AR$, 
a vortex pair occurs at the corner of the vortex ring downstream 
near the nozzle exit, distorting the vortex ring. 
As a result, the positions of the side and corner of the vortex ring exchange, 
resulting in a 45-degree axis switching. 
At that time, turbulence increases at the side and corner of the vortex ring. 
For $AR > 1.0$, further downstream, 
the hairpin part of the vortex ring on the long side of the nozzle develops 
away from the jet center compared to the short side of the nozzle. 
Then, the hairpin part transports the high-velocity fluid near the jet center to the outside, 
causing a 90-degree axis switching. 
As a result, high turbulence occurs over a wide area, 
strengthening the mixing action. 
As $AR$ increases, the position where the axis switching occurs moves downstream.

\item 
When $AR > 1.0$, as $AR$ increases, 
intensive interference between the vortex rings on the upstream and downstream sides 
diffuses the vortices over a wide area, 
causing the jet to spread in the major and minor axis directions downstream. 
Then, as turbulence due to the diffused vortices widely occurs, 
the mixing effect is further strengthened.
\end{enumerate}
%------------------------------------------------------------------------------

In this study, we clarified the occurrence mechanism of axis switching at a low Reynolds number, 
changing the nozzle aspect ratio. 
In the future, we will investigate the effect of excitation frequency 
on axis switching and mixing.

%##############################################################################
%\section*{Acknowledgment}
%##############################################################################

\vspace*{1.0\baselineskip}
\noindent
{\bf Acknowledgements.}
The numerical results in this research were obtained 
using supercomputing resources at Cyberscience Center, Tohoku University. 
This research did not receive any specific grant from funding agencies 
in the public, commercial, or not-for-profit sectors. 
We would like to express our gratitude to Associate Professor Yosuke Suenaga 
of Iwate University for his support of our laboratory. 
%We thank the anonymous reviewers whose comments helped us 
%to improve the clarity of the paper. 
The authors wish to acknowledge the time and effort of everyone involved in this study.

\vspace*{1.0\baselineskip}
\noindent
{\bf Declaration of interests.}
The authors report no conflicts of interest.

\vspace*{1.0\baselineskip}
\noindent
{\bf Author ORCID.} \\
H. Yanaoka \url{https://orcid.org/0000-0002-4875-8174}.

\vspace*{1.0\baselineskip}
\noindent
{\bf Author Contributions}: 
H. Yanaoka considered the content and policy of this research 
and constructed the numerical method and calculation codes. 
Y. Hatakeyama performed the simulations. 
H. Yanaoka and Y. Hatakeyama contributed equally to analyzing data, reaching conclusions, 
and writing the paper.

%##############################################################################
%\section*{References}
%##############################################################################

\bibliographystyle{arXiv_elsarticle-harv}
\bibliography{arXiv2023_hatakeyama_bibfile}

\begin{thebibliography}{39}
\expandafter\ifx\csname natexlab\endcsname\relax\def\natexlab#1{#1}\fi
\providecommand{\url}[1]{\texttt{#1}}
\providecommand{\href}[2]{#2}
\providecommand{\path}[1]{#1}
\providecommand{\DOIprefix}{doi:}
\providecommand{\ArXivprefix}{arXiv:}
\providecommand{\URLprefix}{URL: }
\providecommand{\Pubmedprefix}{pmid:}
\providecommand{\doi}[1]{\href{http://dx.doi.org/#1}{\path{#1}}}
\providecommand{\Pubmed}[1]{\href{pmid:#1}{\path{#1}}}
\providecommand{\bibinfo}[2]{#2}
\ifx\xfnm\relax \def\xfnm[#1]{\unskip,\space#1}\fi
%Type = Article
\bibitem[{Amsden and Harlow(1970)}]{Amsden&Harlow_1970}
\bibinfo{author}{Amsden, A.A.}, \bibinfo{author}{Harlow, F.H.},
  \bibinfo{year}{1970}.
\newblock \bibinfo{title}{A simplified {MAC} technique for incompressible fluid
  flow calculations}.
\newblock \bibinfo{journal}{J. Comput. Phys.} \bibinfo{volume}{6},
  \bibinfo{pages}{322--325}.
\newblock \DOIprefix\doi{https://doi.org/10.1016/0021-9991(70)90029-x}.
%Type = Article
\bibitem[{Azad et~al.(2012)Azad, Quinn and Groulx}]{Azad_et_al_2012}
\bibinfo{author}{Azad, M.}, \bibinfo{author}{Quinn, W.R.},
  \bibinfo{author}{Groulx, D.}, \bibinfo{year}{2012}.
\newblock \bibinfo{title}{Mixing in turbulent free jets issuing from isosceles
  triangular orifices with different apex angles}.
\newblock \bibinfo{journal}{Exp. Therm. Fluid Sci.} \bibinfo{volume}{39},
  \bibinfo{pages}{237--251}.
\newblock \DOIprefix\doi{https://doi.org/10.1016/j.expthermflusci.2012.01.028}.
%Type = Article
\bibitem[{Chen and Yu(2014)}]{Chen&Yu_2014}
\bibinfo{author}{Chen, N.}, \bibinfo{author}{Yu, H.}, \bibinfo{year}{2014}.
\newblock \bibinfo{title}{Mechanism of axis switching in low aspect-ratio
  rectangular jets}.
\newblock \bibinfo{journal}{Comput. Math. with Appl.} \bibinfo{volume}{67},
  \bibinfo{pages}{437--444}.
\newblock \DOIprefix\doi{https://doi.org/10.1016/j.camwa.2013.03.018}.
%Type = Article
\bibitem[{{da Silva} and M\'{e}tais(2002)}]{da_Silva&Metais_2002}
\bibinfo{author}{{da Silva}, C.B.}, \bibinfo{author}{M\'{e}tais, O.},
  \bibinfo{year}{2002}.
\newblock \bibinfo{title}{Vortex control of bifurcating jets: a numerical
  study}.
\newblock \bibinfo{journal}{Phys. Fluids} \bibinfo{volume}{14},
  \bibinfo{pages}{3798--3819}.
\newblock \DOIprefix\doi{https://doi.org/10.1063/1.1506922}.
%Type = Article
\bibitem[{Danaila and Boersma(2002)}]{Danaila&Boersma_2000}
\bibinfo{author}{Danaila, I.}, \bibinfo{author}{Boersma, B.J.},
  \bibinfo{year}{2002}.
\newblock \bibinfo{title}{Direct numerical simulation of bifurcating jets}.
\newblock \bibinfo{journal}{Phys. Fluids} \bibinfo{volume}{12},
  \bibinfo{pages}{1255--1257}.
\newblock \DOIprefix\doi{https://doi.org/10.1063/1.870377}.
%Type = Article
\bibitem[{Gohil et~al.(2012)Gohil, Saha and Muralidhar}]{Gohil_et_al_2012}
\bibinfo{author}{Gohil, T.B.}, \bibinfo{author}{Saha, A.K.},
  \bibinfo{author}{Muralidhar, K.}, \bibinfo{year}{2012}.
\newblock \bibinfo{title}{Numerical study of instability mechanisms in a
  circular jet at low {R}eynolds numbers}.
\newblock \bibinfo{journal}{Comput. Fluids} \bibinfo{volume}{64},
  \bibinfo{pages}{1--18}.
\newblock \DOIprefix\doi{https://doi.org/10.1016/j.compfluid.2012.04.016}.
%Type = Article
\bibitem[{Gohil et~al.(2015)Gohil, Saha and Muralidhar}]{Gohil_et_al_2015}
\bibinfo{author}{Gohil, T.B.}, \bibinfo{author}{Saha, A.K.},
  \bibinfo{author}{Muralidhar, K.}, \bibinfo{year}{2015}.
\newblock \bibinfo{title}{Direct numerical simulation of free and forced square
  jets}.
\newblock \bibinfo{journal}{Int. J. Heat Fluid Flow} \bibinfo{volume}{52},
  \bibinfo{pages}{169--184}.
\newblock
  \DOIprefix\doi{https://doi.org/10.1016/j.ijheatfluidflow.2015.01.003}.
%Type = Article
\bibitem[{Grinstein(2001)}]{Grinstein_2001}
\bibinfo{author}{Grinstein, F.F.}, \bibinfo{year}{2001}.
\newblock \bibinfo{title}{Vortex dynamics and entrainment in rectangular free
  jets}.
\newblock \bibinfo{journal}{J. Fluid Mech.} \bibinfo{volume}{437},
  \bibinfo{pages}{69--101}.
\newblock \DOIprefix\doi{https://doi.org/10.1017/s0022112001004141}.
%Type = Article
\bibitem[{Grinstein et~al.(1995)Grinstein, Gutmark and
  Parr}]{Grinstein_et_al_1995}
\bibinfo{author}{Grinstein, F.F.}, \bibinfo{author}{Gutmark, E.},
  \bibinfo{author}{Parr, T.}, \bibinfo{year}{1995}.
\newblock \bibinfo{title}{Near field dynamics of subsonic free square jets. {A}
  computational and experimental study}.
\newblock \bibinfo{journal}{Phys. Fluids} \bibinfo{volume}{7},
  \bibinfo{pages}{1483--1497}.
\newblock \DOIprefix\doi{https://doi.org/10.1063/1.868534}.
%Type = Article
\bibitem[{Gutmark et~al.(1989)Gutmark, Schadow, Parr, Hanson-Parr and
  Wilson}]{Gutmark_et_al_1989}
\bibinfo{author}{Gutmark, E.}, \bibinfo{author}{Schadow, K.C.},
  \bibinfo{author}{Parr, T.P.}, \bibinfo{author}{Hanson-Parr, D.M.},
  \bibinfo{author}{Wilson, K.J.}, \bibinfo{year}{1989}.
\newblock \bibinfo{title}{Noncircular jets in combustion systems}.
\newblock \bibinfo{journal}{Exp. Fluids} \bibinfo{volume}{7},
  \bibinfo{pages}{248--258}.
\newblock \DOIprefix\doi{https://doi.org/10.1007/bf00198004}.
%Type = Article
\bibitem[{Hama(1962)}]{Hama_1962}
\bibinfo{author}{Hama, F.R.}, \bibinfo{year}{1962}.
\newblock \bibinfo{title}{Progressive deformation of a curved vortex filament
  by its own induction}.
\newblock \bibinfo{journal}{Phys. Fluids} \bibinfo{volume}{5},
  \bibinfo{pages}{1156--1162}.
\newblock \DOIprefix\doi{https://doi.org/10.1063/1.1706500}.
%Type = Article
\bibitem[{Ho and Gutmark(1987)}]{Ho&Gutmark_1987}
\bibinfo{author}{Ho, C.M.}, \bibinfo{author}{Gutmark, E.},
  \bibinfo{year}{1987}.
\newblock \bibinfo{title}{Vortex induction and mass entrainment in a
  small-aspect-ratio elliptic jet}.
\newblock \bibinfo{journal}{J. Fluid Mech.} \bibinfo{volume}{179},
  \bibinfo{pages}{383--405}.
\newblock \DOIprefix\doi{https://doi.org/10.1017/s0022112087001587}.
%Type = Article
\bibitem[{Husain and Hussain(1991)}]{Husain&Hussain_1991}
\bibinfo{author}{Husain, H.S.}, \bibinfo{author}{Hussain, F.},
  \bibinfo{year}{1991}.
\newblock \bibinfo{title}{Elliptic jets. {P}art 2. {D}ynamics of coherent
  structures: pairing}.
\newblock \bibinfo{journal}{J. Fluid Mech.} \bibinfo{volume}{233},
  \bibinfo{pages}{439--482}.
\newblock \DOIprefix\doi{https://doi.org/10.1017/S0022112091000551}.
%Type = Incollection
\bibitem[{Hussain and Zaman(1981)}]{Hussain&Zaman_1981}
\bibinfo{author}{Hussain, A.K.M.F.}, \bibinfo{author}{Zaman, K.B.M.Q.},
  \bibinfo{year}{1981}.
\newblock \bibinfo{title}{The preferred-mode coherent structure in the near
  field of an axisymmetric jet with and without excitation}, in:
  \bibinfo{booktitle}{Unsteady Turbulent Shear Flows}.
  \bibinfo{publisher}{Springer Berlin Heidelberg}, pp.
  \bibinfo{pages}{390--401}.
\newblock \DOIprefix\doi{https://doi.org/10.1007/978-3-642-81732-8_33}.
%Type = Article
\bibitem[{Hussain and Husain(1989)}]{Hussain&Husain_1989}
\bibinfo{author}{Hussain, F.}, \bibinfo{author}{Husain, H.S.},
  \bibinfo{year}{1989}.
\newblock \bibinfo{title}{Elliptic jets. {P}art 1. {C}haracteristics of
  unexcited and excited jets}.
\newblock \bibinfo{journal}{J. Fluid Mech.} \bibinfo{volume}{208},
  \bibinfo{pages}{257--320}.
\newblock \DOIprefix\doi{https://doi.org/10.1017/s0022112089002843}.
%Type = Article
\bibitem[{Jiang et~al.(2007)Jiang, Guo, Chan and Lin}]{Jiang_et_al_2007}
\bibinfo{author}{Jiang, P.}, \bibinfo{author}{Guo, Y.C.},
  \bibinfo{author}{Chan, C.K.}, \bibinfo{author}{Lin, W.Y.},
  \bibinfo{year}{2007}.
\newblock \bibinfo{title}{Frequency characteristics of coherent structures and
  their excitations in small aspect-ratio rectangular jets using large eddy
  simulation}.
\newblock \bibinfo{journal}{Comput. Fluids} \bibinfo{volume}{36},
  \bibinfo{pages}{611--621}.
\newblock \DOIprefix\doi{https://doi.org/10.1016/j.compfluid.2006.05.001}.
%Type = Article
\bibitem[{Koshigoe et~al.(1989)Koshigoe, Gutmark and
  Schadow}]{Koshigoe_et_al_1989}
\bibinfo{author}{Koshigoe, S.}, \bibinfo{author}{Gutmark, E.},
  \bibinfo{author}{Schadow, K.C.}, \bibinfo{year}{1989}.
\newblock \bibinfo{title}{Initial development of noncircular jets leading to
  axis switching}.
\newblock \bibinfo{journal}{AIAA J.} \bibinfo{volume}{27},
  \bibinfo{pages}{411--419}.
\newblock \DOIprefix\doi{https://doi.org/10.2514/3.10128}.
%Type = Article
\bibitem[{Krothapalli et~al.(1981)Krothapalli, Baganoff and
  Karamcheti}]{Krothapalli_et_al_1981}
\bibinfo{author}{Krothapalli, A.}, \bibinfo{author}{Baganoff, D.},
  \bibinfo{author}{Karamcheti, K.}, \bibinfo{year}{1981}.
\newblock \bibinfo{title}{On the mixing of a rectangular jet}.
\newblock \bibinfo{journal}{J. Fluid Mech.} \bibinfo{volume}{107},
  \bibinfo{pages}{201--220}.
\newblock \DOIprefix\doi{https://doi.org/10.1017/s0022112081001730}.
%Type = Article
\bibitem[{Michalke and Hermann(1982)}]{Michalke&Hermann_1982}
\bibinfo{author}{Michalke, A.}, \bibinfo{author}{Hermann, G.},
  \bibinfo{year}{1982}.
\newblock \bibinfo{title}{On the inviscid instability of a circular jets with
  external flow}.
\newblock \bibinfo{journal}{J. Fluid Mech.} \bibinfo{volume}{114},
  \bibinfo{pages}{343--359}.
\newblock \DOIprefix\doi{https://doi.org/10.1017/S0022112082000196}.
%Type = Article
\bibitem[{Miller et~al.(1995)Miller, Madnia and Givi}]{Miller_et_al_1995}
\bibinfo{author}{Miller, R.S.}, \bibinfo{author}{Madnia, C.K.},
  \bibinfo{author}{Givi, P.}, \bibinfo{year}{1995}.
\newblock \bibinfo{title}{Numerical simulation of non-circular jets}.
\newblock \bibinfo{journal}{Comput. Fluids} \bibinfo{volume}{24},
  \bibinfo{pages}{1--25}.
\newblock \DOIprefix\doi{https://doi.org/10.1016/0045-7930(94)00019-U}.
%Type = Article
\bibitem[{Quinn(1992)}]{Quinn_1992}
\bibinfo{author}{Quinn, W.R.}, \bibinfo{year}{1992}.
\newblock \bibinfo{title}{Turbulent free jet flows issuing from sharp-edged
  rectangular slots: {T}he influence of slot aspect ratio}.
\newblock \bibinfo{journal}{Exp. Therm. Fluid Sci.} \bibinfo{volume}{5},
  \bibinfo{pages}{203--215}.
\newblock \DOIprefix\doi{https://doi.org/10.1016/0894-1777(92)90007-r}.
%Type = Article
\bibitem[{Quinn(2007)}]{Quinn_2007}
\bibinfo{author}{Quinn, W.R.}, \bibinfo{year}{2007}.
\newblock \bibinfo{title}{Experimental study of the near field and transition
  region of a free jet issuing from a sharp-edged elliptic orifice plate}.
\newblock \bibinfo{journal}{Eur. J. Mech. B Fluids} \bibinfo{volume}{26},
  \bibinfo{pages}{583--614}.
\newblock \DOIprefix\doi{https://doi.org/10.1016/j.euromechflu.2006.10.005}.
%Type = Article
\bibitem[{Rembold et~al.(2002)Rembold, Adams and Kleiser}]{Rembold_et_al_2002}
\bibinfo{author}{Rembold, B.}, \bibinfo{author}{Adams, N.A.},
  \bibinfo{author}{Kleiser, L.}, \bibinfo{year}{2002}.
\newblock \bibinfo{title}{Direct numerical simulation of a transitional
  rectangular jet}.
\newblock \bibinfo{journal}{Int. J. Heat Fluid Flow} \bibinfo{volume}{23},
  \bibinfo{pages}{547--553}.
\newblock \DOIprefix\doi{https://doi.org/10.1016/s0142-727x(02)00150-9}.
%Type = Article
\bibitem[{Reynolds et~al.(2003)Reynolds, Parekh, Juvet and
  Lee}]{Reynolds_et_al_2003}
\bibinfo{author}{Reynolds, W.C.}, \bibinfo{author}{Parekh, D.E.},
  \bibinfo{author}{Juvet, P.J.D.}, \bibinfo{author}{Lee, M.J.D.},
  \bibinfo{year}{2003}.
\newblock \bibinfo{title}{Bifurcating and blooming jets}.
\newblock \bibinfo{journal}{Annu. Rev. Fluid Mech.} \bibinfo{volume}{35},
  \bibinfo{pages}{295--315}.
\newblock
  \DOIprefix\doi{https://doi.org/10.1146/annurev.fluid.35.101101.161128}.
%Type = Article
\bibitem[{Sfeir(1976)}]{Sfeir_1976}
\bibinfo{author}{Sfeir, A.A.}, \bibinfo{year}{1976}.
\newblock \bibinfo{title}{The velocity and temperature fields of rectangular
  jets}.
\newblock \bibinfo{journal}{Int. J. Heat Mass Transf.} \bibinfo{volume}{19},
  \bibinfo{pages}{1289--1297}.
\newblock \DOIprefix\doi{https://doi.org/10.1016/0017-9310(76)90081-8}.
%Type = Article
\bibitem[{Sforza et~al.(1966)Sforza, Steiger and
  Trentacoste}]{Sforza_et_al_1966}
\bibinfo{author}{Sforza, P.M.}, \bibinfo{author}{Steiger, M.H.},
  \bibinfo{author}{Trentacoste, N.}, \bibinfo{year}{1966}.
\newblock \bibinfo{title}{Studies on three-dimensional viscous jets}.
\newblock \bibinfo{journal}{AIAA J.} \bibinfo{volume}{4},
  \bibinfo{pages}{800--806}.
\newblock \DOIprefix\doi{https://doi.org/10.2514/3.3549}.
%Type = Article
\bibitem[{Straccia and Farnsworth(2021)}]{Straccia&Farnsworth_2021}
\bibinfo{author}{Straccia, J.C.}, \bibinfo{author}{Farnsworth, J.A.N.},
  \bibinfo{year}{2021}.
\newblock \bibinfo{title}{Axis switching in low to moderate aspect ratio
  rectangular orifice synthetic jets}.
\newblock \bibinfo{journal}{Phys. Rev. Fluid} \bibinfo{volume}{6},
  \bibinfo{pages}{054702}.
\newblock \DOIprefix\doi{https://doi.org/10.1103/PhysRevFluids.6.054702}.
%Type = Article
\bibitem[{Suto et~al.(2002)Suto, Matsubara and Kobayashi}]{Suto_et_al_2002}
\bibinfo{author}{Suto, H.}, \bibinfo{author}{Matsubara, K.},
  \bibinfo{author}{Kobayashi, M.}, \bibinfo{year}{2002}.
\newblock \bibinfo{title}{Direct numerical simulation of a spatially developing
  round jet ({V}alidity of numerical results and mechanisms of {R}eynolds
  stress transport}.
\newblock \bibinfo{journal}{JSME Ser. B} \bibinfo{volume}{68},
  \bibinfo{pages}{777--784}.
\newblock \DOIprefix\doi{https://doi.org/10.1299/kikaib.68.777}.
  \bibinfo{note}{(in Japanese)}.
%Type = Article
\bibitem[{Toyoda et~al.(1992)Toyoda, Shirahama and Kotani}]{Toyoda_et_al_1992}
\bibinfo{author}{Toyoda, K.}, \bibinfo{author}{Shirahama, Y.},
  \bibinfo{author}{Kotani, K.}, \bibinfo{year}{1992}.
\newblock \bibinfo{title}{Manipulation of vortical structures in noncircular
  jets}.
\newblock \bibinfo{journal}{JSME Ser. B} \bibinfo{volume}{58},
  \bibinfo{pages}{7--13}.
\newblock \DOIprefix\doi{https://doi.org/10.1299/kikaib.58.7}.
  \bibinfo{note}{(in Japanese)}.
%Type = Article
\bibitem[{Tsuchiya and Horikoshi(1986)}]{Tsuchiya&Horikoshi_1986}
\bibinfo{author}{Tsuchiya, Y.}, \bibinfo{author}{Horikoshi, C.},
  \bibinfo{year}{1986}.
\newblock \bibinfo{title}{On the spread of rectangular jets}.
\newblock \bibinfo{journal}{Exp. Fluids} \bibinfo{volume}{4},
  \bibinfo{pages}{197--204}.
\newblock \DOIprefix\doi{https://doi.org/10.1007/bf00717815}.
%Type = Article
\bibitem[{Tyliszczak(2018)}]{Tyliszczak_2018}
\bibinfo{author}{Tyliszczak, A.}, \bibinfo{year}{2018}.
\newblock \bibinfo{title}{Parametric study of multi-armed jets}.
\newblock \bibinfo{journal}{Int. J. Heat Fluid Flow} \bibinfo{volume}{73},
  \bibinfo{pages}{82--100}.
\newblock
  \DOIprefix\doi{https://doi.org/10.1016/j.ijheatfluidflow.2018.07.002}.
%Type = Article
\bibitem[{Tyliszczak and Wawrzak(2022)}]{Tyliszczak&Wawrzak_2022}
\bibinfo{author}{Tyliszczak, A.}, \bibinfo{author}{Wawrzak, A.},
  \bibinfo{year}{2022}.
\newblock \bibinfo{title}{A numerical study of a lifted h$_2$/n$_2$ flame
  excited by an axial and flapping forcing}.
\newblock \bibinfo{journal}{Sci. Rep.} \bibinfo{volume}{12},
  \bibinfo{pages}{2753}.
\newblock \DOIprefix\doi{https://doi.org/10.1038/s41598-022-06740-4}.
%Type = Article
\bibitem[{Yanaoka(2023)}]{Yanaoka_2023}
\bibinfo{author}{Yanaoka, H.}, \bibinfo{year}{2023}.
\newblock \bibinfo{title}{Influences of conservative and non-conservative
  {L}orentz forces on energy conservation properties for incompressible
  magnetohydrodynamic flows}.
\newblock \bibinfo{journal}{J. Comput. Phys.} \bibinfo{volume}{491},
  \bibinfo{pages}{112372 (36 pages)}.
\newblock \DOIprefix\doi{https://doi.org/10.1016/j.jcp.2023.112372}.
%Type = Article
\bibitem[{Yanaoka and Inafune(2023)}]{Yanaoka&Inafune_2023}
\bibinfo{author}{Yanaoka, H.}, \bibinfo{author}{Inafune, R.},
  \bibinfo{year}{2023}.
\newblock \bibinfo{title}{Frequency response of three-dimensional natural
  convection of nanofluids under microgravity environments with gravity
  modulation}.
\newblock \bibinfo{journal}{Numer. Heat Tr. A-Appl.} \bibinfo{volume}{83},
  \bibinfo{pages}{745--769}.
\newblock \DOIprefix\doi{https://doi.org/10.1080/10407782.2022.2161437}.
%Type = Article
\bibitem[{Yanaoka et~al.(2007a)Yanaoka, Inamura and
  Kawabe}]{Yanaoka_et_al_2007c}
\bibinfo{author}{Yanaoka, H.}, \bibinfo{author}{Inamura, T.},
  \bibinfo{author}{Kawabe, S.}, \bibinfo{year}{2007}a.
\newblock \bibinfo{title}{Turbulence and heat transfer of a hairpin vortex
  formed behind a cube in a laminar boundary layer}.
\newblock \bibinfo{journal}{Numer. Heat Tr., A-Appl.} \bibinfo{volume}{52},
  \bibinfo{pages}{973--990}.
\newblock \DOIprefix\doi{https://doi.org/10.1080/10407780701389590}.
%Type = Article
\bibitem[{Yanaoka et~al.(2007b)Yanaoka, Inamura, Suenaga and
  Kobayashi}]{Yanaoka_et_al_2007b}
\bibinfo{author}{Yanaoka, H.}, \bibinfo{author}{Inamura, T.},
  \bibinfo{author}{Suenaga, Y.}, \bibinfo{author}{Kobayashi, Y.},
  \bibinfo{year}{2007}b.
\newblock \bibinfo{title}{Numerical simulation of vortex structures and heat
  transfer behind a hill in a laminar boundary layer}.
\newblock \bibinfo{journal}{JSME, Ser. B} \bibinfo{volume}{73},
  \bibinfo{pages}{2357--2544}.
\newblock \DOIprefix\doi{https://doi.org/10.1299/kikaib.73.2537}.
  \bibinfo{note}{(in Japanese)}.
%Type = Article
\bibitem[{Yanaoka et~al.(2008)Yanaoka, Inamura, Suenaga and
  Kobayashi}]{Yanaoka_et_al_2008b}
\bibinfo{author}{Yanaoka, H.}, \bibinfo{author}{Inamura, T.},
  \bibinfo{author}{Suenaga, Y.}, \bibinfo{author}{Kobayashi, Y.},
  \bibinfo{year}{2008}.
\newblock \bibinfo{title}{Numerical simulation of vortex structures and heat
  transfer behind a hill in a laminar boundary layer}.
\newblock \bibinfo{journal}{Heat Trans. Asian Res.} \bibinfo{volume}{37},
  \bibinfo{pages}{398--411}.
\newblock \DOIprefix\doi{https://doi.org/10.1002/htj.20217}.
%Type = Article
\bibitem[{Zaman(1996)}]{Zaman_1996}
\bibinfo{author}{Zaman, K.B.M.Q.}, \bibinfo{year}{1996}.
\newblock \bibinfo{title}{Axis switching and spreading of an asymmetric jet:
  the role of coherent structure dynamics}.
\newblock \bibinfo{journal}{J. Fluid Mech.} \bibinfo{volume}{316},
  \bibinfo{pages}{1--27}.
\newblock \DOIprefix\doi{https://doi.org/10.1017/s0022112096000420}.
%Type = Article
\bibitem[{Zhang and Chua(2012)}]{Zhang&Chua_2012}
\bibinfo{author}{Zhang, Z.K.}, \bibinfo{author}{Chua, L.P.},
  \bibinfo{year}{2012}.
\newblock \bibinfo{title}{Mixing due to a heated elliptic air jet}.
\newblock \bibinfo{journal}{Int. J. Heat Mass Transf.} \bibinfo{volume}{55},
  \bibinfo{pages}{4566--4579}.
\newblock
  \DOIprefix\doi{https://doi.org/10.1016/j.ijheatmasstransfer.2012.04.001}.

\end{thebibliography}

\end{document}